\documentclass{iopart}
\usepackage{amssymb}
\usepackage{iopams}

\usepackage{graphicx}

\newtheorem{theorem}{Theorem}

\newtheorem{definition}[theorem]{Definition}

\begin{document}
\title{Hamiltonian motions of plane curves and formation of singularities and bubbles}
\author{B.G.Konopelchenko $^1$ and G.Ortenzi $^2$}
\address{$^1$ Dipartimento di Fisica, Universit\`{a} del Salento 
and INFN, Sezione di Lecce, 73100 Lecce, Italy }
\address{ $^2$ Dipartimento di Matematica Pura ed Applicazioni, 
Universit\`{a} di Milano Bicocca, 20125 Milano, Italy}
\ead{\mailto{Boris.Konopeltchenko@le.infn.it} and \mailto{giovanni.ortenzi@unimib.it}}
\begin{abstract}
A class of Hamiltonian deformations of plane curves is defined and studied. Hamiltonian deformations of conics and cubics are considered as illustrative examples. These deformations are described by systems of hydrodynamical type equations. It is shown that solutions of these systems describe processes of formation of singularities (cusps, nodes), bubbles, and change of genus of a curve.
\end{abstract}
\section{Introduction}
\par Dynamics of curves and interfaces is a key ingredient in various important phenomena in physics and theories in  mathematics (see e.g. \cite{Aetal,BCC,Lan,Ols}).
 Special class of deformations of curves described by integrable equations has attracted a particular interest. It has been studied in a number of papers (see \cite{BBEIM,BAZW,DS,DN,GP,Has,Kono,KMAM,KO,KK,K1,K2,KMWZ,Lamb,Lak,LTW,MAM,MWZ,
NSW,TBAZW,WZ} and references therein).\\
Approaches used in these papers differ, basically, by the way how the evolution of a curve is fixed. In papers \cite{DS,GP,Has,Lak,Lamb,NSW} the motion of the curve is defined by requirement that the time derivative of position vector of a curve is a linear superposition of tangent and (bi)normal vectors with a certain specification of coefficients in this superposition.  Deformations considered in papers \cite{BBEIM,DN,K1,K2,TBAZW} are defined, basically, by the corresponding Lax pair. Within the study of Laplacian growth problem \cite{BAZW,KMWZ,LTW,MWZ,TBAZW,WZ,MAM} the time evolution of the interface is characterized by the dynamics of the Schwartz function of a curve prescribed by the Darcy law. 
Semiclassical deformations of algebraic curves analyzed in \cite{Kono,KK,KMAM} are fixed by the existence of a specific generating function constructed $via$ a Lenard scheme. 
Finally, the coisotropic deformations introduced in \cite{KO} are defined by the requirement that the ideal of the deformed curve is a Poisson ideal.\par
In the present paper a novel class of deformations of plane curves is considered. For such deformations the function $f(p_1,p_2)$ which defines a curve $f(p_1,p_2)=0$ in the plane $(p_1,p_2)$ obeys an inhomogeneous Liouville equation with $p_1,p_2$ and deformation parameters $x_1,x_2,t$ playing the role of independent variables. These deformations are referred as the Hamiltonian deformations (motions) of a curve since the coefficients in the linear equation are completely fixed by the single function $H(p_1,p_2;x_1,x_2,t)$ playing a role of Hamiltonian for time $t$ dynamics. Equivalently Hamiltonian deformations can be defined as those for which an ideal $\mathcal{I}$ of a deformed curve is invariant under the action of the vector field $\partial_t + \nabla_H$. For algebraic plane curves Hamiltonian deformations represent a particular class of coisotropic deformations introduced in \cite{KO}.\par
Here we concentrate on the study of Hamiltonian dynamics of plane quadrics (conics) and cubics. This dynamics is described by integrable systems of hydrodynamical type for the coefficients of polynomials defining algebraic curves with the dKdV, dDS, dVN, (2+1)-dimensional one-layer Benney system and other equations among them. It is shown that particular solutions of these systems describe formation of singularities (cusps, nodes) and bubbles on the real plane.\par
The paper is organized as follows. General definition and interpretations are given in section \ref{sect-defdef}. Hamiltonian deformations of plane quadrics are studied in section \ref{sect-quadr}. As particular examples one has deformations of circle described by dispersionless Veselov-Novikov (dVN) equation and deformations of an ellipse given by a novel system of equations. Section \ref{sect-cub} is devoted to Hamiltonian deformations of plane cubics. An analysis of formation of singularities for cubics and genus transition is given. In section $5$ a connection between singular cubics and Burgers-Hopf equation is considered. In section $6$ we discuss briefly deformations of a quintic.
Possible extensions of the approach presented in the paper are noted in section $7$.
\section{Definition and interpretation}
\label{sect-defdef}
\par So let the plane curve $\Gamma$ be defined by the equation
\begin{equation}
 \label{genGamma}
f(p_1,p_2)=0
\end{equation}
 where $f$ is a function of local coordinates $p_1,p_2$ on the complex plane $\mathbb{C}^2$. To introduce deformations of the curve (\ref{genGamma}) we assume that $f$ depends also on the deformation parameters  $x_1,x_2,t$.
\begin{definition}
 If the function $f(p_1,p_2;x_1,x_2,t)$ obeys the equations
\begin{equation}
 \label{defdefcurve}
\partial_t f +\{f,H\} = \alpha(p_1,p_2;x_1,x_2,t)f
\end{equation}
with certain function $H( p_1,p_2;x_1,x_2,t)$  and a function $\alpha$ , where $\{f,H\}$ is the Poisson bracket 
\begin{equation}
\label{stdPoiss}
 \{f,H\}=\sum_{i=1}^2 \left( \partial_{x_i}f\partial_{p_i}H-\partial_{p_i}f\partial_{x_i}H \right),
\end{equation}
then it is said that the function $f$ as the function of $x_1$, $x_2$, and $t$ defines Hamiltonian deformations of the curve (\ref{genGamma}). 
\end{definition}
To justify this definition we first observe that equation (\ref{defdefcurve}) is nothing else than inhomogeneous Liouville equation which is well-known in classical mechanics. On $\Gamma $ the r.h.s. of equation (\ref{defdefcurve}) vanishes and, hence, 
\begin{equation}
\label{gendefLiouv}
 (\partial_t+\nabla_H) f |_\Gamma =0 
\end{equation}
where $\nabla_H$ is the Hamiltonian vector field in the space with coordinates $p_1,p_2,x_1,x_2$. Thus Hamiltonian deformations of the curve (\ref{genGamma}) generated by $H$ can be equivalently defined as those for which the function $f$ obeys the condition (\ref{gendefLiouv}).\par
Following the standard interpretation of the Liouville equation in the classical mechanics one can view equations (\ref{defdefcurve}) and (\ref{gendefLiouv}) as
\begin{equation}
 \frac{df}{dt} =\alpha f 
\end{equation}
and
\begin{equation}
\left. \frac{df}{dt} \right|_{\Gamma}=0
\end{equation}
where the total derivative $\frac{d}{dt}$ is calculated along the characteristics of equation (\ref{defdefcurve}) defined by Hamiltonian equations
\begin{equation}
\label{eqn-Ham}
\frac{d p_i}{dt} =-\partial_{x_i}H, \qquad \frac{d x_i}{dt} =\partial_{p_i}H, \quad i=1,2 .
\end{equation}
\par For a given irreducible curve $\Gamma$ ($f(p_1,p_2;x_1,x_2,t)=0$) a function $\alpha$ is not defined uniquely. Indeed, for any function $\beta(p_1,p_2;x_1,x_2,t)$  the function $\hat{f}=\beta f$ defines the same curve (\ref{genGamma}). Then, equation (\ref{defdefcurve}) takes the form
\begin{equation}
 \partial_t \hat{f}+\{\hat{f},H\}=\hat{\alpha} \hat{f}
\end{equation}
 where
\begin{equation}
\label{alpha-gauge}
\beta \hat{\alpha}=\alpha+\partial_t  \beta + \{ \beta,H\}
\end{equation}
while the condition (\ref{gendefLiouv}) remains unchanged. Thus, the freedom in form of $\alpha$ given by (\ref{alpha-gauge}) is just a gauge type freedom.
For given $\alpha$ the choice of $\beta$ as a solution of the equation
\begin{equation}
 \beta \alpha=-\partial_t  \beta -\{ \beta ,H\}
\end{equation}
gives $\hat{\alpha}=0$. So there exists a gauge at which equation (\ref{defdefcurve}) is homogeneous one. The drawback of such ``gauge'' is that in many cases equation (\ref{genGamma}) has no simple, standard form in this gauge.\par
Gauge freedom observed above becomes very natural in the formulation based on the notion of ideal of a curve. Ideal $\mathcal{I}=\langle f \rangle$ of an irreducible curve (\ref{genGamma}) consists of all functions of the form $\beta f $ (see e.g. \cite{Har}). In these terms Hamiltonian deformations of a curve can be characterized equivalently by
\begin{definition}
 Hamiltonian deformations of a curve generated by a function $H$ are those for which the ideal $\mathcal{I}$ of the deformed curve is closed under the action of the vector field $\partial_t + \nabla_H$, i.e.
\begin{equation}
\label{defdefideal}
 (\partial_t + \nabla_H) \mathcal{I} \subset \mathcal{I}.
\end{equation}
\end{definition}
\par Definition of the Hamiltonian deformation in this form clearly shows the irrelevance of the concrete form of function $\alpha$ in equation (\ref{defdefcurve}). Moreover such formulation reveals also the nonuniqueness in the form of $H$ for the same deformation. Indeed, for the family of Hamiltonians $\widetilde{H}=H+G$ where $G$ is arbitrary element of $\mathcal{I}$ (i.e. $g\cdot f$) one has equation (\ref{defdefcurve}) with the family of functions $\hat{\alpha}$ of the form $\widetilde{\alpha}=\alpha+\{f,g\}$. Hence, equations $\left.(\partial_t+\nabla_{\widetilde{H}})f\right|_\Gamma=0$ and $\left.(\partial_t+\nabla_{H})f\right|_\Gamma=0$ and corresponding equations (\ref{defdefideal}) coincide. This freedom ($H \to H+gf$ where $g(p_1,p_2;x_1,x_2,t)$ is an arbitrary function) in the choice of Hamiltonians is associated with different possible evaluations of $H$ on the curve $\Gamma$. In practice it corresponds to different possible resolutions of the concrete equation (\ref{genGamma}) with respect to one of the variables ($p_1$ or $p_2$). This freedom serves to choose simplest and most convenient form of Hamiltonian $H$.
\par For algebraic curves, i.e. for functions $f$ polynomial in $p_1,p_2$, Hamiltonian deformations of plane curves represent themselves a special subclass of coisotropic deformations of algebraic curves in three dimensional space considered in \cite{KO}.\par
In three dimensional space the curve $\Gamma$ is defined by the system of two equations
\begin{equation}
 f_1(p_1,p_2,p_3,x_1,x_2,x_3)=0, \qquad f_2(p_1,p_2,p_3,x_1,x_2,x_3)=0
\end{equation}
where $x_1,x_2,x_3$ are deformation parameters. Its coisotropic deformation is fixed by the condition \cite{KO} 
\begin{equation}
 \label{KO-cond}
\left.\{f_1,f_2\}\right|_\Gamma=0
\end{equation}
where $\{\ ,\ \}$ is the canonical Poisson bracket in $\mathbb{R}^6$. It is easy to see that with the choices 
$ f_1=f(p_1,p_2;x_1,x_2,x_3)$, and  $f_2=p_3+H(p_1,p_2;x_1,x_2,x_3)$ the condition (\ref{KO-cond}) is reduced to (\ref{gendefLiouv}) where $t=x_3$.\\
In general, Hamiltonian deformations of the plane curve are parameterized by three independent variables $x_1,x_2$, and $t$. If one variable is cyclic (say $x_2$, i.e. $\partial_{x_2}H=0$ and, hence, $p_2=const$) then deformations are described by $(1+1)$ dimensional equations.
\section{Deformations of plane quadrics}
\label{sect-quadr}
\par
\begin{figure}[h!]
\centering
\begin{tabular}{ccccc}
\includegraphics[width=4cm, height=5cm]{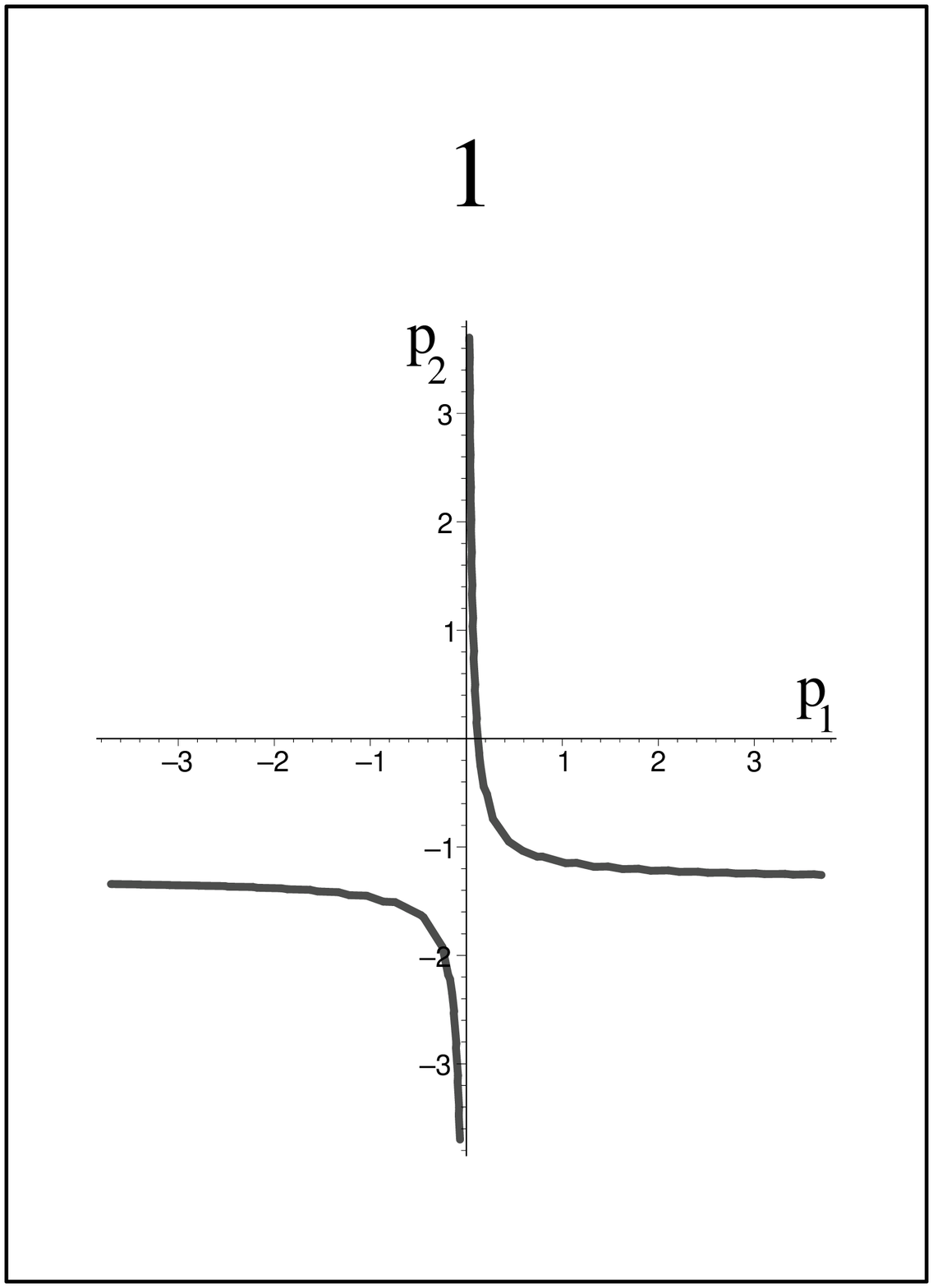}
& &
\includegraphics[width=4cm, height=5cm]{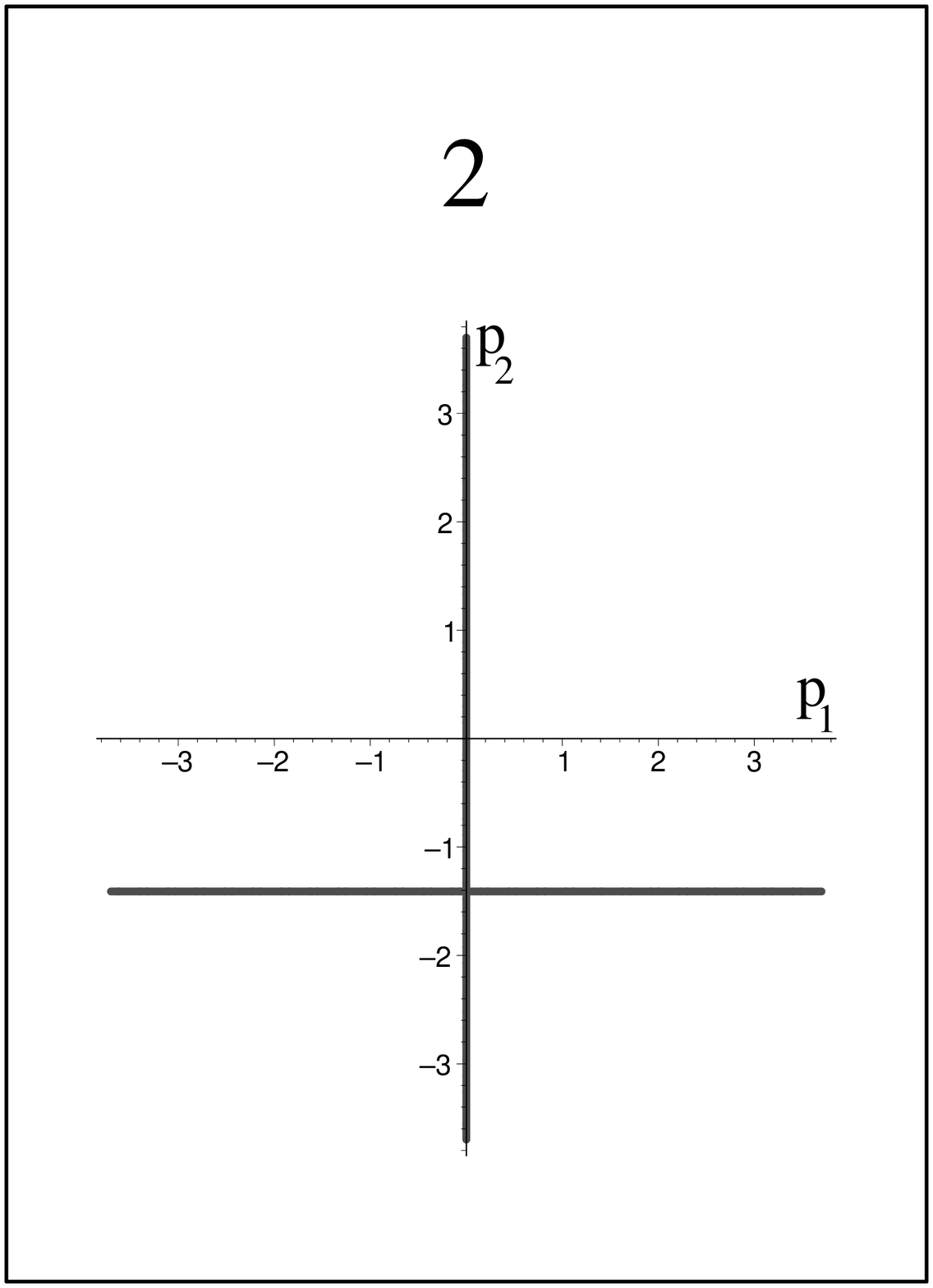}
& &
\includegraphics[width=4cm, height=5cm]{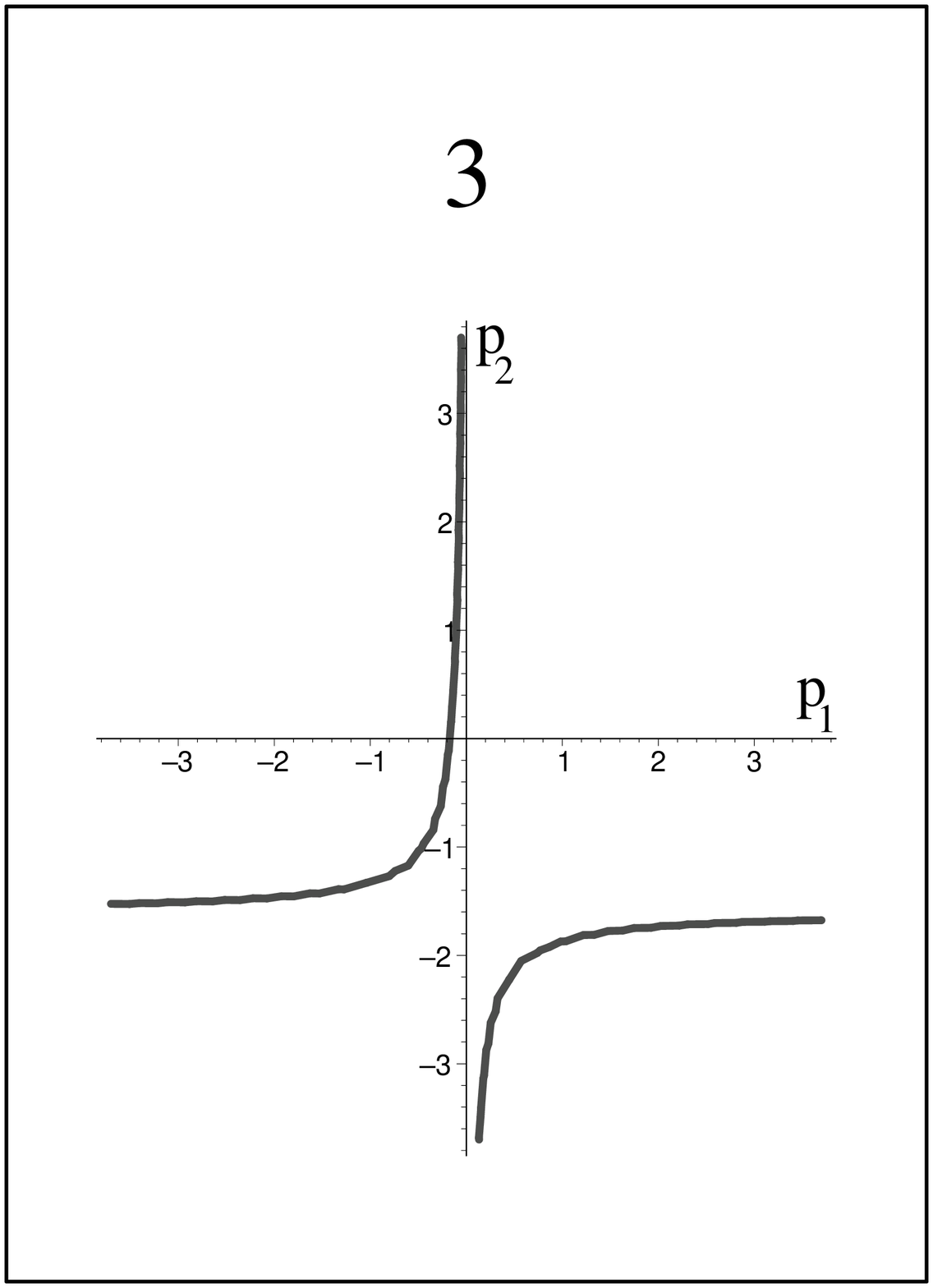}
\end{tabular}
\caption{First plot: $x_2=-1, t=1.3$, second plot: $x_2=-1, t=\sqrt{2}=1.4142$, third plot: $x_2=-1, t=1.6$.}
\label{fig-hyp}
\end{figure}

 In the rest of the paper we will consider Hamiltonian deformations of complex algebraic plane curves. We begin with quadrics defined by equation
\begin{equation}
\label{genquadrcurve}
 f=a{p_1}^2+b{p_2}^2+c{p_1}p_2+d{p_1}+e{p_2}+h=0
\end{equation}
where $a,b,\dots,h$ are functions of the deformation parameters $x_1,x_2,t$.
Choosing
\begin{equation}
\label{genquadrHam}
 H=\alpha{p_1}^2+\beta{p_2}^2+\gamma{p_1}p_2+\delta{p_1}+\mu{p_2}+\nu,
\end{equation}
one gets the system of six equations of hydrodynamical type. This system has several interesting reductions. One is associated with the constraint $a=b=e=0$, $c=1$, $\gamma=\mu=0$, and $\alpha$ and $\beta$ are constants. It is given by the system
\begin{equation}
\label{quadquadconstr}
\eqalign{
d_t+\delta d_{x_1}-\beta(d^2)_{x_2}+2\alpha h_{x_1} -\nu_{x_2}=0, \\
h_t+(h\delta)_{x_1}-2\beta(dh)_{x_2}=0, \\
\delta_{x_2}-2\alpha d_{x_1}=0, \\
\nu_{x_1}-2\beta h_{x_2}=0.
}
\end{equation}
At $\delta=0$, $\alpha=0$, $\beta=\frac{1}{2}$ it is the $(2+1)$ dimensional generalization of the one-layer Benney system proposed in \cite{K2,Z}
\begin{equation}
\label{1layerB-def}
\eqalign{
d_{x_1t}-\frac{1}{2}(d^2)_{x_1x_2} -h_{x_2x_2}=0, \\
h_t-(dh)_{x_2}=0. 
}
\end{equation}
The curve $\Gamma$ is the hyperbola ${p_1}p_2+d{p_1}+h=0$ and the solutions of the system (\ref{1layerB-def}) describe deformations of the hyperbola. The simplest solution of the system is given by
\begin{equation}
 d=t, \qquad h= x_2+\frac{{t}^2}{2}.
\end{equation}
The corresponding evolution of the hyperbola is shown in fig. \ref{fig-hyp}.\par
At $\beta=-\alpha=\frac{1}{2}$, the system (\ref{quadquadconstr}) becomes the dispersionless Davey-Stewardson (dDS) system considered in \cite{Kono}. It describes a class of Hamiltonian motions of hyperbola. \par
For cubic $H$ given by 
\begin{equation}
\label{gencubHam}
 H=\alpha_1{p_1}^3+\alpha_{2}{p_2}^3+\alpha_{3}{p_1}^2p_2+\alpha_{4}{p_2}^2p_1+\alpha_{5}{p_1}^2+\alpha_{6}{p_2}^2+\alpha_{7}p_1p_2+\alpha_{8}p_1+\alpha_{9}p_2+\alpha_{10}
\end{equation}
 the condition (\ref{gendefLiouv}) gives rise to a pretty long system of equations. This system admits several distinguished reductions. \par 
Under the constraint 
$b=c=d=0$, $a=e=1$, $\alpha_2=\alpha_3=\alpha_4=\alpha_5=\alpha_6=\alpha_7=\alpha_9=0$, $\alpha_1=1$ we obtain
\begin{equation}
\label{m-nm-dKP}
\eqalign{
h_{t}+\frac{3}{2}h_{x_{1}}h-{\alpha_{10}}_{x_{2}}  =0, \\
3h_{x_{2}}+4{\alpha_{10}}_{x_{1}}  =0.
}
\end{equation}
which is the well-known dispersionless Kadomtsev-Petviashvilii (dKP) equation (see e.g. \cite{K1,KG,Z}). It describes Hamiltonian deformations of a parabola ${p_1}^2+p_1+h=0$. \par
For the reduction $a=b=1$, $c=d=e=0$, $\alpha_2=\alpha_4=\alpha_5=\alpha_6=\alpha_7=\alpha_{10}=0$, $\alpha_1=-\frac{1}{3}\alpha_3=1$ one has the system 
\begin{equation}
\label{dVN}
\eqalign{
h_t+(h\alpha_8)_{x_1}+(h\alpha_9)_{x_2} =0,  \\
3h_{x_1}-{\alpha_8}_{x_1}+{\alpha_9}_{x_2} =0,\\
3h_{x_2}+{\alpha_8}_{x_2}+{\alpha_9}_{x_1}=0
}
\end{equation}
that is the dispersionless Veselov-Novikov equation \cite{K1}. It gives us Hamiltonian deformations of a circle ${p_1}^2+{p_2}^2+h=0$. \par
Another interesting reduction describes deformations of an ellipse
\begin{equation}
 \label{ellipse}
\frac{{p_1}^2}{a^2}+\frac{{p_2}^2}{b^2}-1=0,
\end{equation}
i.e. the quadric (\ref{genquadrcurve}) with $a \to \frac{1}{a^2}$, $b \to \frac{1}{b^2}$, $c=d=e=0$, and $h=-1$.
Considering particular Hamiltonian (\ref{gencubHam}) with $\alpha_{3}=\alpha_{4}=\alpha_{5}=\alpha_{6}=\alpha_{7}=\alpha_{10}=0$, one has the following new system of equations
\begin{equation}
 \label{defellipsecub}
\eqalign{
a_t+(\alpha_8 a+\alpha_1 a^3)_{x_1}+\alpha_9 a_{x_2}=0,\\
b_t+(\alpha_9 b+\alpha_2 b^3)_{x_2}+\alpha_8 b_{x_1}=0,\\
a^4 {\alpha_1}_{x_2}-b^4 {\alpha_2}_{x_1}=0,\\
a^4 {\alpha_1}_{x_2}+a^2 {\alpha_8}_{x_2}+b^2 {\alpha_9}_{x_1}=0,\\
a(\alpha_2 b^3)_{x_2}+b(\alpha_1 a^3)_{x_1}-3\alpha_1a^3b_{x_1}-3\alpha_2b^3a_{x_2}=0.
}
\end{equation}
For the system (\ref{defellipsecub}) equation (\ref{defdefcurve}) is of the form
\begin{equation}
 \label{Liouvellipse}
f_t+\{f,H\}=(A{p_1}^2+B{p_2}^2+Cp_1p_2)f
\end{equation}
where
\begin{equation}
A=-2{\alpha_1}_{x_1}-6 \frac{\alpha_1 a_{x_1}}{a}, \qquad
B=-2{\alpha_2}_{x_2}-6 \frac{\alpha_2 b_{x_2}}{b}, \qquad
C=-2\frac{a^2}{b^2}{\alpha_1}_{x_1}.
\end{equation}
Solutions of the system (\ref{defellipsecub}) describe various regimes in Hamiltonian motions of the ellipse. In particular, for the initial data given by $a(x_1,x_2,t=0)=b(x_1,x_2,t=0)$ one has deformations of a circle into ellipse.\par
For $\alpha_1=const$ and $\alpha_2=const$ the system (\ref{defellipsecub}) admits the reduction $a=b$. At $\alpha_1=4$, $\alpha_2=0$ the corresponding system coincides with dVN equation (\ref{dVN}) modulo the substitutions $h=-a^2$, and $\alpha_8 \to 3h+\alpha_8$. We note that the last change of $\alpha_8$ is suggested by the transformation of the type $H \to H+gf$ discussed in section $2$ from the Hamiltonian $H$ for the system (\ref{defellipsecub}) to the Hamiltonian for the dVN equation.\par
In the case of cyclic variable $x_2$ the last three equations (\ref{defellipsecub}) give
\begin{equation}
\eqalign{
\alpha_2=const, \qquad \alpha_9=const, \qquad  \alpha_1=\frac{b^3}{a^3},\\ 
}
\end{equation}
while the first two equations take the form ($x=x_1$)
\begin{equation}
\label{11elpsa8}
\eqalign{
 a_t+(\alpha_8a+b^3)_{x}=0,\\
b_t+\alpha_8 b_{x}=0.
}
\end{equation}
The Hamiltonian becomes $H=\frac{b^3}{a^3}{p_1}^3+\alpha_8 p_1$. 
Without loss of generality we choose $\alpha_2=\alpha_9=0$.
Introducing new variables  $u=\frac{a}{b}$ and $v=b^2$, one rewrites the (1+1) dimensional system (\ref{11elpsa8}) as
\begin{equation}
\label{11epsa8rapp}
\eqalign{
u_t+\left( \alpha_8 u+\frac{3}{2} v \right)_{x}=0,\\
v_t+\alpha_8 v_{x}=0.
}
\end{equation}
Finally using the eccentricity $\epsilon=\sqrt{1-u^2}$ of an ellipse, one gets the system
\begin{equation}
\eqalign{
\epsilon_t-\frac{\sqrt{1-\epsilon^2}}{\epsilon}\left( \alpha_8 \sqrt{1-\epsilon^2}+\frac{3}{2} v \right)_{x}=0,\\
v_t+\alpha_8 v_{x}=0,
}
\end{equation}
which describes Hamiltonian deformations of an ellipse in a pure geometrical terms.\par
The system (\ref{11epsa8rapp}) represents a particular example of the two component (1+1)-dimensional systems of hydrodynamical type (with arbitrary function $\alpha_8(u,v)$). It is well known that such systems are linearizable by a hodograph transformation $x=x(u,v)$, $t=t(u,v)$ (see e.g. \cite{DN}). In our case the corresponding linear system is 
\begin{equation}
 \eqalign{ 
x_u-\alpha_8t_u=0,\\
x_v+\left(\frac{3}{2}+u{\alpha_8}_v\right)t_u-\left(\alpha_8+u{\alpha_8}_u\right)t_v=0.
}
\end{equation}
This fact allows us to construct explicitly wide class of solutions for the system (\ref{11epsa8rapp}).\par
\begin{figure}[h!]
\centering
\begin{tabular}{ccccc}
\includegraphics[width=4cm, height=5cm]{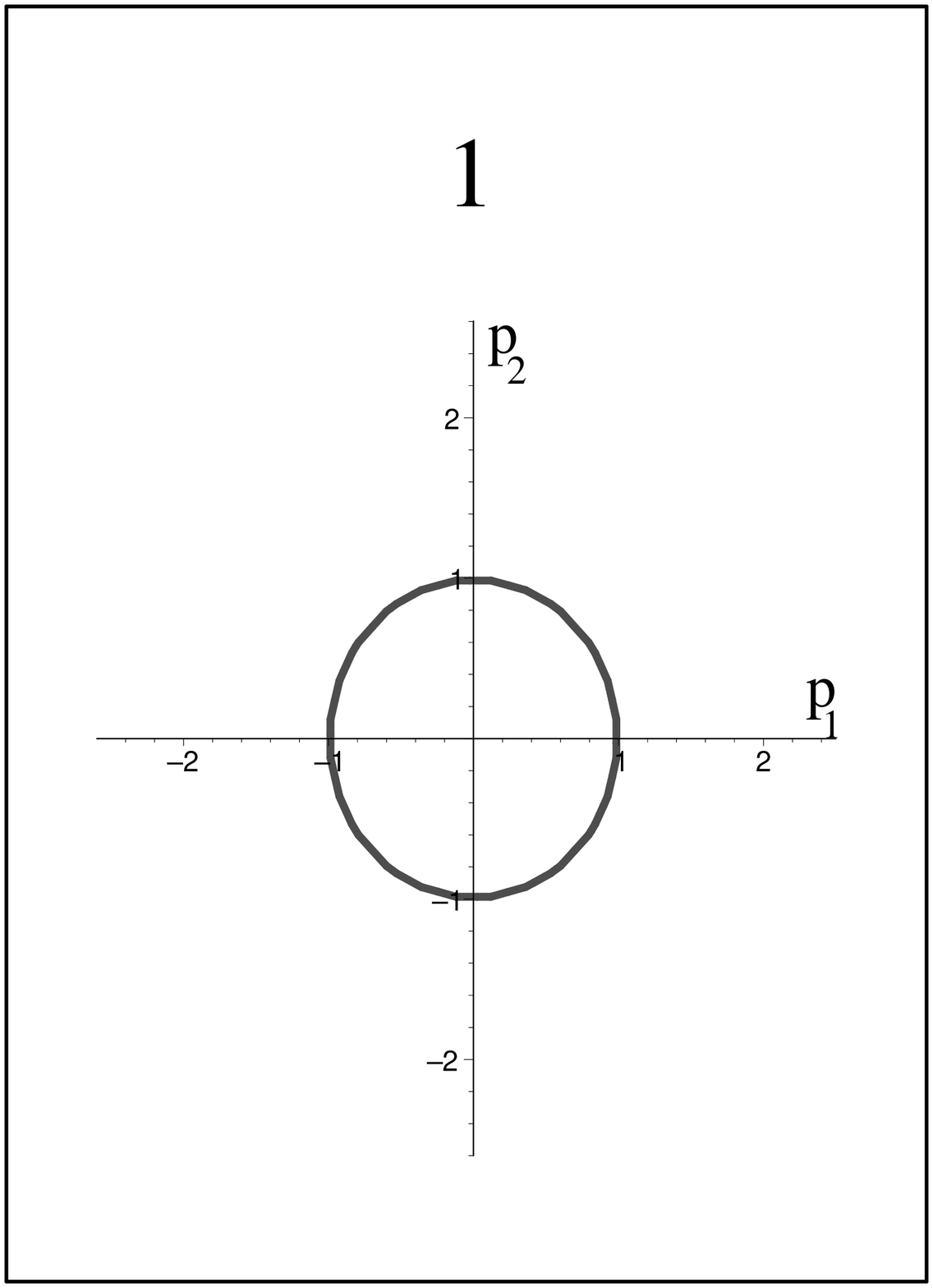}
& &
\includegraphics[width=4cm, height=5cm]{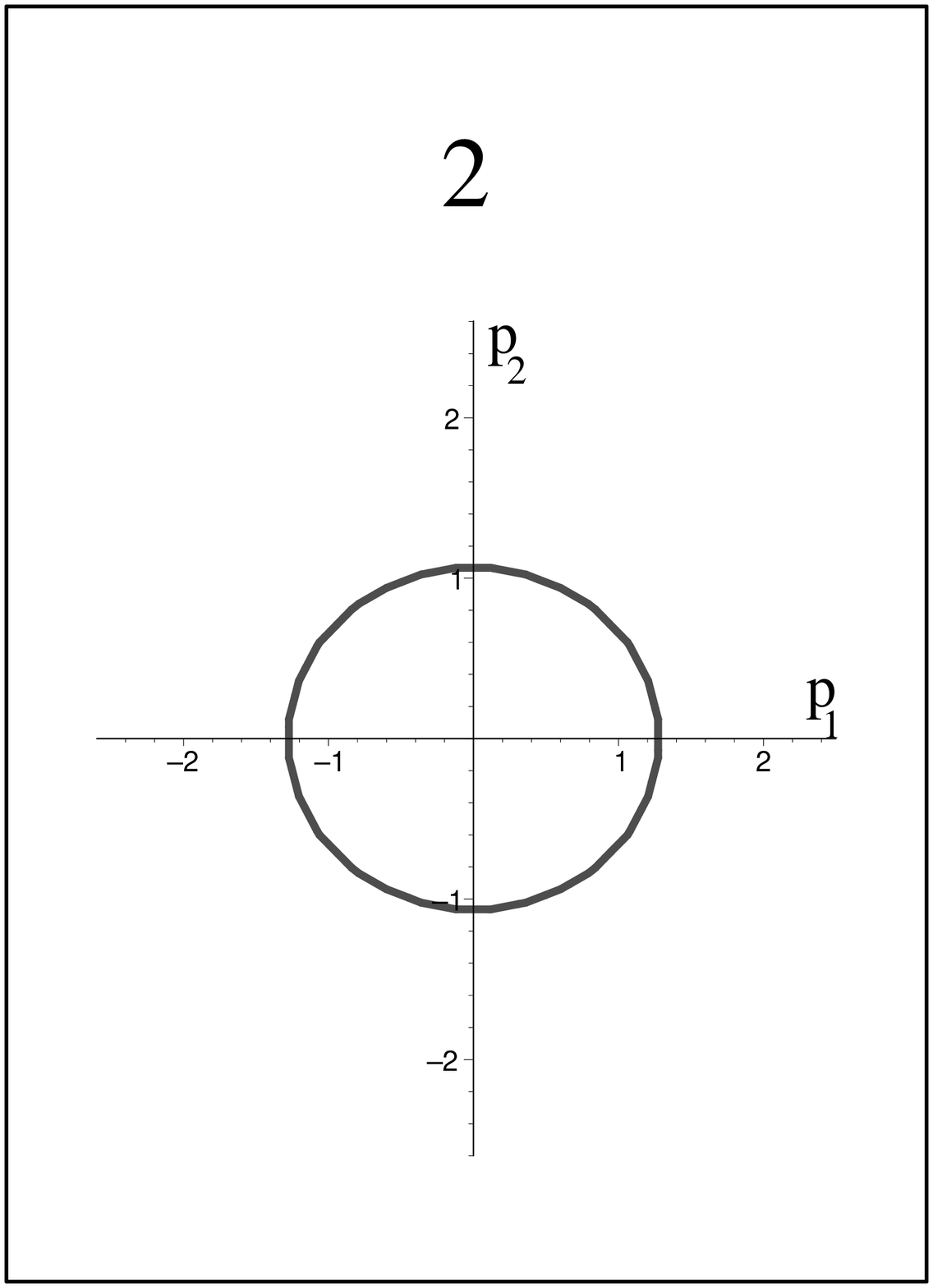}
& &
\includegraphics[width=4cm, height=5cm]{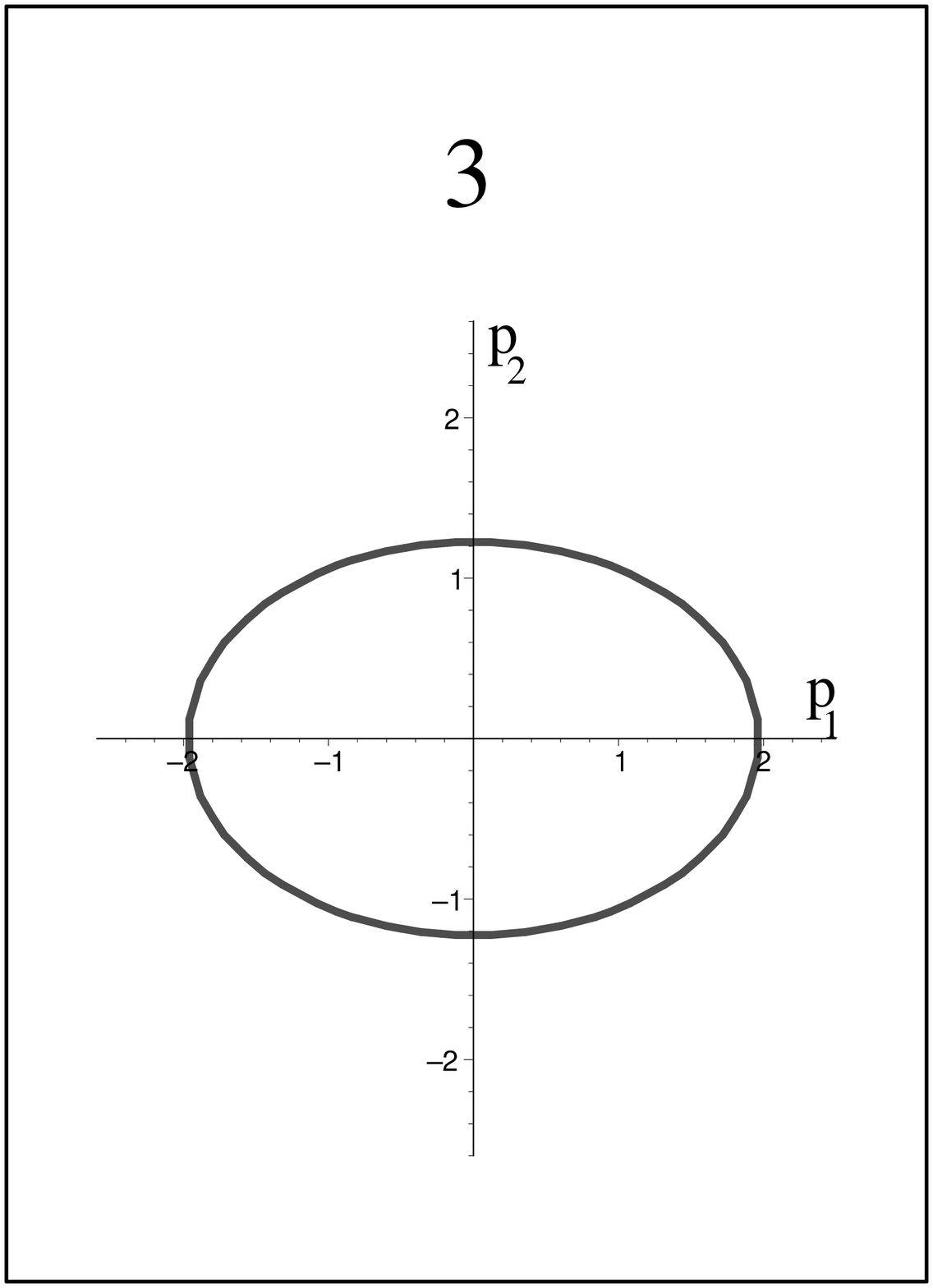}
\end{tabular}
\caption{First plot: $x_1=-1, t=1$, second plot: $x_1=-1, t=1.2$, third plot: $x_1=-1, t=1.6$.}
\label{fig-ellipse}
\end{figure}

At $\alpha_8=u$ the system (\ref{11epsa8rapp}) is the dispersionless limit of the Hirota-Satsuma system \cite{HS}. It has a simple polynomial solution 
\begin{equation}
 \eqalign{ 
u=t, \\
v=-\frac{2}{3} x +\frac{1}{3} t^2.
}
\end{equation}
which provide us with deformation of a circle ($t=1$) into an ellipse shown in figure \ref{fig-ellipse}.
\section{Deformations of plane cubic curve}
\label{sect-cub}
\par
\begin{figure}[h!]
\centering
\includegraphics[width=8cm, height=8cm]{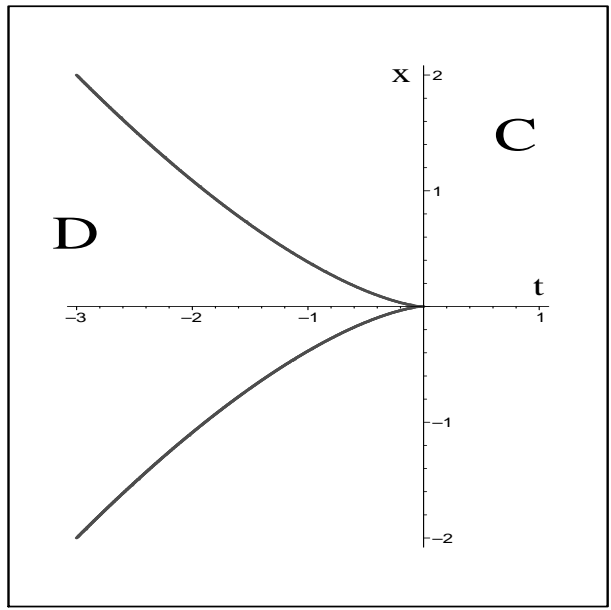}
\caption{On the curve the genus of the cubic (\ref{ellcurvetx}) is $0$, outside the curve it is $1$. In the D region the real section of the elliptic curve is disconnected, in the C region it is connected.}\label{fig:phtrtriv}
\end{figure}

 Hamiltonian deformations of plane cubics exhibit much richer and interesting structure.\\
The general form of the plane cubic is (see e.g \cite{Har})
\begin{equation}
 f={p_3}^2-{p_2}^3-u_4p_3p_2-u_3{p_2}^2-u_2p_3-u_1p_2-u_0.
\end{equation}
$(1+1)$ and $(2+1)$ dimensional coisotropic deformations of plane cubics have been considered in the paper \cite{KO}
where corresponding hydrodynamical type system have been derived. Here we will concentrate on the analysis of the formation of singularities and bubbles for the real section of cubic curves described by particular solutions of these systems. To this end we restrict ourself to the $(1+1)$ dimensional deformations which correspond to the case of cyclic variable $x_2$ and the reduction $u_4=u_2=0$. Thus, denoting $p_2=z$ and $p_1=p$, one has the cubic curve
\begin{equation}
\label{ellcurve}
 p^2-(z^3+u_3z^2+u_1z+u_0)=0.
\end{equation}
The particular choice of $H$ as
\begin{equation}
 H=\left(\frac{u_3}{2}-z\right)p
\end{equation}
gives rise to the following system $(x=x_1)$
\begin{equation}
 \label{KdV-g1}
\eqalign{
{u_3}_t={u_1}_x-\frac{3}{2}u_3{u_3}_x, \\
{u_1}_t={u_0}_x-u_1{u_3}_x-\frac{1}{2}u_3{u_1}_x, \\
{u_0}_t=-u_0{u_3}_x-\frac{1}{2}u_3{u_0}_x 
}
\end{equation}
which is the well known three-component dispersionless KdV system (see \cite{FP,KK}). The system (\ref{KdV-g1}) describes the Hamiltonian (coisotropic) deformations of the curve (\ref{ellcurve}). Recall that the moduli of the elliptic curve (\ref{ellcurve}) are given by \cite{S}
\begin{equation}
 \label{moduli}
\eqalign{
g_2=u_1-\frac{1}{3}{u_3}^2,\\
g_3=u_{{0}}+{\frac {2}{27}}\,{u_{{3}}}^{3}-\frac{1}{3}\,u_{{3}}u_{{1}}
}
\end{equation}
 while the discriminant $\Delta$ is
\begin{equation}
 \label{discr}
\Delta=-16(4{g_2}^3+27{g_3}^2)=16\,{u_{{3}}}^{2}{u_{{1}}}^{2}-64\,{u_{{1}}}^{3}-432\,{u_{{0}}}^{2}-64
\,{u_{{3}}}^{3}u_{{0}}+288\,u_{{0}}u_{{3}}u_{{1}}.
\end{equation}
The elliptic curve has singular points and, hence, become rational (genus zero) at the points where $\Delta=0$.\\
We note also that in this case equation (\ref{defdefcurve}) is
\begin{equation}
 \partial_t f + \{f,H\}=-{u_3}_x f.
\end{equation}
We will analyze polynomial solutions of the system (\ref{KdV-g1}). The simples one, linear in $x$ and $t$, is
\begin{equation}
 \label{linear-soln-g1}
\eqalign{
u_3=2a,\\ 
u_1=a^2+2b+2d\ t,\\
u_0=2(ab+c)+2d\ x-2ad\ t
}
\end{equation}
where $a,b,c,d$ are arbitrary constants. With the choice $a=b=c=0$, and $d=\frac{1}{2}$ one has
\begin{equation}
\label{triv-soln}
 u_3= 0, \quad u_1= t, \quad u_0=x.
\end{equation}
 Equation (\ref{ellcurve}) in this case is
\begin{equation}
\label{ellcurvetx}
 p^2-z^3-tz-x=0
\end{equation}
and discriminant becomes
\begin{equation}
 \Delta=-16(4t^3+27x^2).
\end{equation}
So for this solution deformation parameters $t$ and $x$ coincide with the moduli $g_2$ and $g_3$.\\
On the curve  
\begin{equation}
\label{phdiagrtriv}
 4t^3+27x^2=0
\end{equation}
in the plane $(x,t)$ (fig. \ref{fig:phtrtriv})
which has the standard parameterization
\begin{equation}
 x=2s^3, \qquad t=-3s^2,
\end{equation}
the elliptic curve (\ref{ellcurve}) is singular, i.e.
\begin{equation}
\label{ell-sing-par}
 p^2-(z-s)^2(z+2s)=0.
\end{equation}
  The curve (\ref{phdiagrtriv}) divides the plane $(t,x)$ in two different domains (fig. \ref{fig:phtrtriv}). In the domain D the discriminant $\Delta > 0$ and, hence, the real section of the elliptic curve is disconnected while in the domain C ($\Delta < 0$) it is connected. The genus of the curve in both domains is equal to $1$. On the curve (\ref{phdiagrtriv}) it vanishes. \par
\begin{figure}[h!]
\centering
\begin{tabular}{ccccc}
\includegraphics[width=4cm, height=5cm]{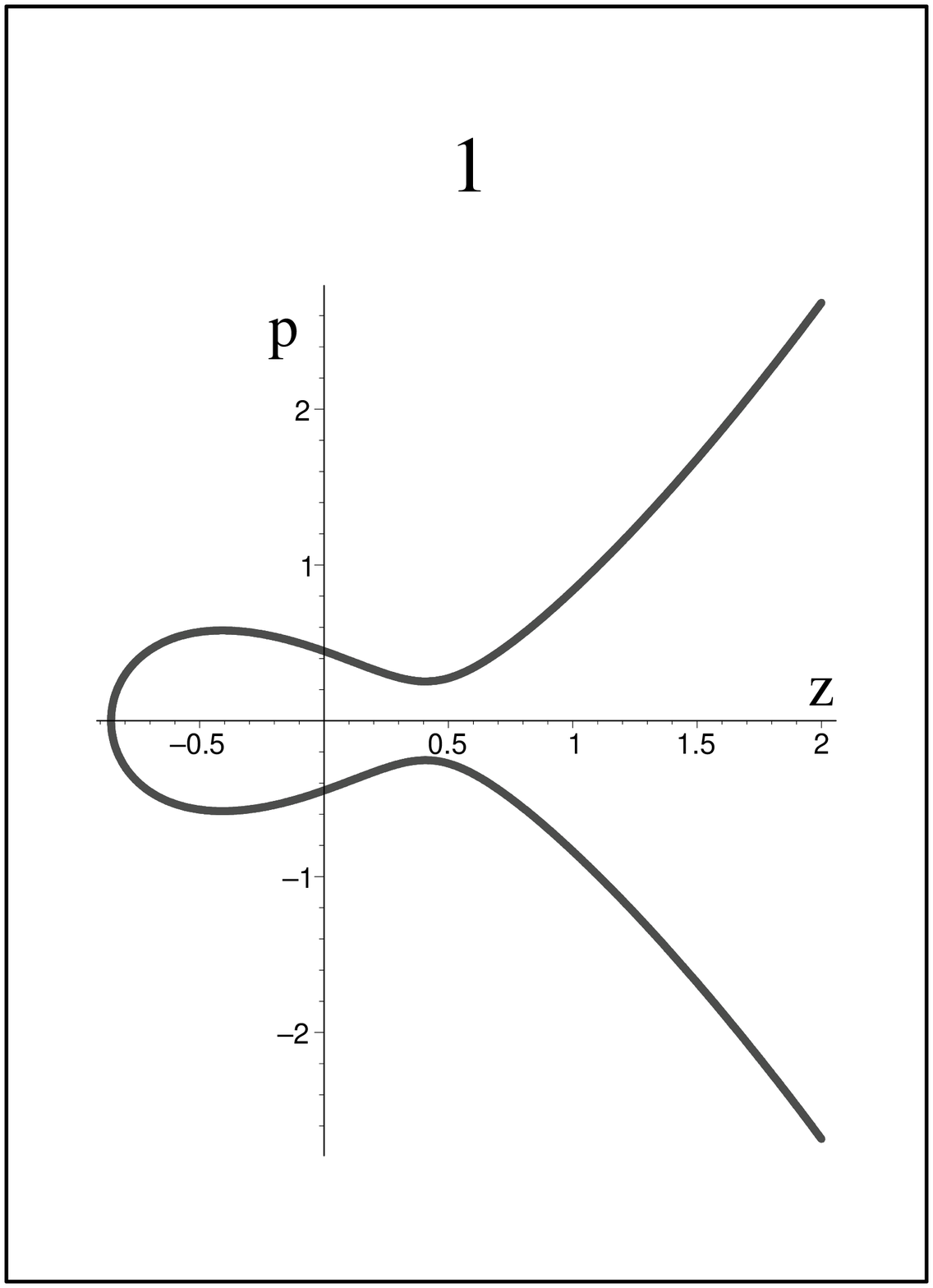}
& &
\includegraphics[width=4cm, height=5cm]{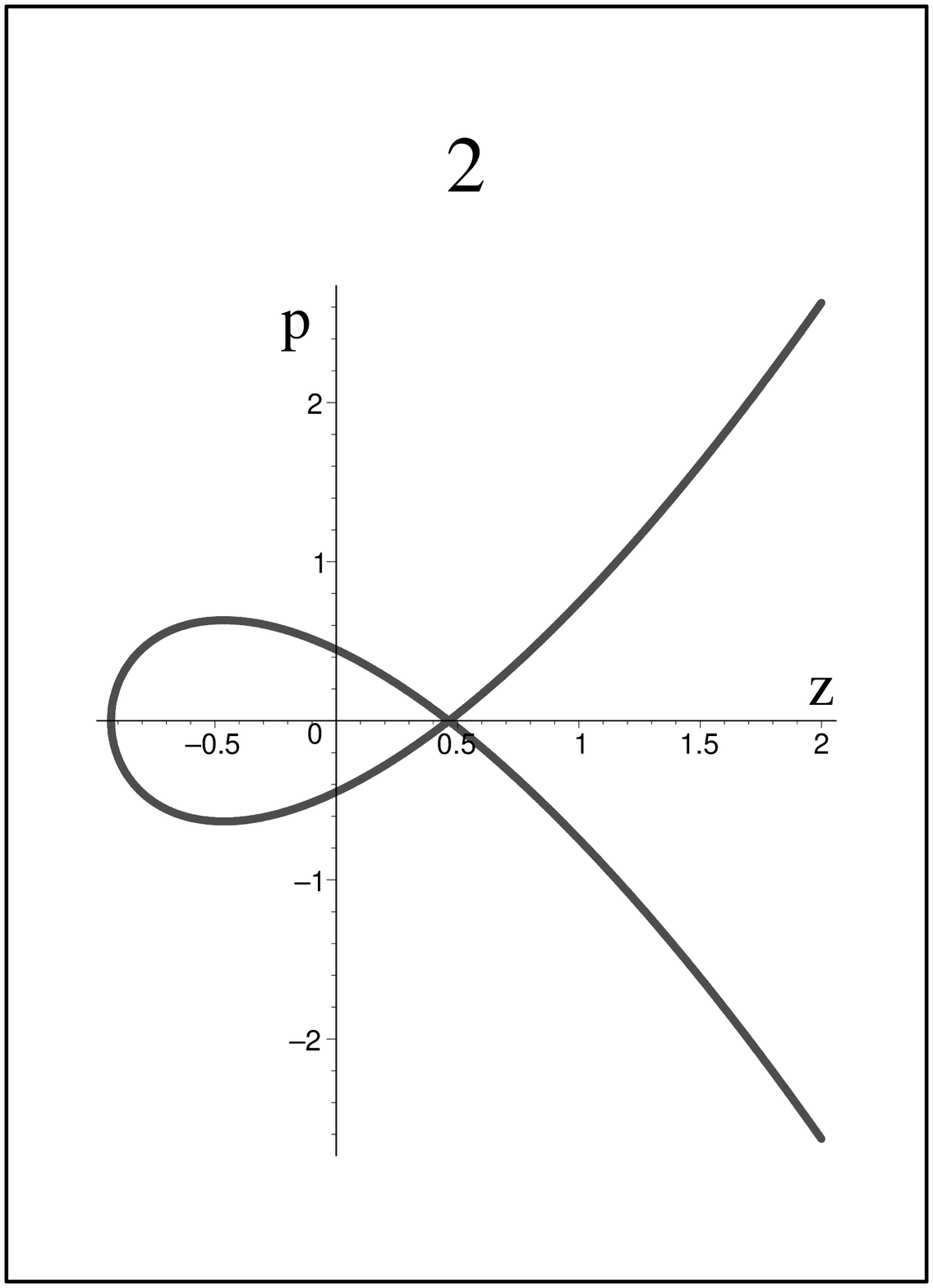}
& &
\includegraphics[width=4cm, height=5cm]{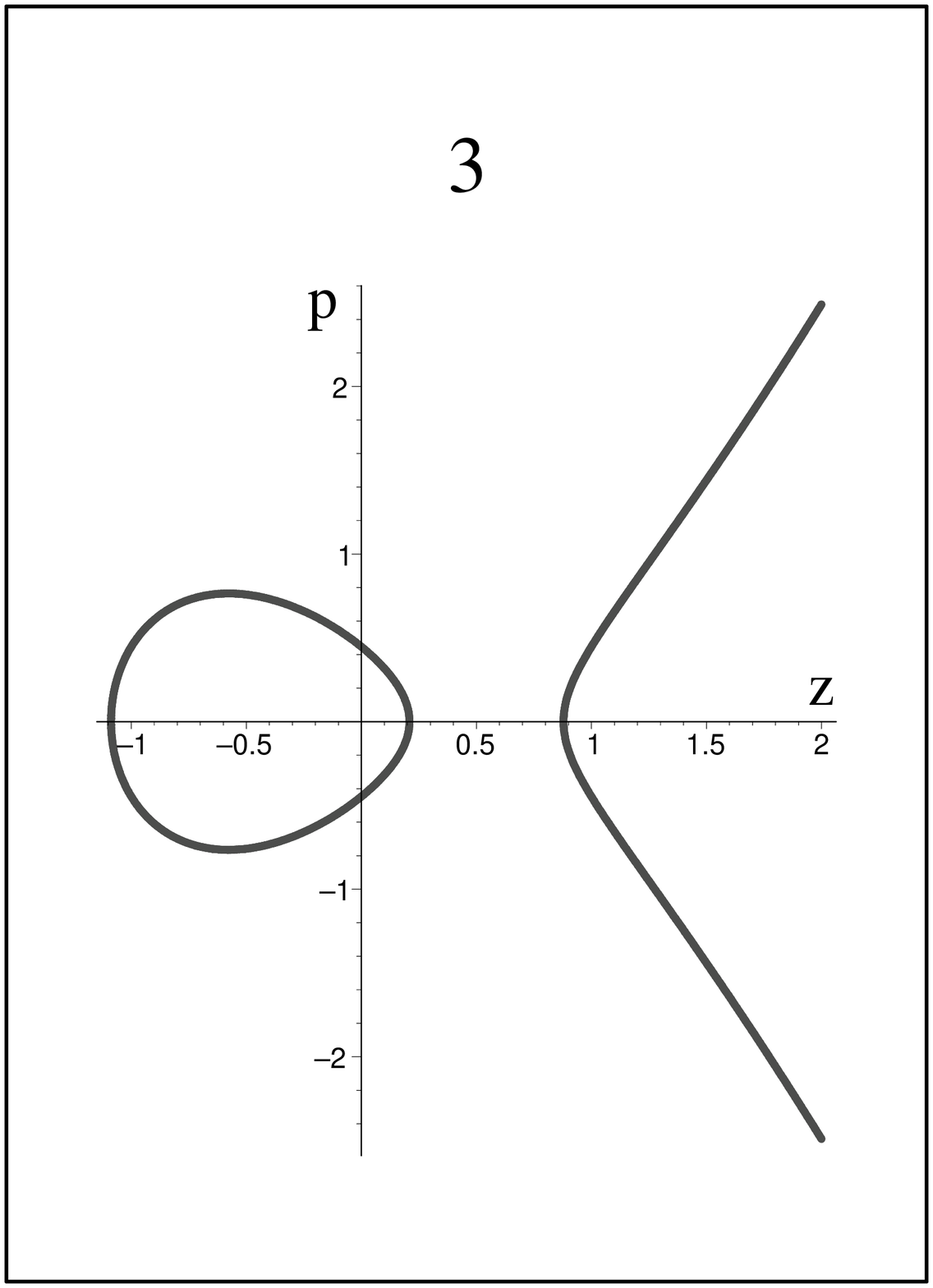}
\end{tabular}
\caption{First plot: $x=0.2, t=-0.5$, second plot: $x=0.2, t=-3 \times 10^{-2/3}=-0.64633$, third plot: $x=0.2, t=-1$.}
\label{PE+triv}
\end{figure}
\begin{figure}[h!]
\centering
\begin{tabular}{ccccc} 
\includegraphics[width=4cm, height=5cm]{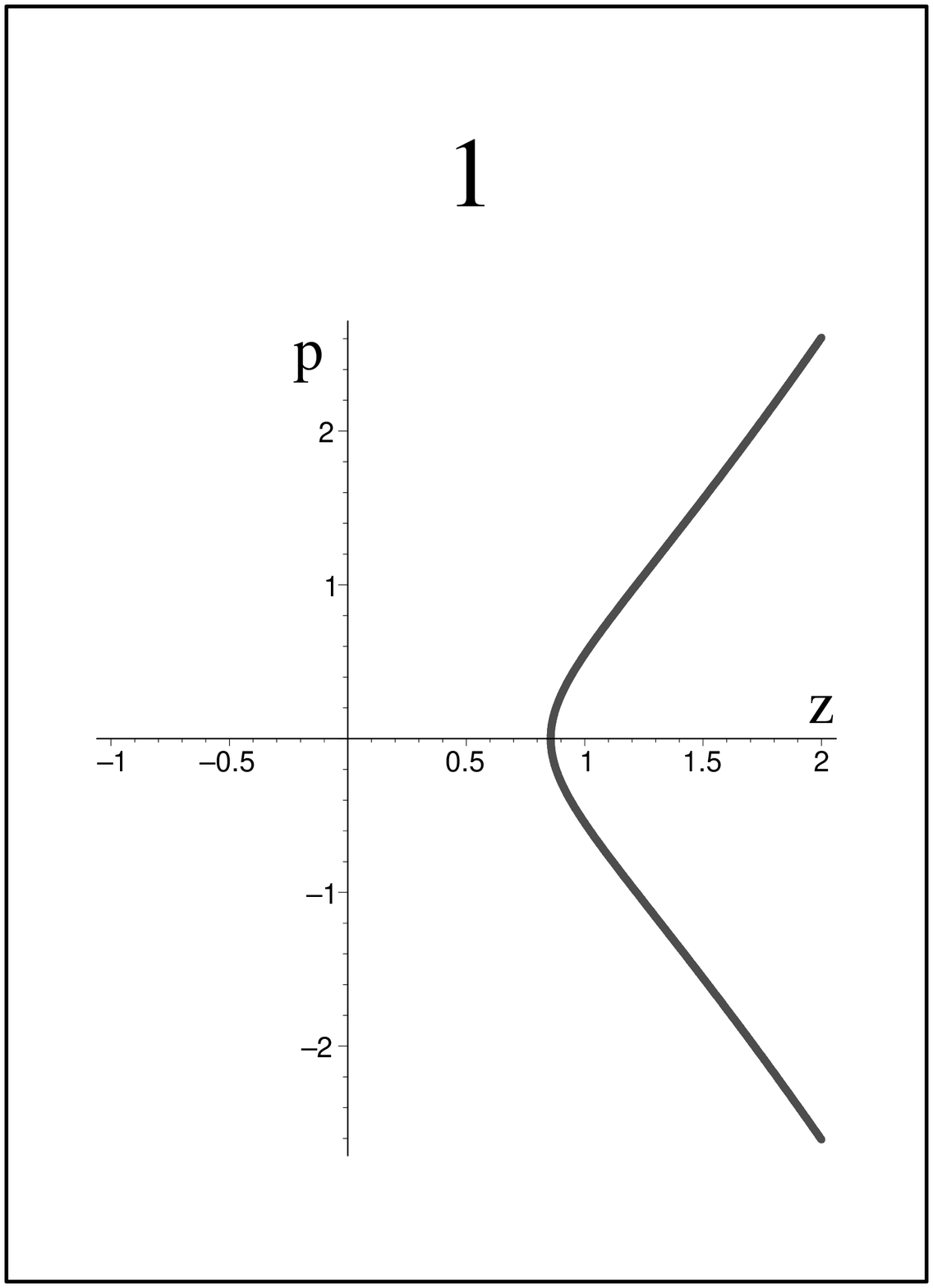}
& &
\includegraphics[width=4cm, height=5cm]{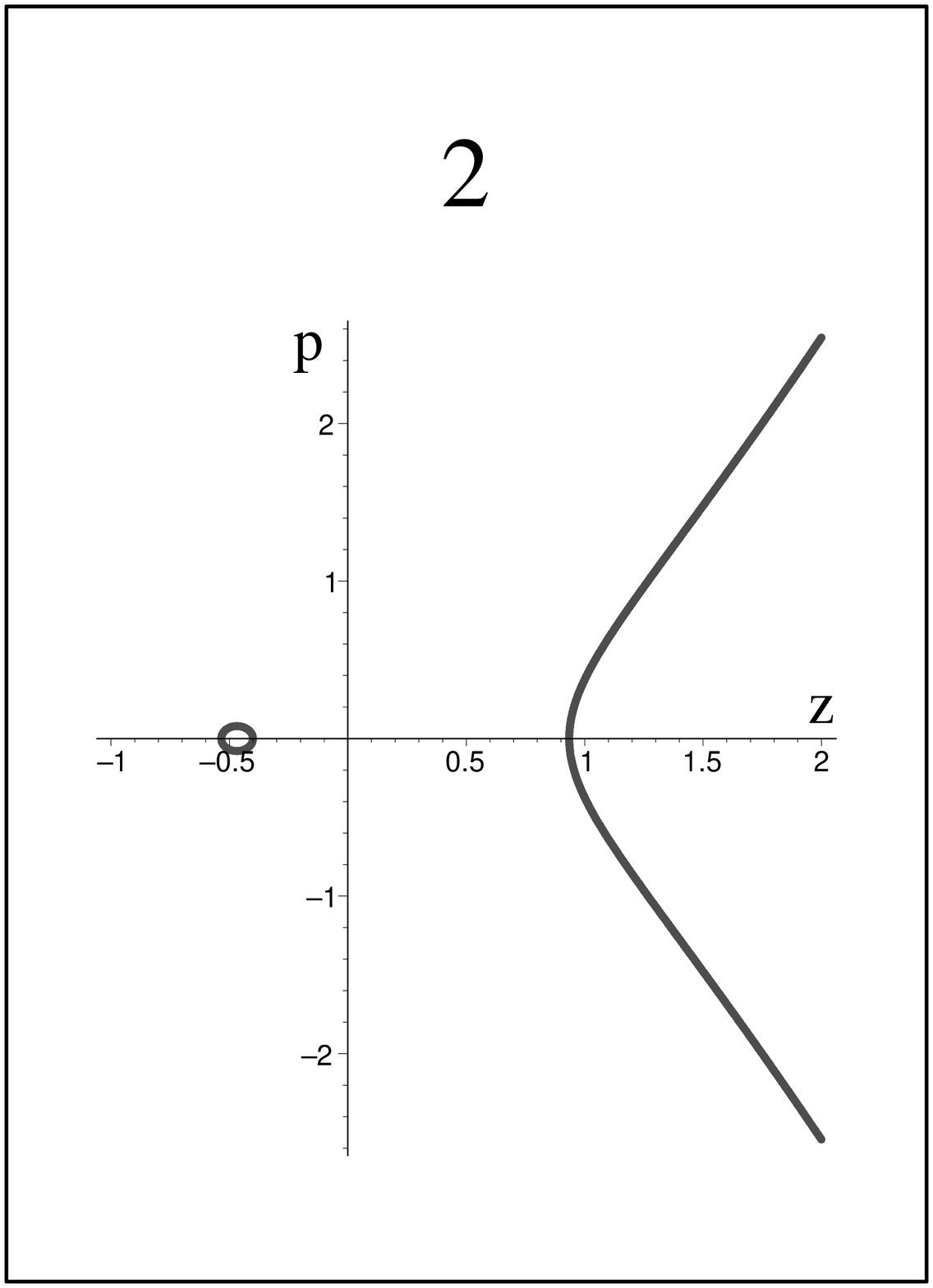}
& &
\includegraphics[width=4cm, height=5cm]{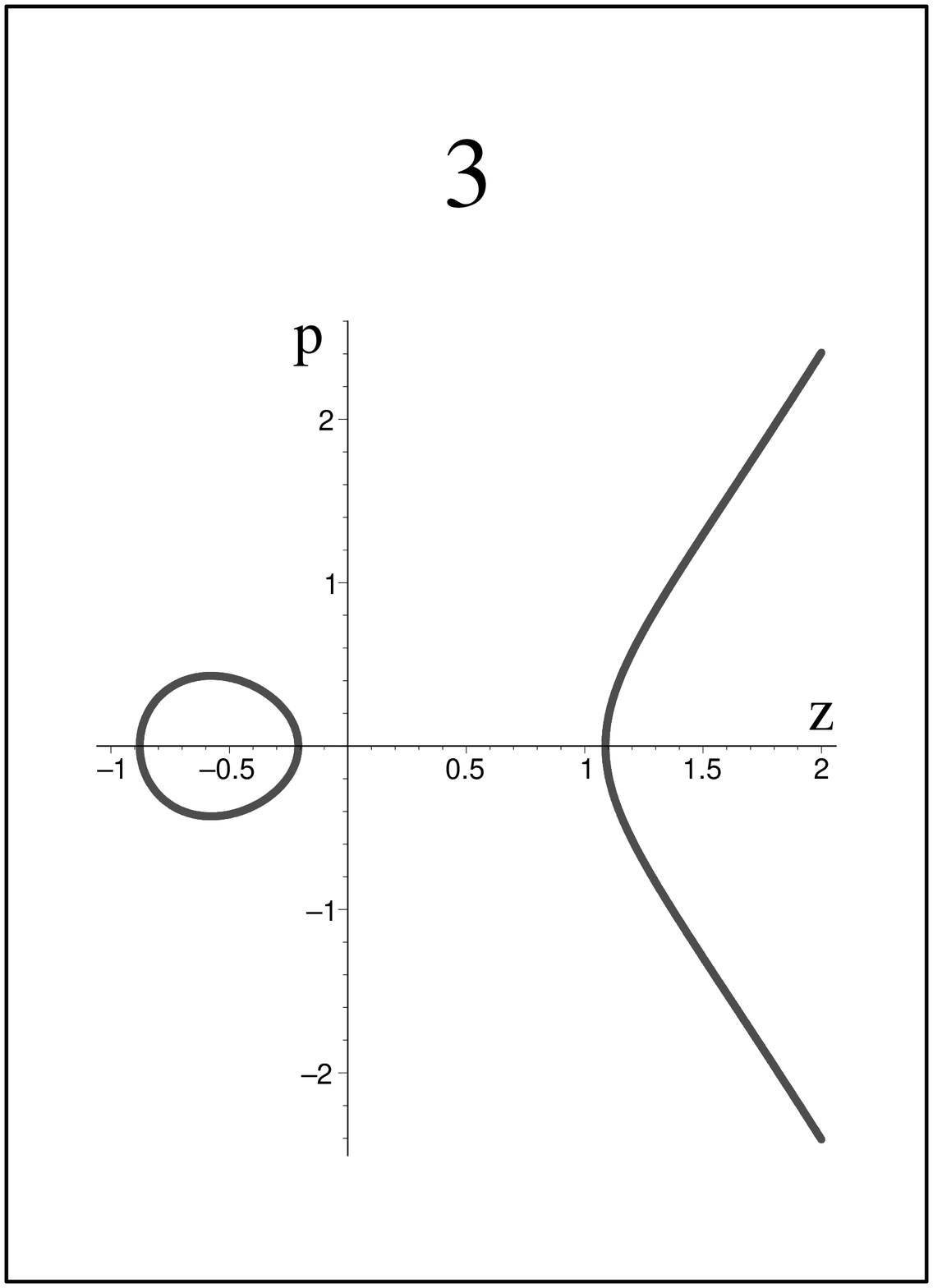}
\end{tabular}
\caption{First plot: $x=-0.2, t=-0.5$, second plot: $x=-0.2, t=-0.66$, third plot: $x=-0.2, t=-1$. The bubble becomes a point in $z=-0.46416$ at $t=-3 \times 10^{-2/3}.$}
\label{PE-triv}
\end{figure}
\begin{figure}[h!]
\centering
\begin{tabular}{ccccc}
\includegraphics[width=4cm, height=5cm]{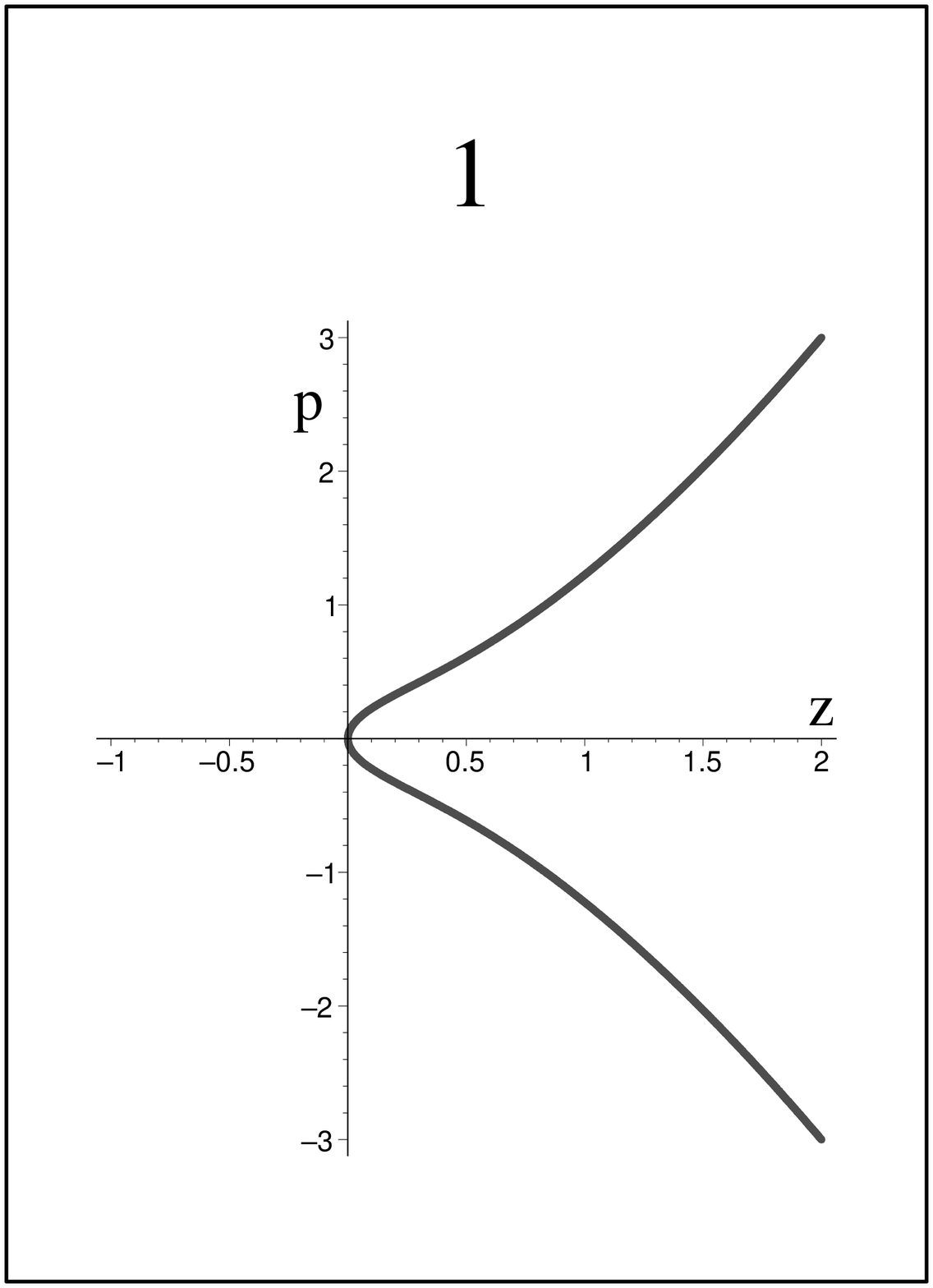}
& &
\includegraphics[width=4cm, height=5cm]{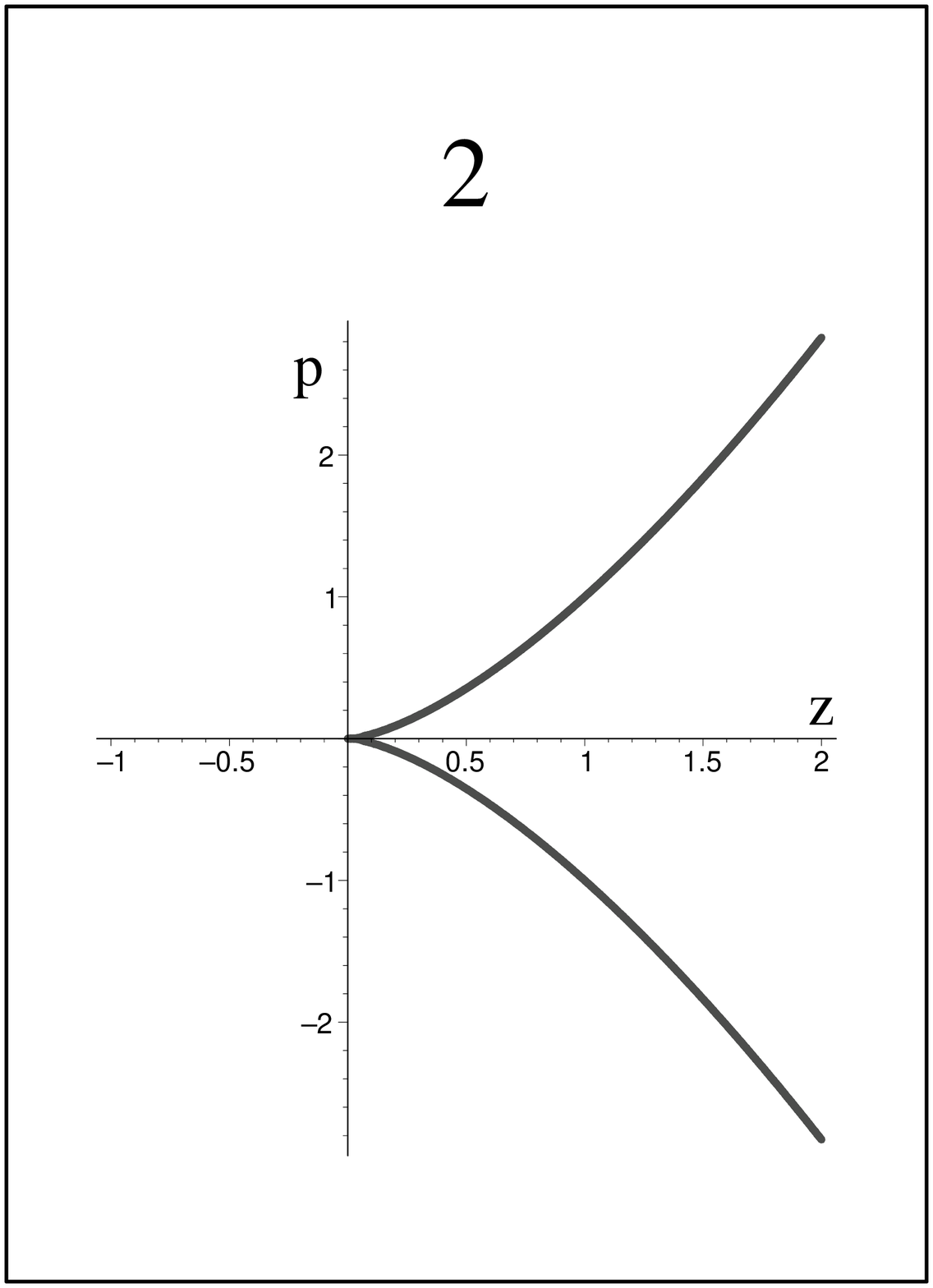}
& &
\includegraphics[width=4cm, height=5cm]{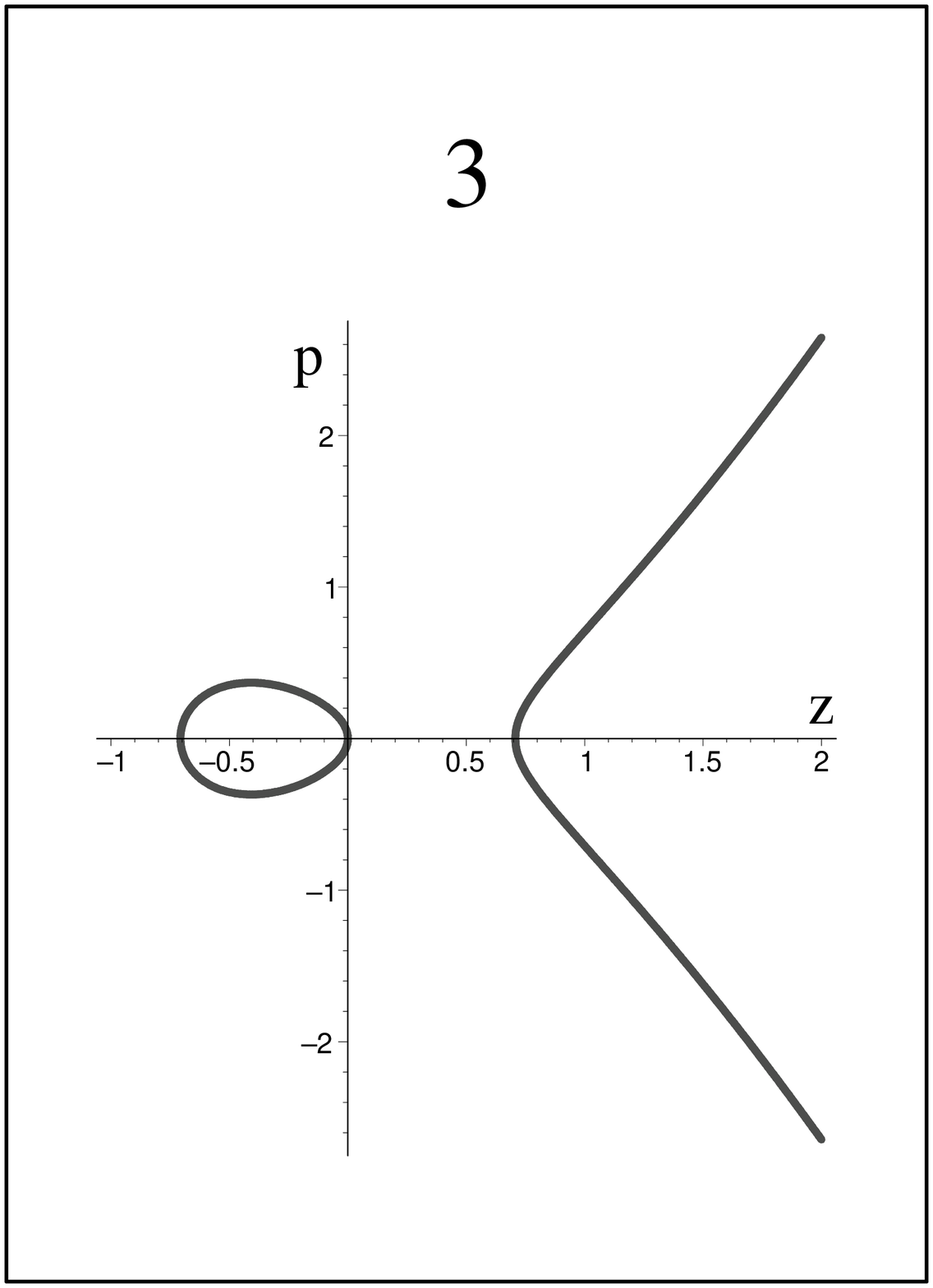}
\end{tabular}
\caption{First plot: $x=0, t=0.5$, second plot: $x=0, t=0$, third plot: $x=0, t=-0.5$.}
\label{PE0triv}
\end{figure}

Numerical analysis shows that the transition from the connected to disconnected curve (and viceversa) may happen in three qualitatively different ways shown in the figures (\ref{PE+triv}), (\ref{PE0triv}), and 
(\ref{PE-triv}) depending on the sign of $x$ on the curve (\ref{phdiagrtriv}).
Figure (\ref{PE+triv}) represents the evolution of the real section of the curve (\ref{ellcurvetx}) in the case when deformation parameters $t,x$ change along the line $x=0.2$. Transition from connected to disconnected (with bubble) curve goes through the formation of a double point (node) at $t=-0.64633$. At $x=-0.2$ (fig. \ref{PE-triv}) the bubble grows up from the point which appears at  $t=3 \times 10^{-2/3}$. From the complex viewpoint the above two regimes are the same. They are related to each other by the involution $x \to -x$ of the curve (\ref{ellcurvetx}). Transition through the line $x=0$ at $t=0$ is the limiting case of the above transitions. The formation (or annihilation) of a bubble goes through the formation of the cusp $p^2-z^3=0$ (figure \ref{PE0triv}). In the figures \ref{PE+triv},\ref{PE-triv},\ref{PE0triv} and others the plots are ordered in the way to describe transition from connected to disconnected curve. The processes of annihilation and creation of a bubble are converted to each other by a simple change of sign of Hamiltonian.\par
\begin{figure}[h!]
\centering
\includegraphics[width=8cm, height=8cm]{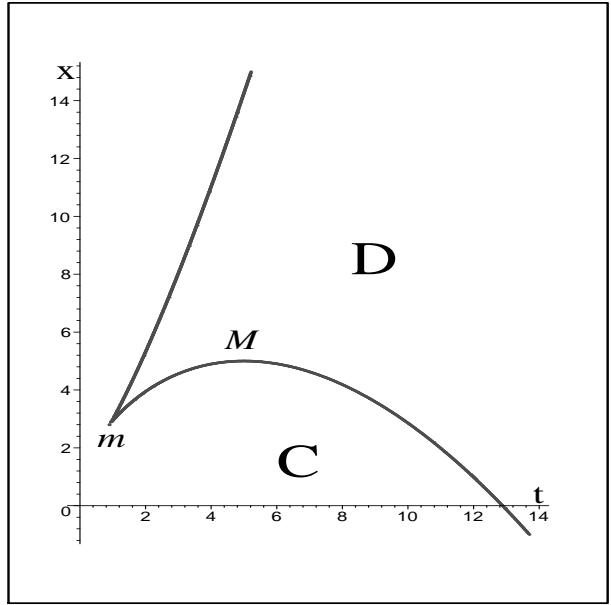}
\caption{On the line the genus of the  curve (\ref{Ecp}) is $0$, outside the line is $1$. In the D region the real part of the elliptic curve is disconnected, in the C region it is connected. The point $M$ is the local maximum  of the curve and the point $m$ is a local minimum (cusp).}\label{fig:phtr}
\end{figure}
\begin{figure}[h!]
\centering
\begin{tabular}{ccccc}
\includegraphics[width=4cm, height=5cm]{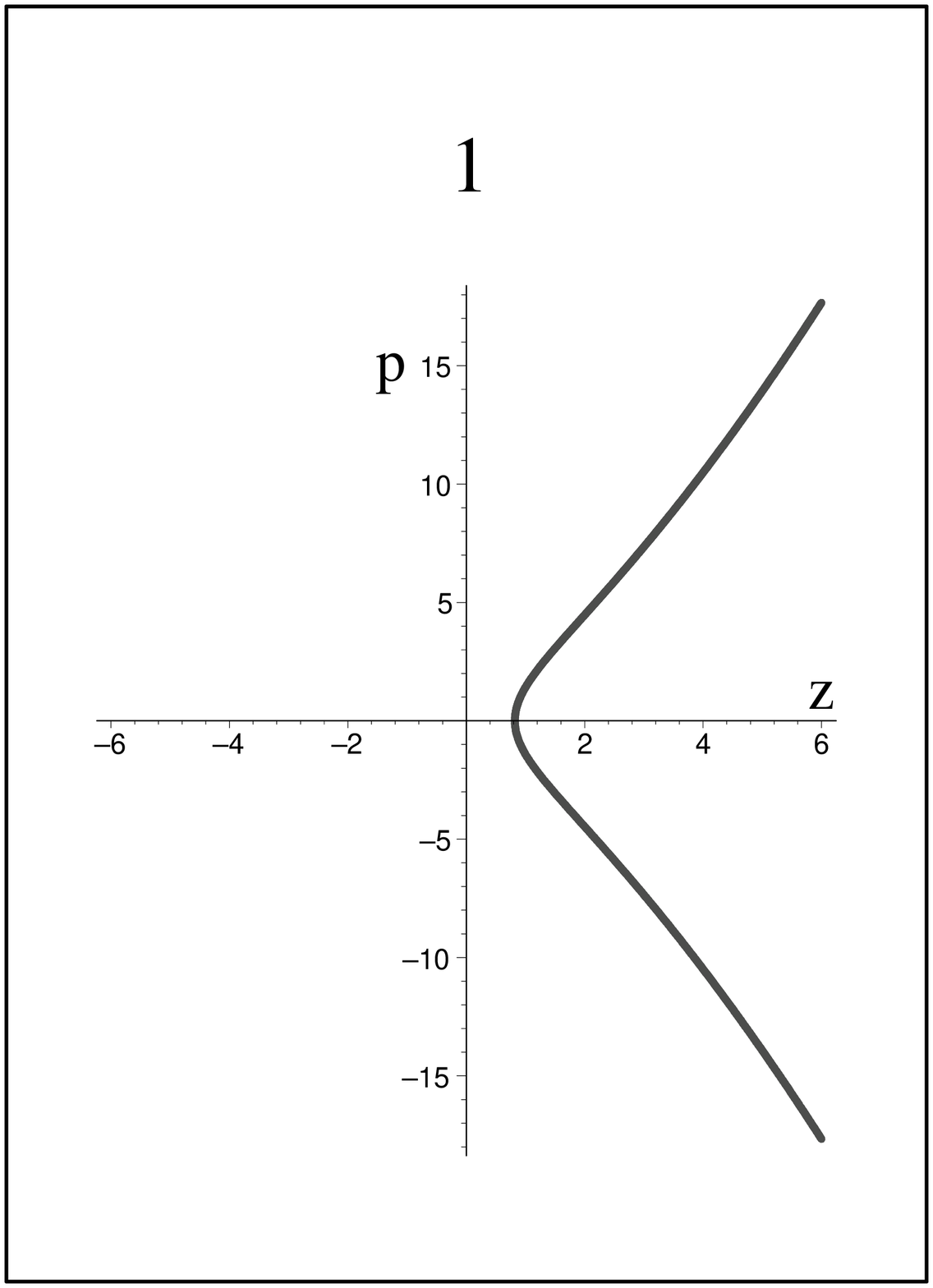}
& &
\includegraphics[width=4cm, height=5cm]{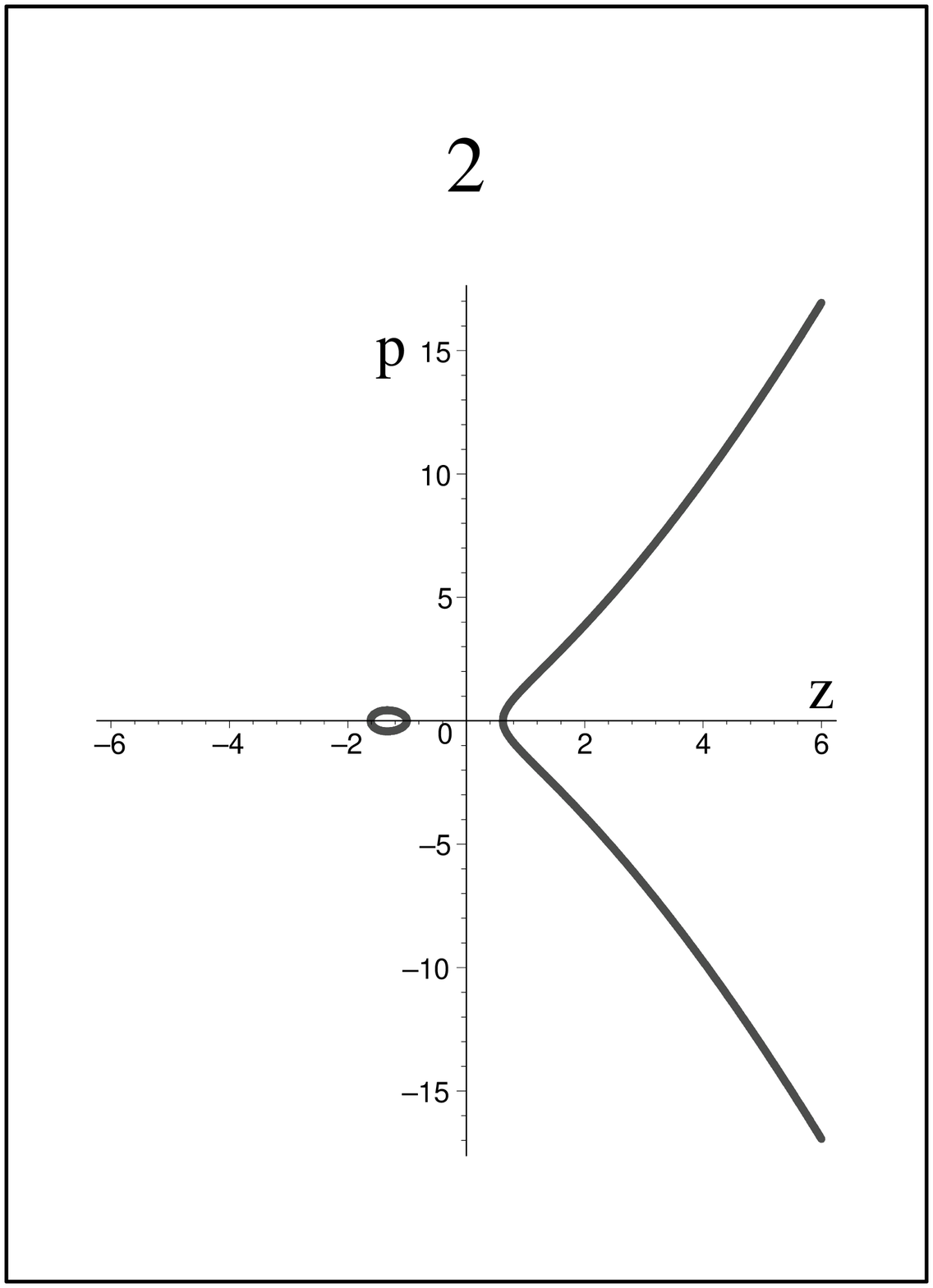}
& &
\includegraphics[width=4cm, height=5cm]{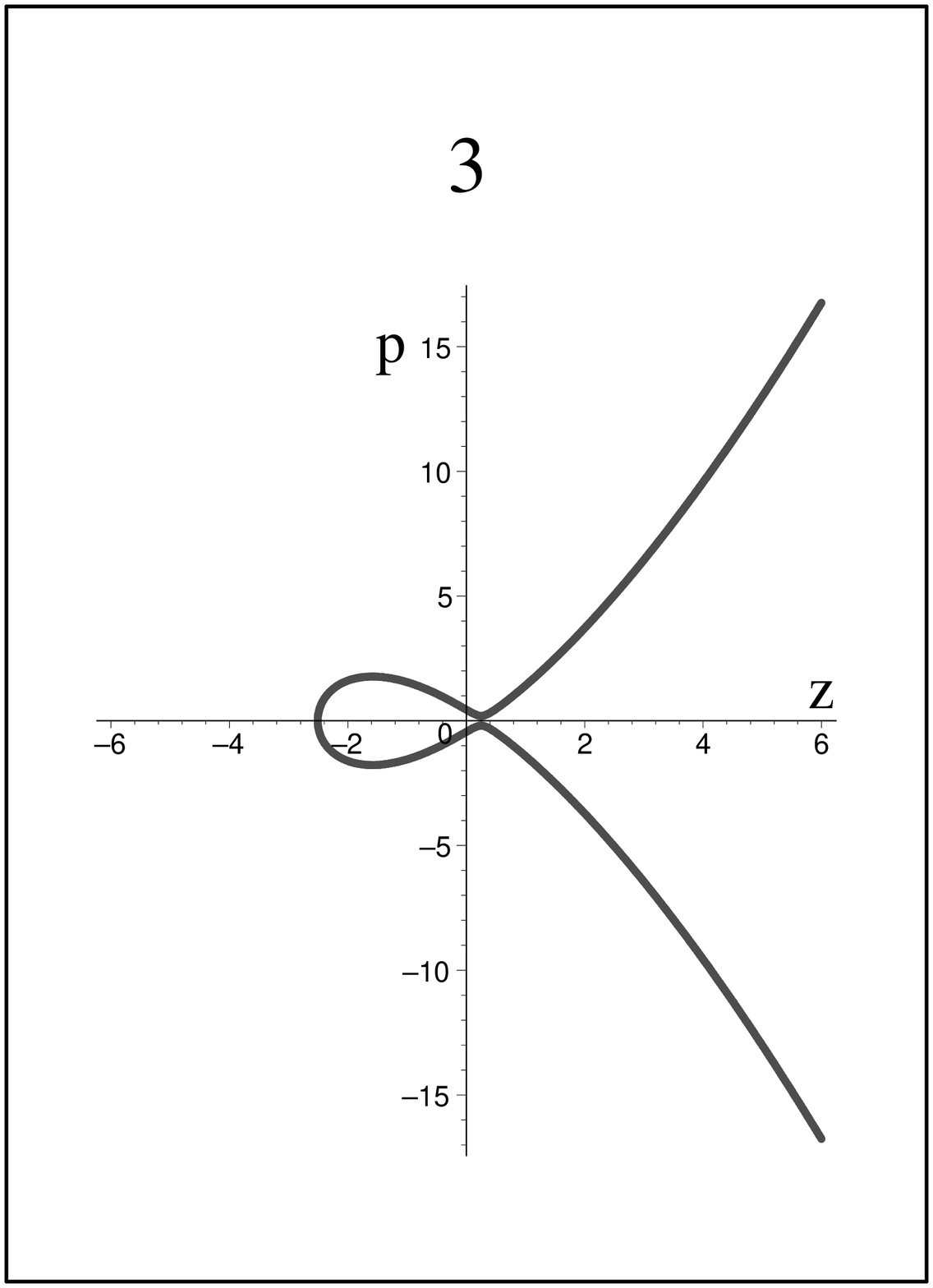}
\\
\includegraphics[width=4cm, height=5cm]{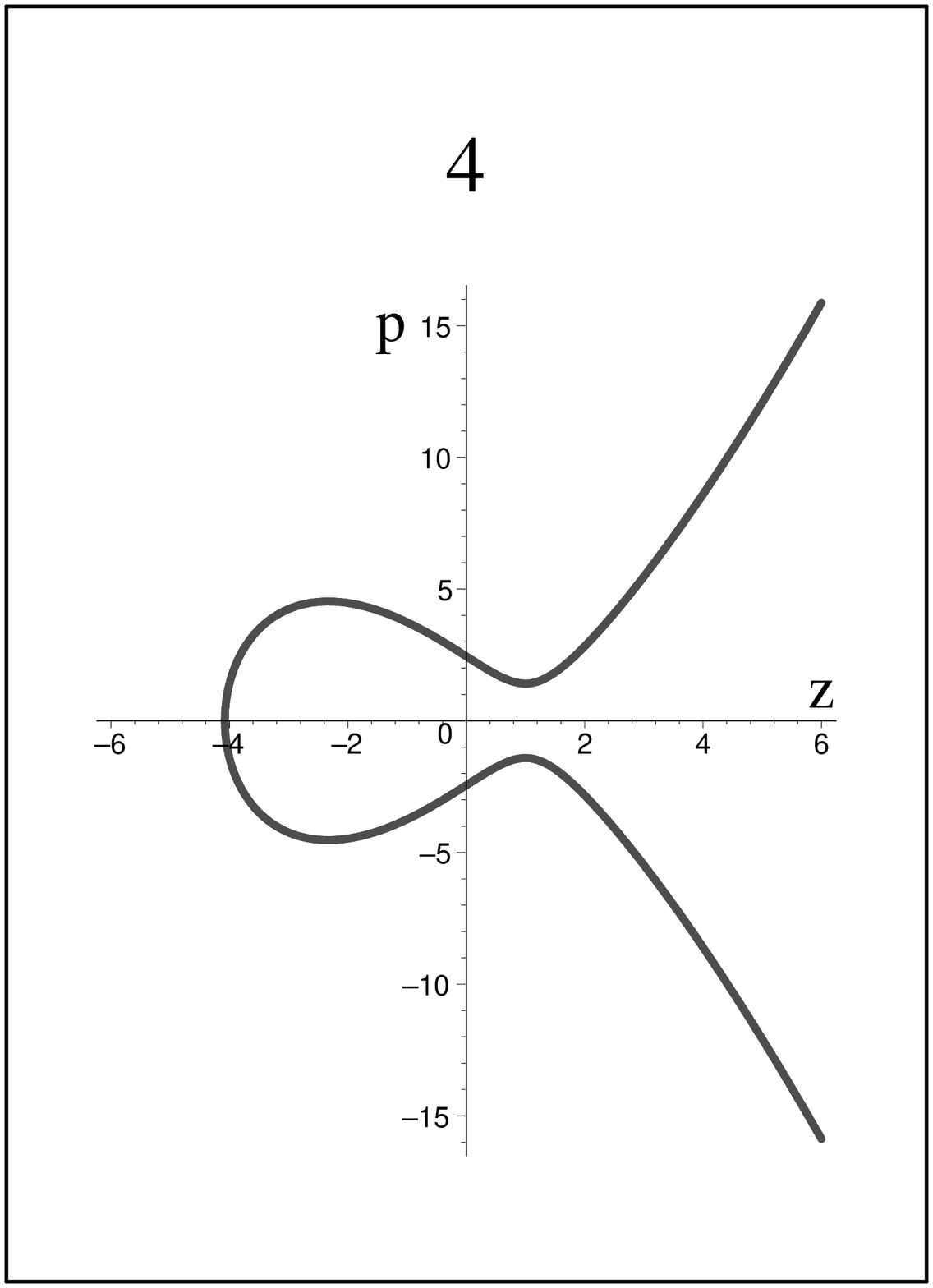}
& &
\includegraphics[width=4cm, height=5cm]{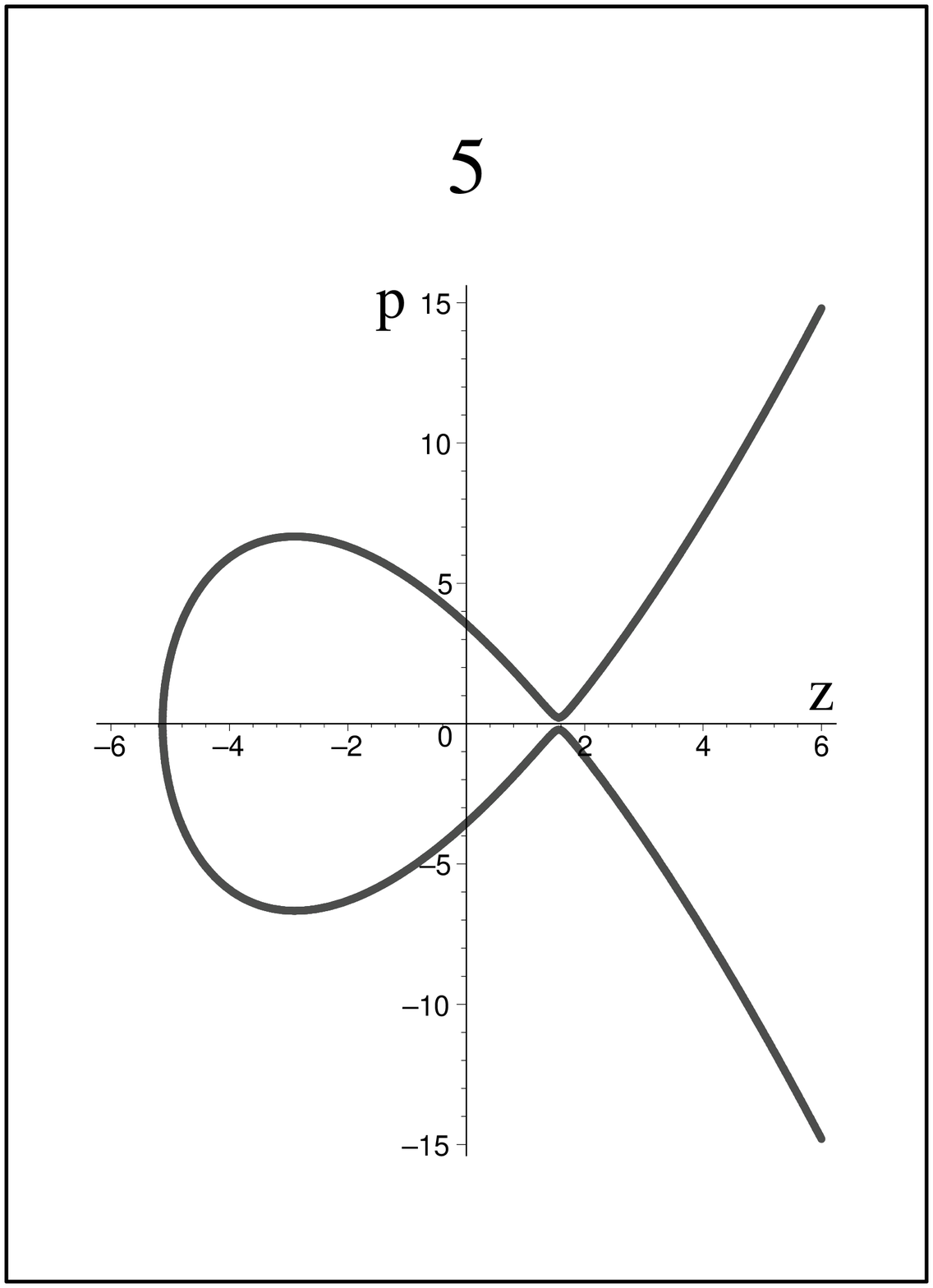}
& &
\includegraphics[width=4cm, height=5cm]{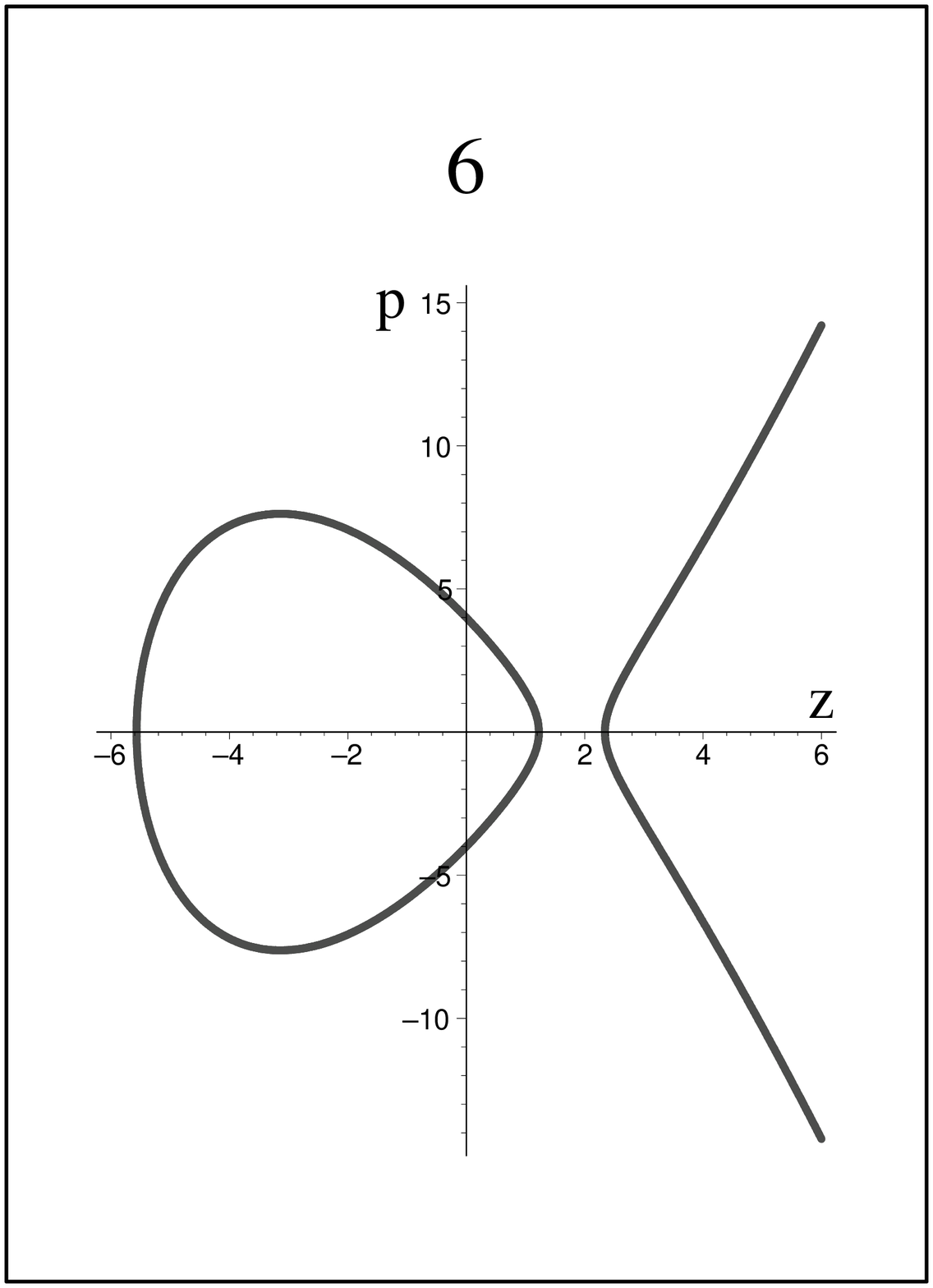}
\end{tabular}
\caption{First plot: $x=4, t=-1$, second plot: $x=4, t=1.5$, third plot: $x=4, t=2.1$, fourth plot: $x=4, t=5$, fifth plot: $x=4, t=8.3$, sixth plot: $x=4, t=10$.}
\label{fig:Elin1oscill}
\end{figure}

Richer transition phenomena are observed  for the solution (\ref{linear-soln-g1}) with $a=b=c=-d=1$, i.e.
\begin{equation}
u_3=2, \qquad u_1=3-2t, \qquad u_0=4+2(t-x).   
\end{equation}
In this case the curve is given by
\begin{equation}
\label{Ecp}
 {p_{{3}}}^{2}-{z}^{3}-2\,{z}^{2}- \left( 3-2\,t \right) z-2\,t-4+2\,x=0
\end{equation}
and the discriminant is
\begin{equation}
\Delta=-64(50+95\,{t}^{2}-70\,x+100\,t-90\,xt-8\,{t}^{3}+27\,{x}^{2}).
\end{equation}
The ``phase diagram'' (with the curve $\Delta=0$) is given in figure (\ref{fig:phtr}).
Transition across the curve $\Delta=0$ at $x>x_M$ and $x<x_m$ are qualitatively the same as shown in figures \ref{PE+triv} and \ref{PE-triv}. New regime occurs for the transition points with $x_m<x<x_M$. In this case changing deformation parameters along the line $x=const$ one has the process of formation of the bubble then its absorption and again formation fig. \ref{fig:Elin1oscill}. 
A new behavior appears in this evolution. Starting from a connected curve, a bubble borns and grows far from the boundary until it is  connected with the wall through a node. After that the node is desingularized and the (real section) of the curve returns connected. Then, always through a node, a bubble is generated again by the boundary and never comes back. \par
Higher order polynomial solution of the system (\ref{KdV-g1}) is given by
\begin{equation}
 \label{2oscill}
\eqalign{
u_3=2A+2E\,t,\\
u_1=-E^2\ t^2 +2D\ t +2E\ x +A^2+2B, \\
u_0=-(AE^2+ED)\ t^2 -(2AD+2A^2E)\ t +2AE\ x +2D\ x +2C+2AB
}
\end{equation}
where $A,B,C,D,E$ are constants.\par
\begin{figure}[h!]
\centering
\includegraphics[width=8cm, height=8cm]{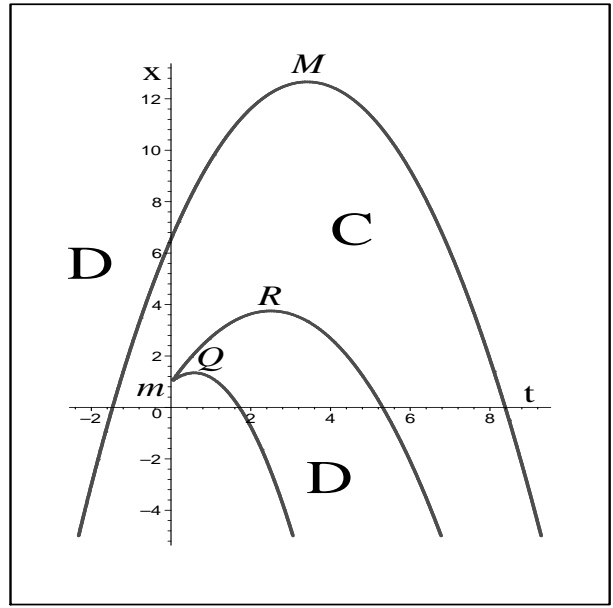}
\caption{On the curve the genus of the  cubic (\ref{Ecp}) is $0$, outside the curve it is $1$. In the D regions the real section of the elliptic curve is disconnected, in the C region it is connected. The point $M$ is the global maximum  of the curve, the points $Q$ and $R$ are local maxima and the point $m$ is a local minimum (cusp).}\label{fig:phtr2}
\end{figure}
\begin{figure}[h!]
\centering
\begin{tabular}{ccccc}
\includegraphics[width=4cm, height=5cm]{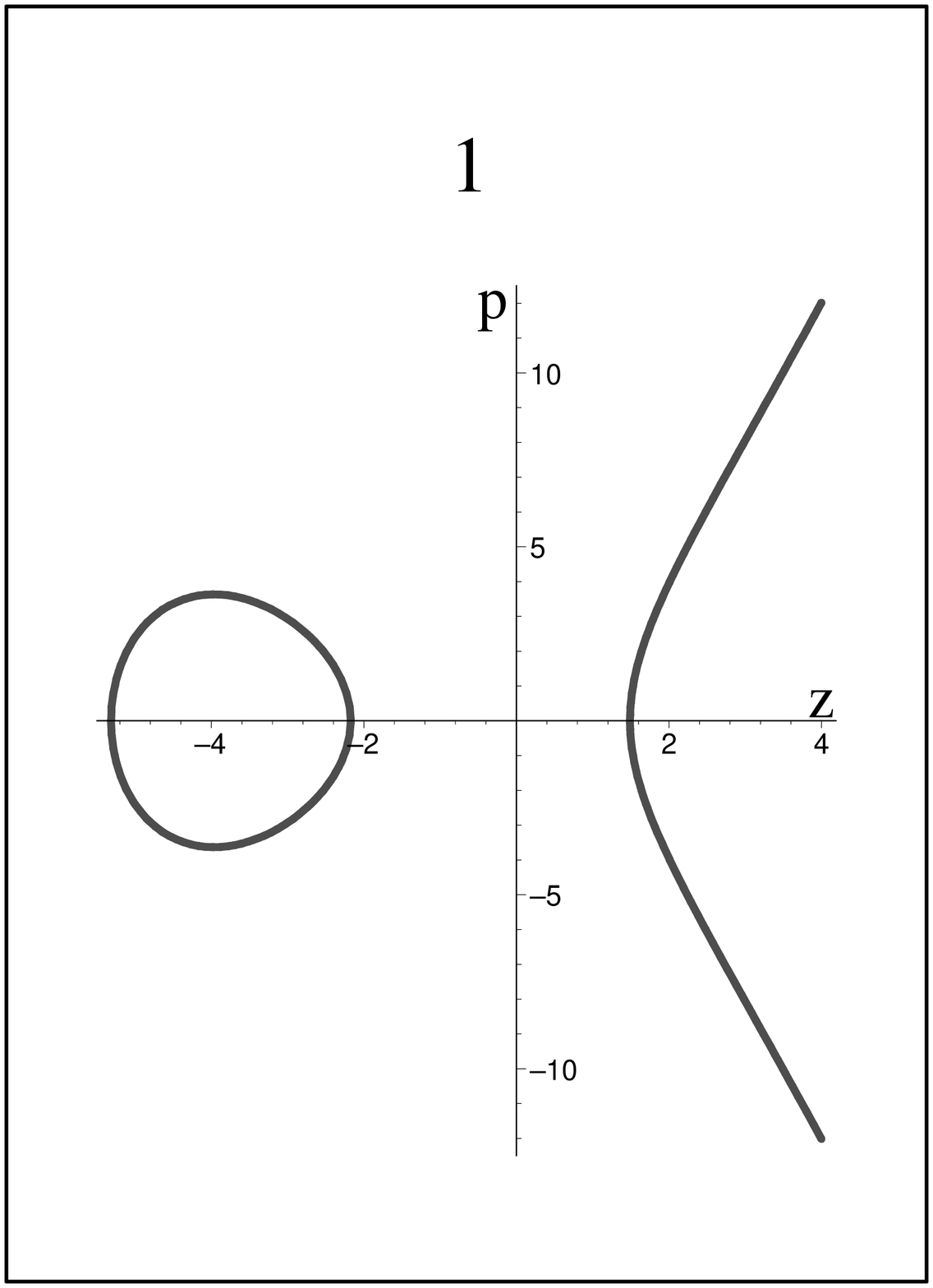}
& &
\includegraphics[width=4cm, height=5cm]{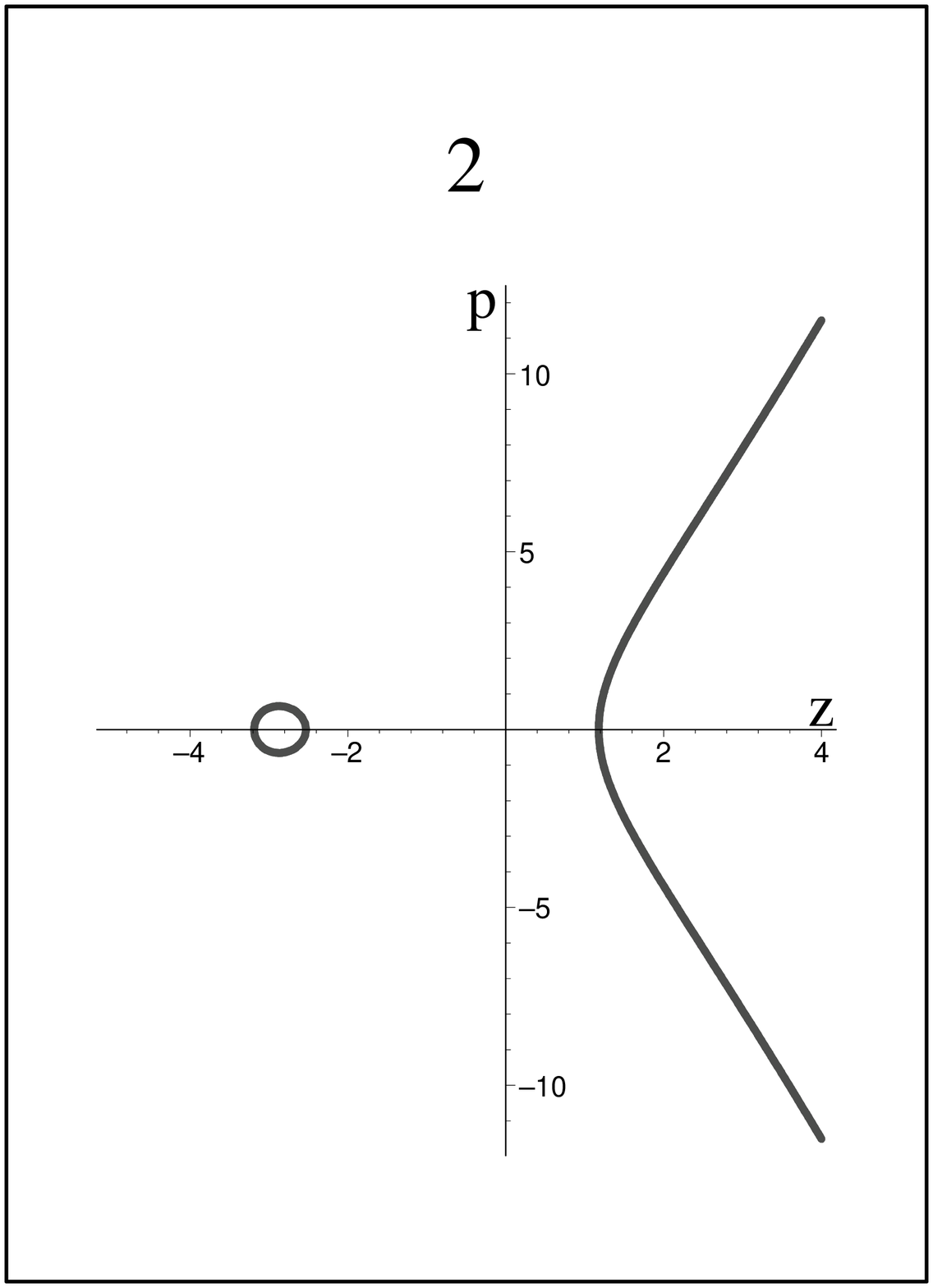}
& &
\includegraphics[width=4cm, height=5cm]{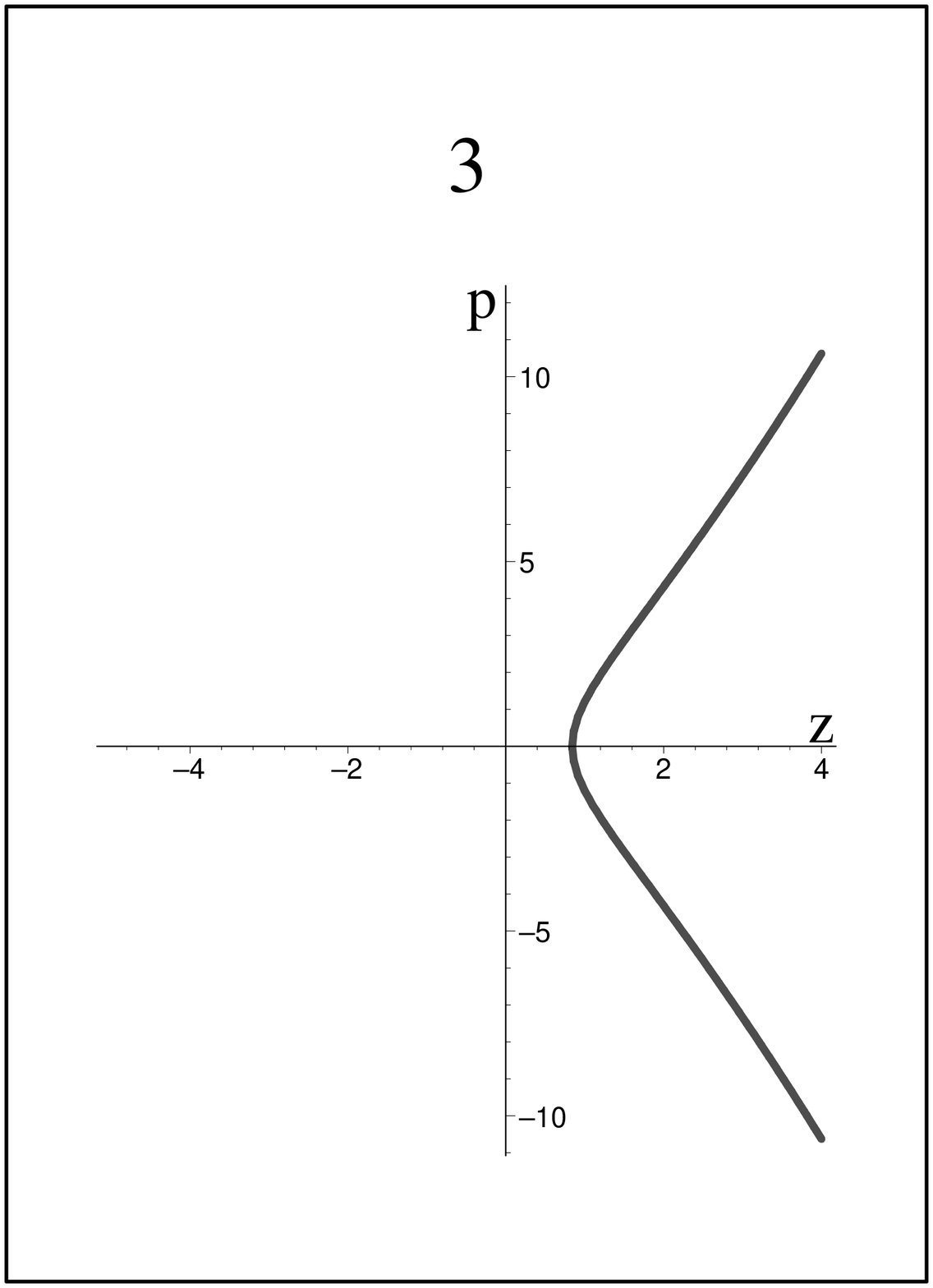}
\\
\includegraphics[width=4cm, height=5cm]{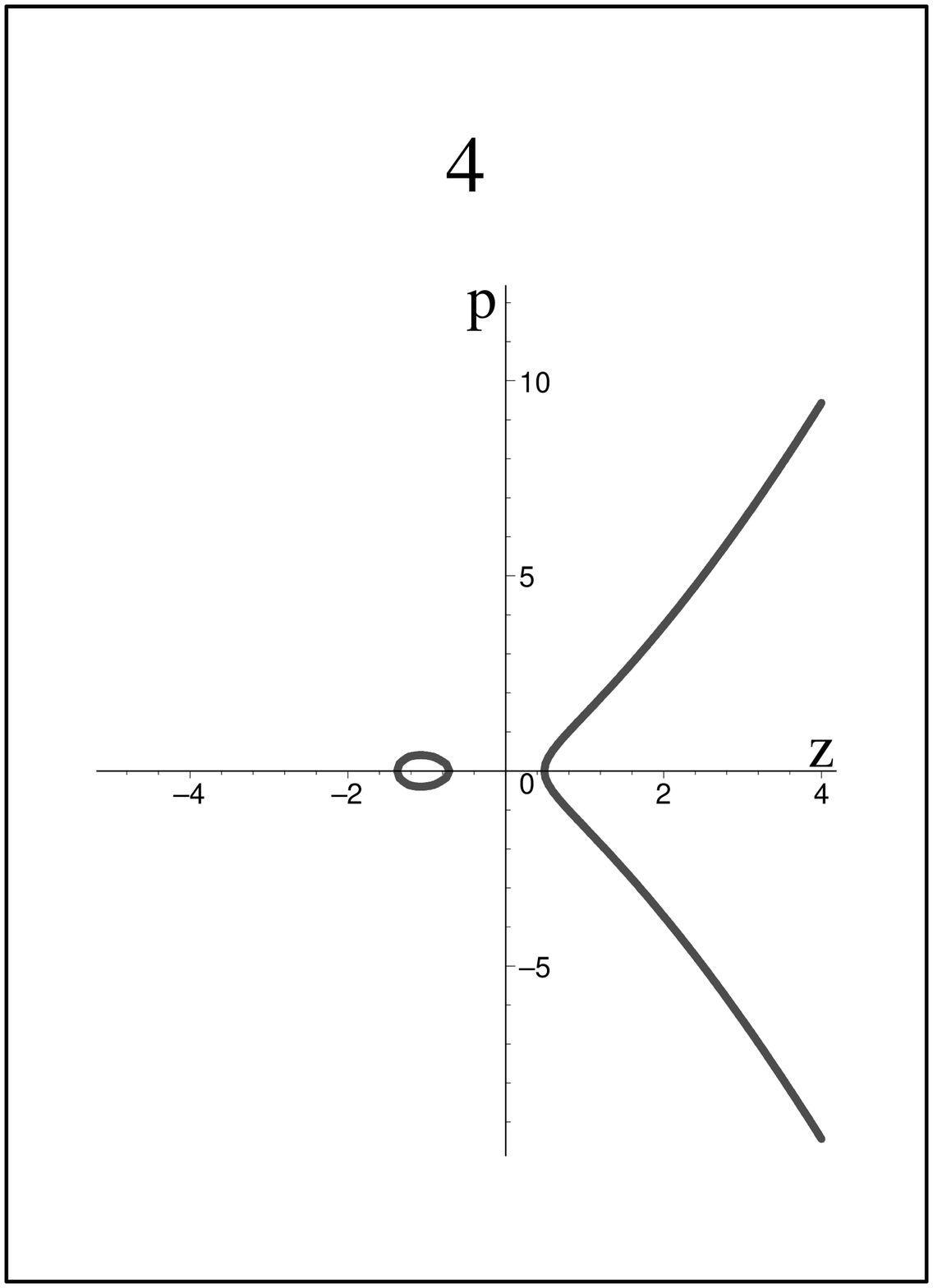}
& &
\includegraphics[width=4cm, height=5cm]{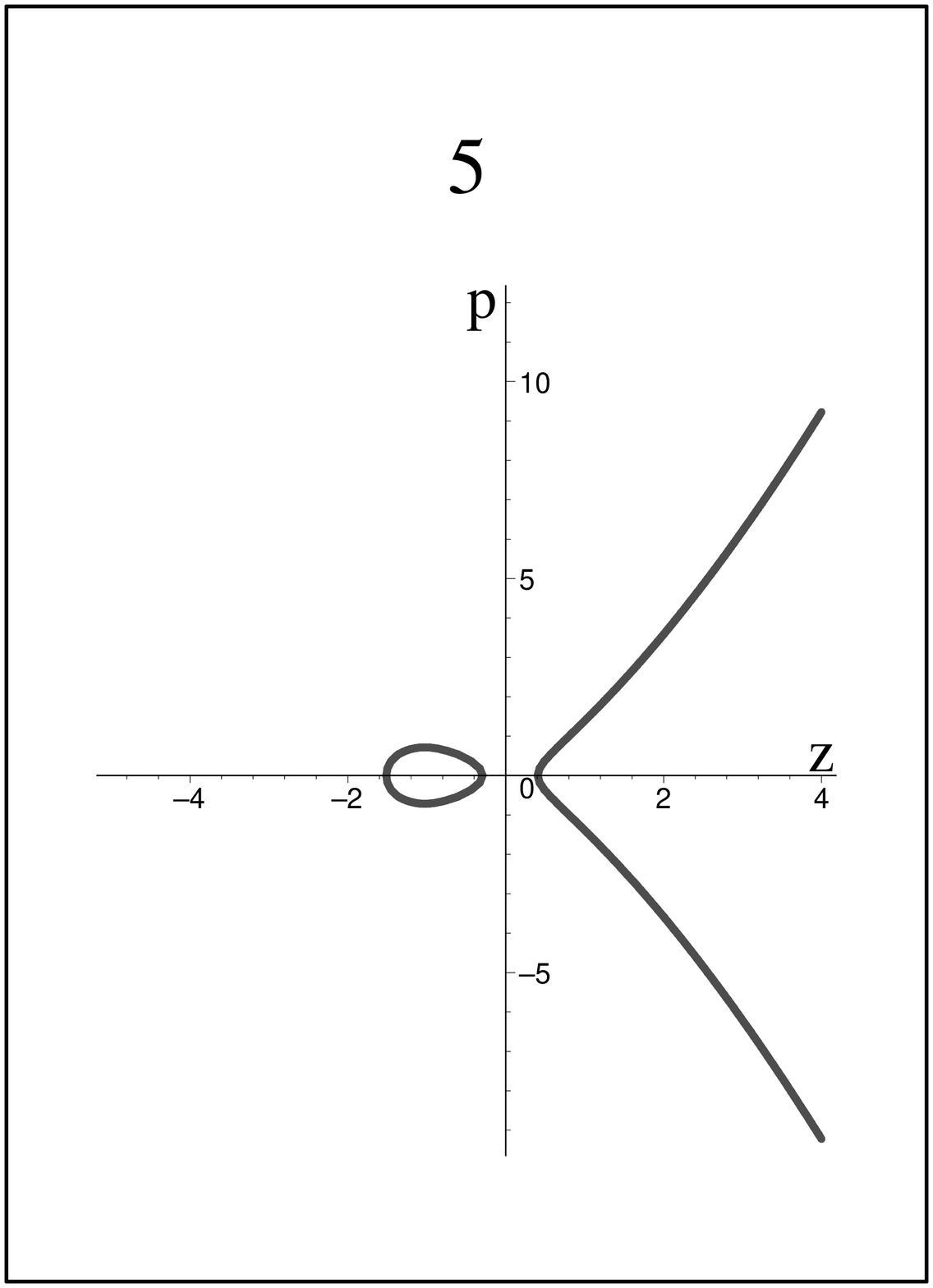}
& &
\includegraphics[width=4cm, height=5cm]{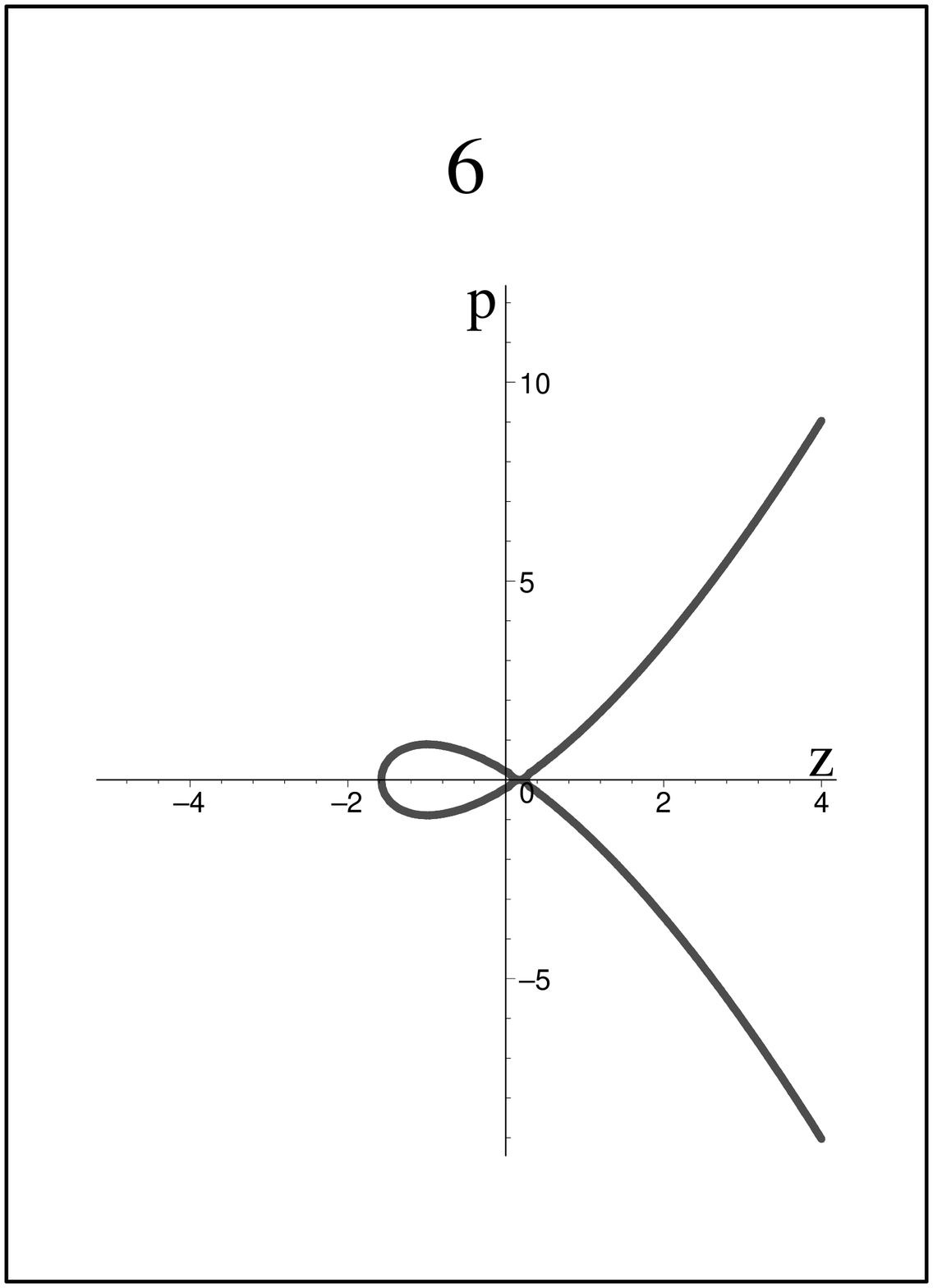}
\\
\includegraphics[width=4cm, height=5cm]{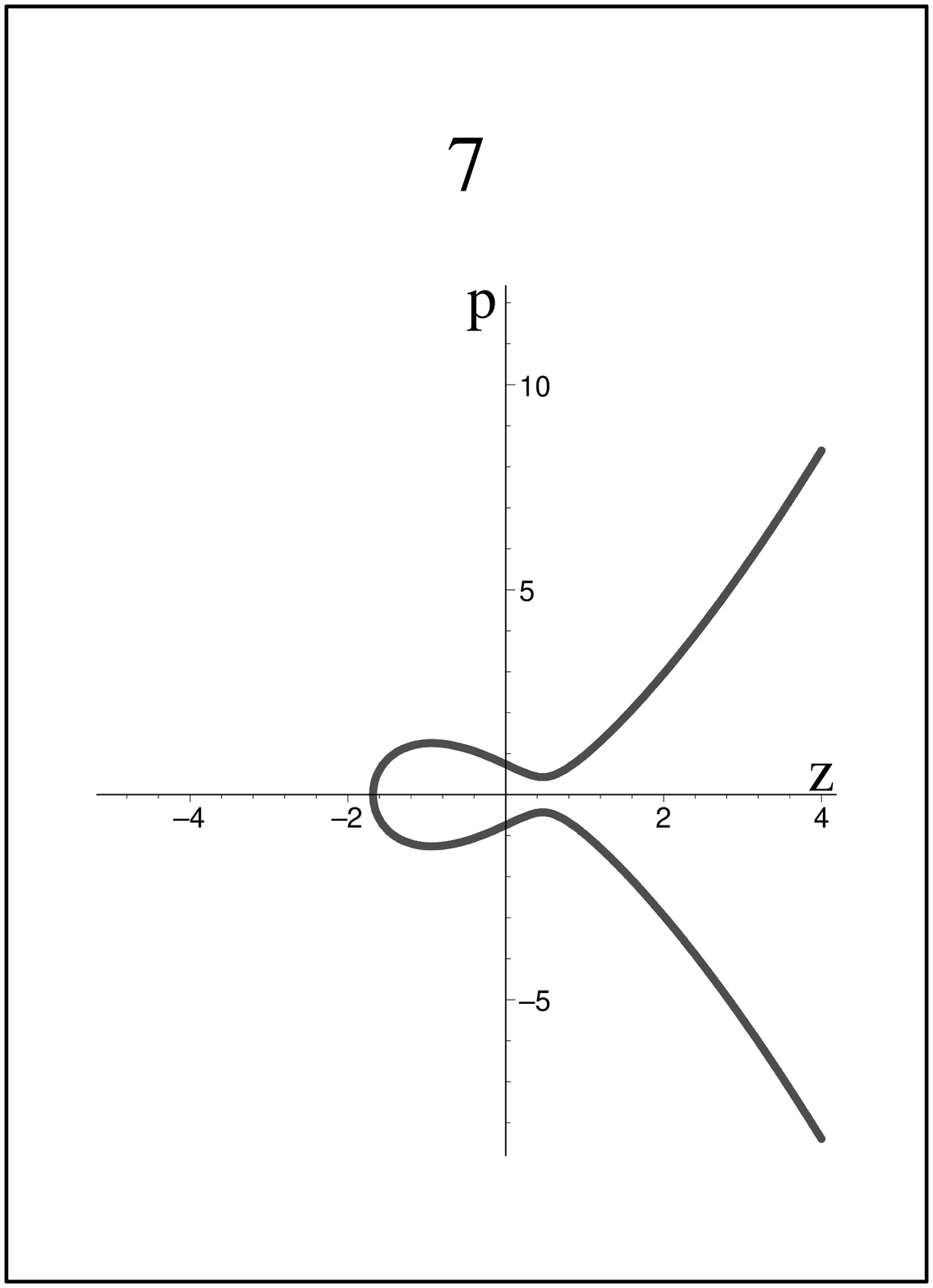}
& &
\includegraphics[width=4cm, height=5cm]{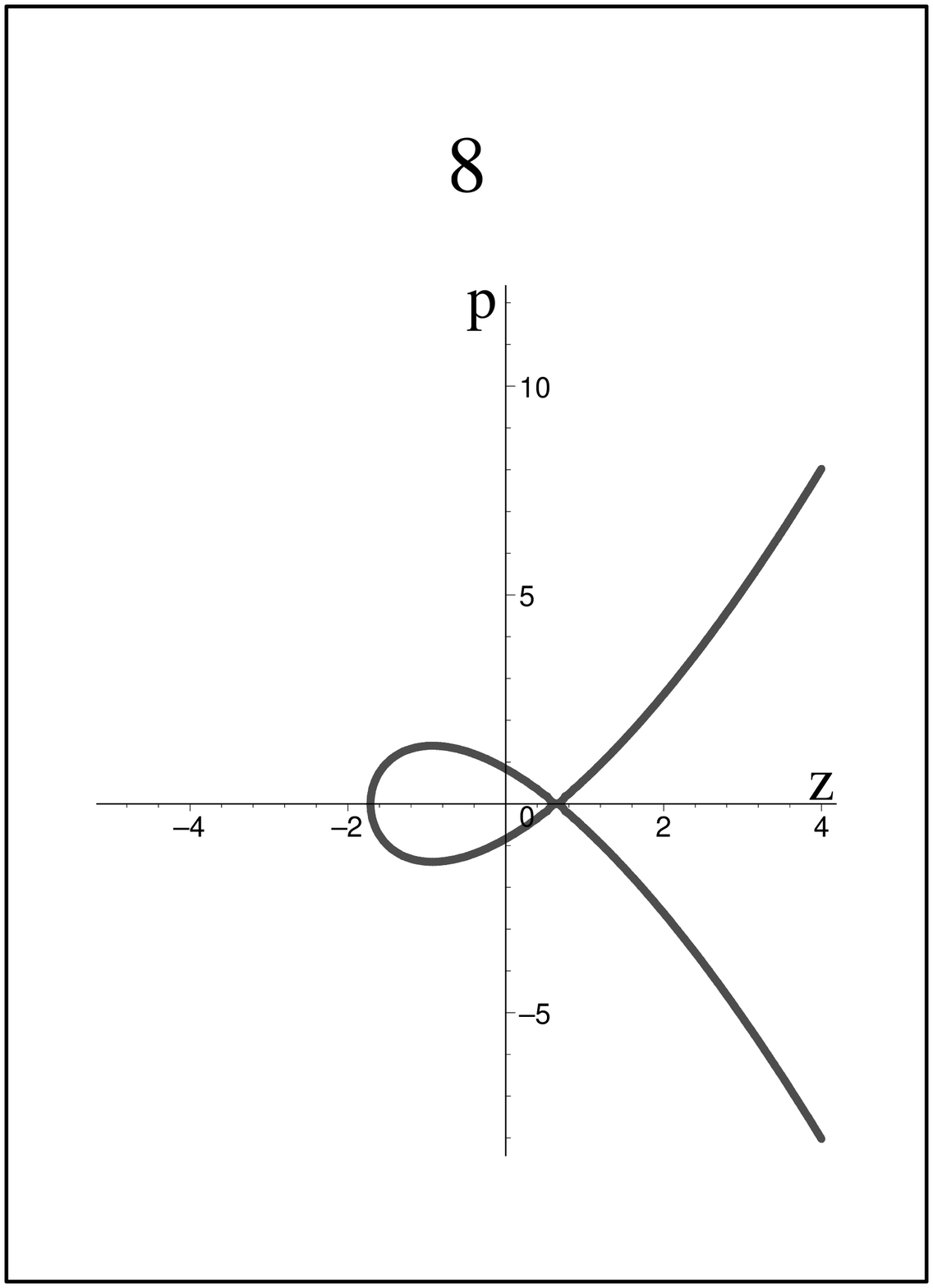}
& &
\includegraphics[width=4cm, height=5cm]{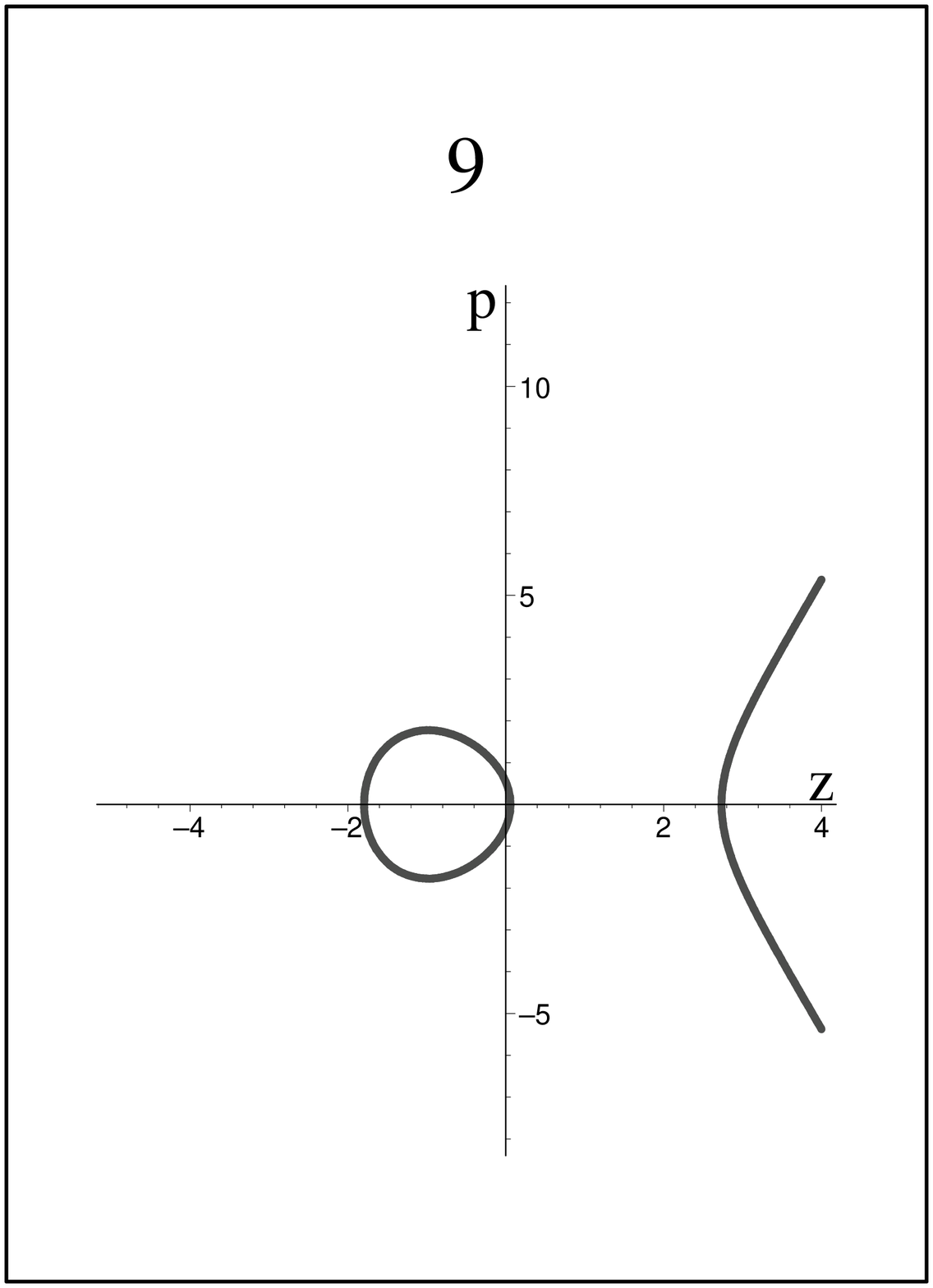}
\\
\includegraphics[width=4cm, height=5cm]{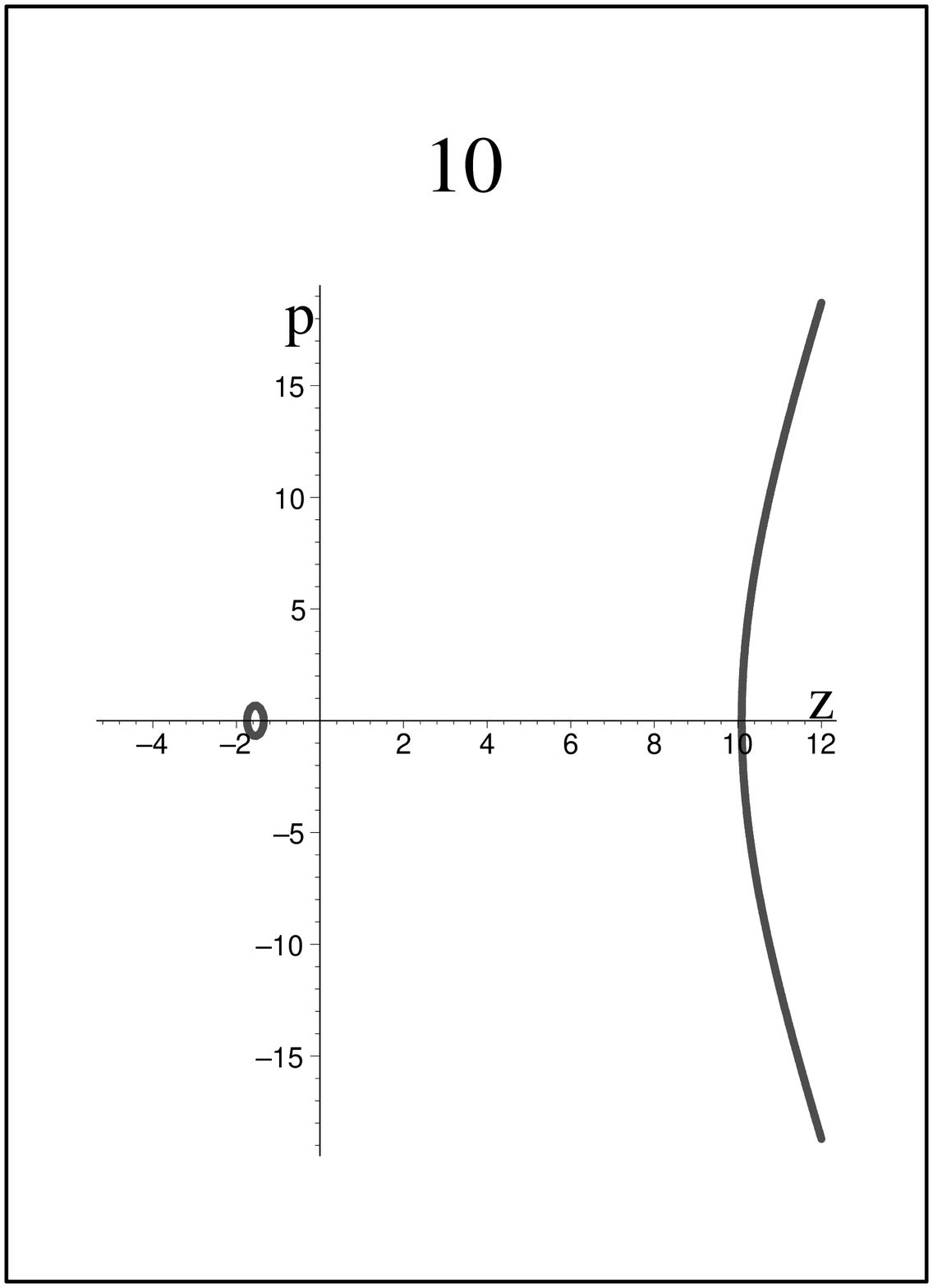}
& &
\includegraphics[width=4cm, height=5cm]{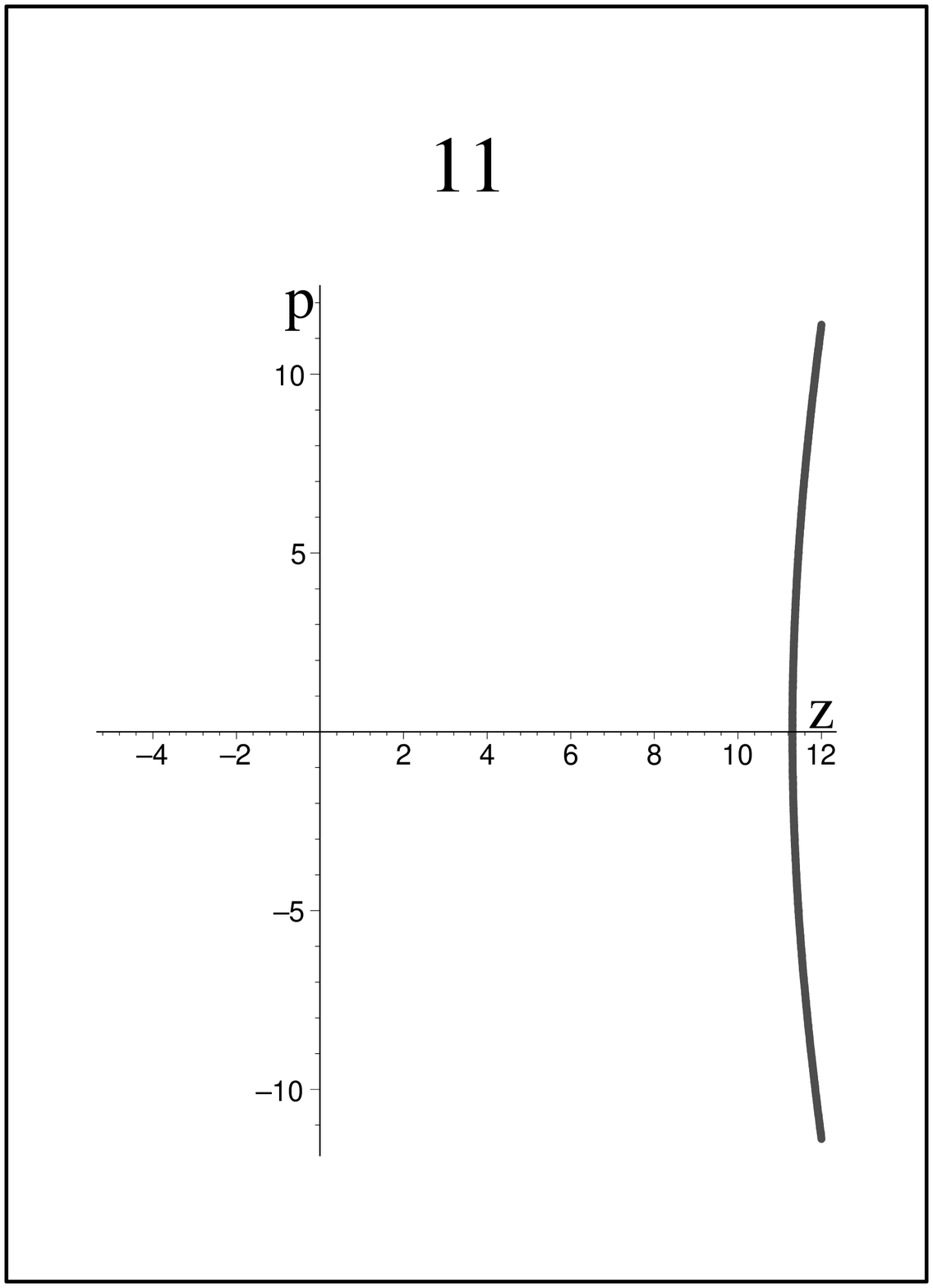}
& &
\includegraphics[width=4cm, height=5cm]{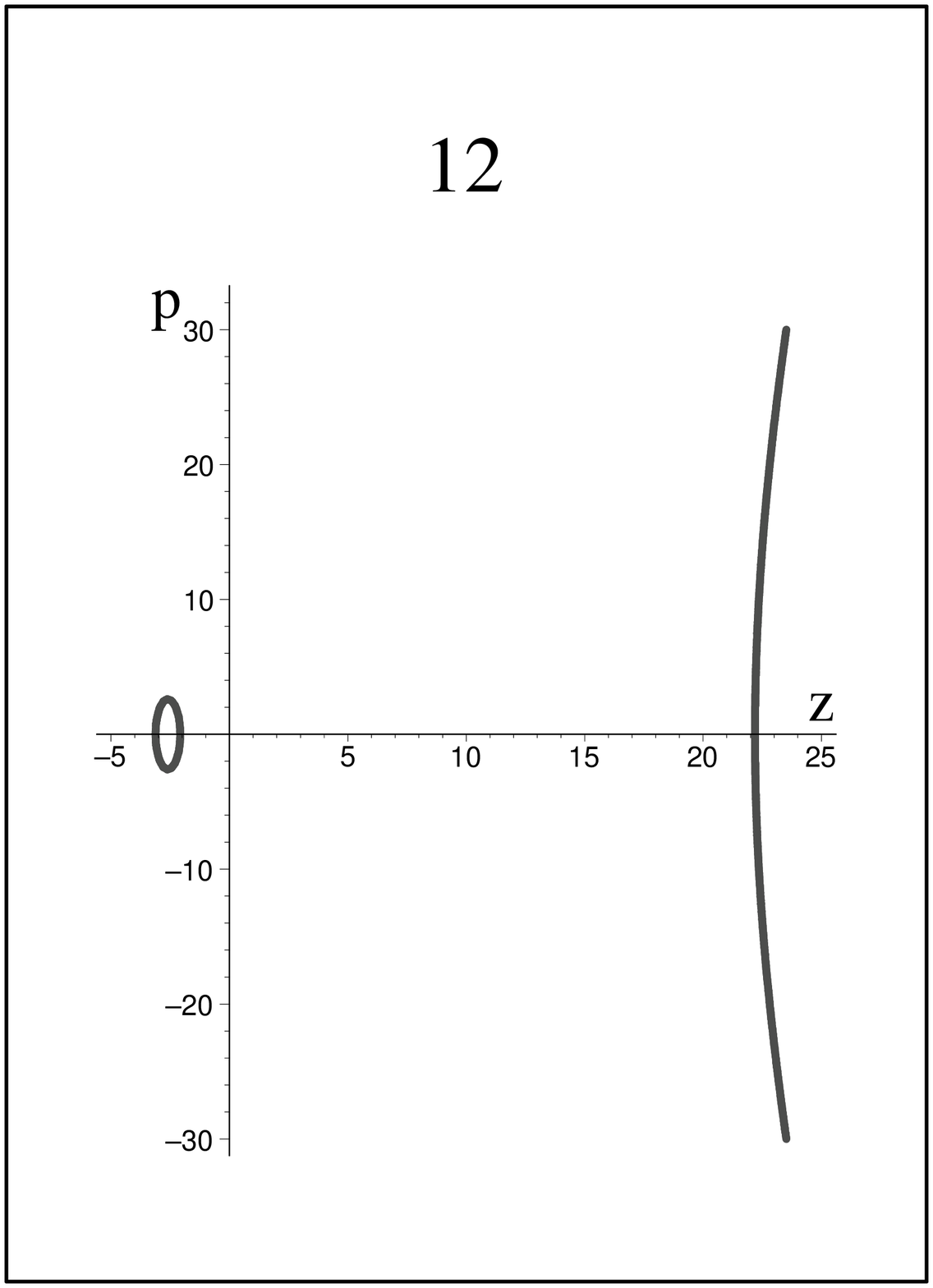}
\end{tabular}
\caption{First plot: $x=1.3, t=-2$, second plot: $x=1.3, t=-1.27$, third plot: $x=1.3, t=-0.5$, 
fourth plot: $x=1.3, t=0.2$, fifth plot: $x=1.3, t=0.3$, sixth plot: $x=1.3, t=0.3871$, 
seventh plot: $x=1.3, t=0.65$, eighth plot: $x=1.3, t=0.786$, ninth plot: $x=1.3, t=1.5$, 
tenth plot: $x=1.3, t=4.5$, eleventh plot: $x=1.3, t=5$, twelveth plot: $x=1.3, t=9.5$.}
\label{fig:2oscill1bubble}
\end{figure}

For the particular choice $A=B=C=-D=-E=1$ 
the elliptic curve becomes
\begin{equation}
p^2-z^3-(2-2t)z^2-(-t^2-2t-2x+3)z-(-2t^2+4t-4x+4)=0 
\end{equation}
and the discriminant is
\begin{equation}
\eqalign{
\Delta=&16\left(-200+260\,{t}^{4}-272\,{x}^{2}-800\,t+440\,x-104\,{t}^{5}+64\,{t}^{2}{
x}^{2}+8\,{t}^{6}+40\,{t}^{4}x\right.\\&\left.-320\,{t}^{3}x-224\,{x}^{2}t+32\,{x}^{3}
+1040\,xt+96\,{t}^{2}x-640\,{t}^{2}+504\,{t}^{3}\right).
}
 \end{equation}
The ``phase diagram'' containing the six order curve $\Delta=0$ is presented in figure (\ref{fig:phtr2}).\\
For such deformation of the elliptic curve one observes much richer structure of transition regimes. At $x>x_{M}$ the curve remain disconnected for all values of deformation parameters. At $x_{M}>x>x_R$ the transition involves the absorption and the recreation of a bubble. For $x_{R}>x>x_Q$ the transition involves a double absorption and creation of a bubble. At $x<x_m$ one has the replication of this regime. For $x_{Q}>x>x_m$ the transition process  goes through a complicated oscillation depicted in figure \ref{fig:2oscill1bubble} of the bubble interacting with the wall by means of many different nodal critical points. 
 The above simple examples demonstrate the richness of possible transition processes between connected and disconnected (with a bubble) real sections of elliptic curve described by the system (\ref{KdV-g1}).
\section{Singular elliptic curve and Burgers-Hopf equation}
\label{sect-singevo}
\par For all processes of formation and annihilation of bubbles described in the previous section the elliptic curve passes through the ``singular point'' $x^*,t^*$ with $\Delta(x^*,t^*)=0$ at which the curve becomes degenerated (rational). For example, for the simplest solution (\ref{triv-soln}) the elliptic curve assumes the form (\ref{ell-sing-par}) for the deformation parameters obeying equation (\ref{phdiagrtriv}). The cubic curve (\ref{ell-sing-par}) is degenerated and possesses a double point for $s \neq 0$ and a cusp for $s=0$ (see e.g. \cite{S}).\par
More generally, an elliptic curve written in terms of moduli, i.e. 
\begin{equation}
\label{ellcurveg2g3}
 p^2-(z^2+g_2z+g_3)=0
\end{equation}
is degenerated if $4{g_2}^3+27{g_3}^2=0$. In this case the curve (\ref{ellcurveg2g3}) can be presented in the form
\begin{equation}
 \label{singcubicBH}
p^2-(z+2u)(z-u)^2=0
\end{equation}
where $u$ is the uniformizing variable defined by $g_2=-3u^2$ and $g_3=2u^3$.\\
Singular cubic curve (\ref{singcubicBH}) has a well known rational parameterization 
\begin{equation}
\label{ratparsingcurveBH}
 \eqalign{
z=q^2-2u,\\
p=q^3-3uq.
}
\end{equation}
Indeed, introducing the variable $q$ by relation
\begin{equation}
 \label{pqratpar}
p=q(z-u),
\end{equation}
one represents equation (\ref{singcubicBH}) as
\begin{equation}
 \label{q2uratpar}
q^2-(z+2u)=0.
\end{equation}
Two equations (\ref{pqratpar}) and (\ref{q2uratpar}) are obviously equivalent to (\ref{ratparsingcurveBH}).\par
\begin{figure}[h!]
\centering
\begin{tabular}{ccccc} 
\includegraphics[width=4cm, height=5cm]{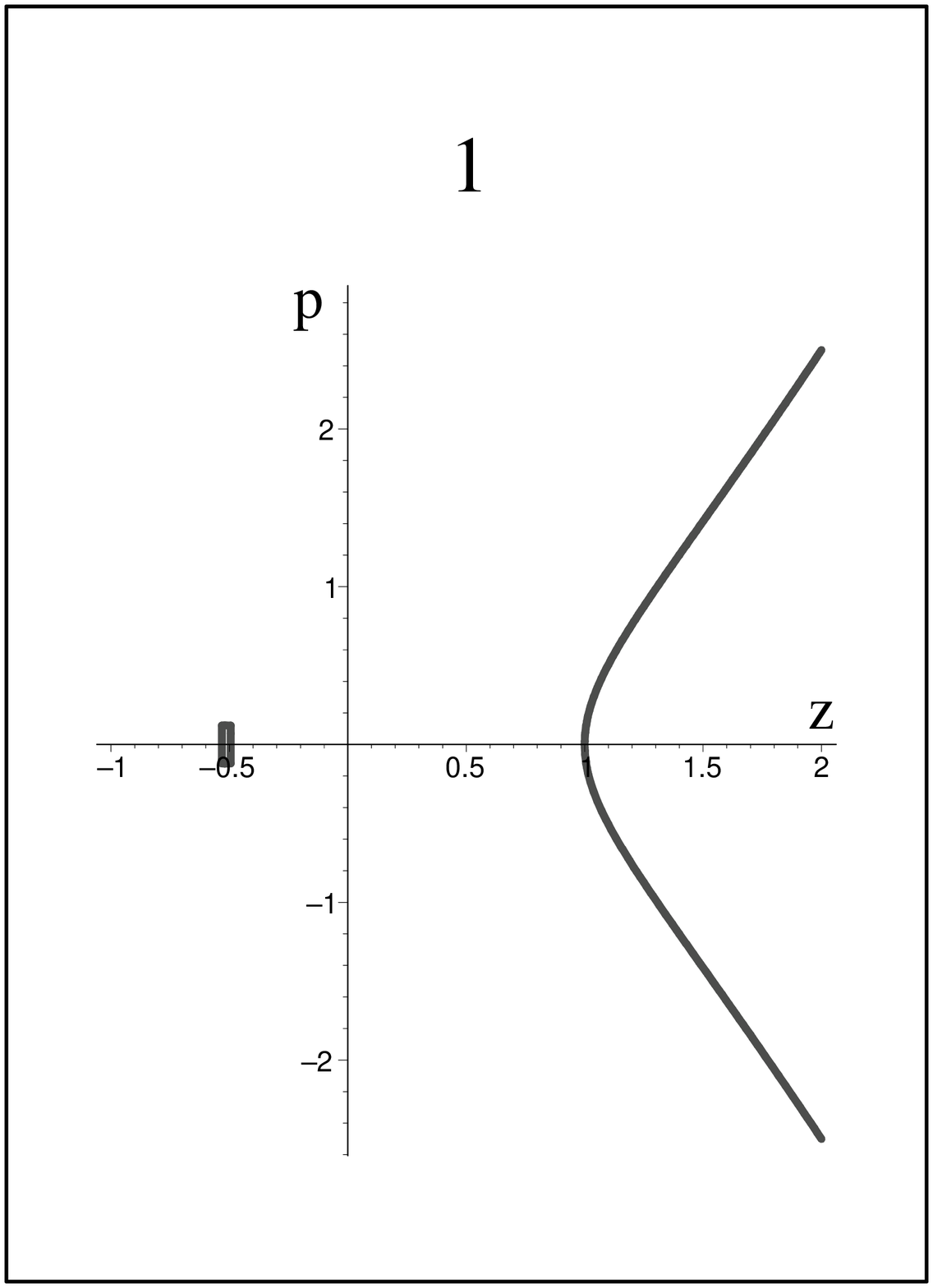}
& &
\includegraphics[width=4cm, height=5cm]{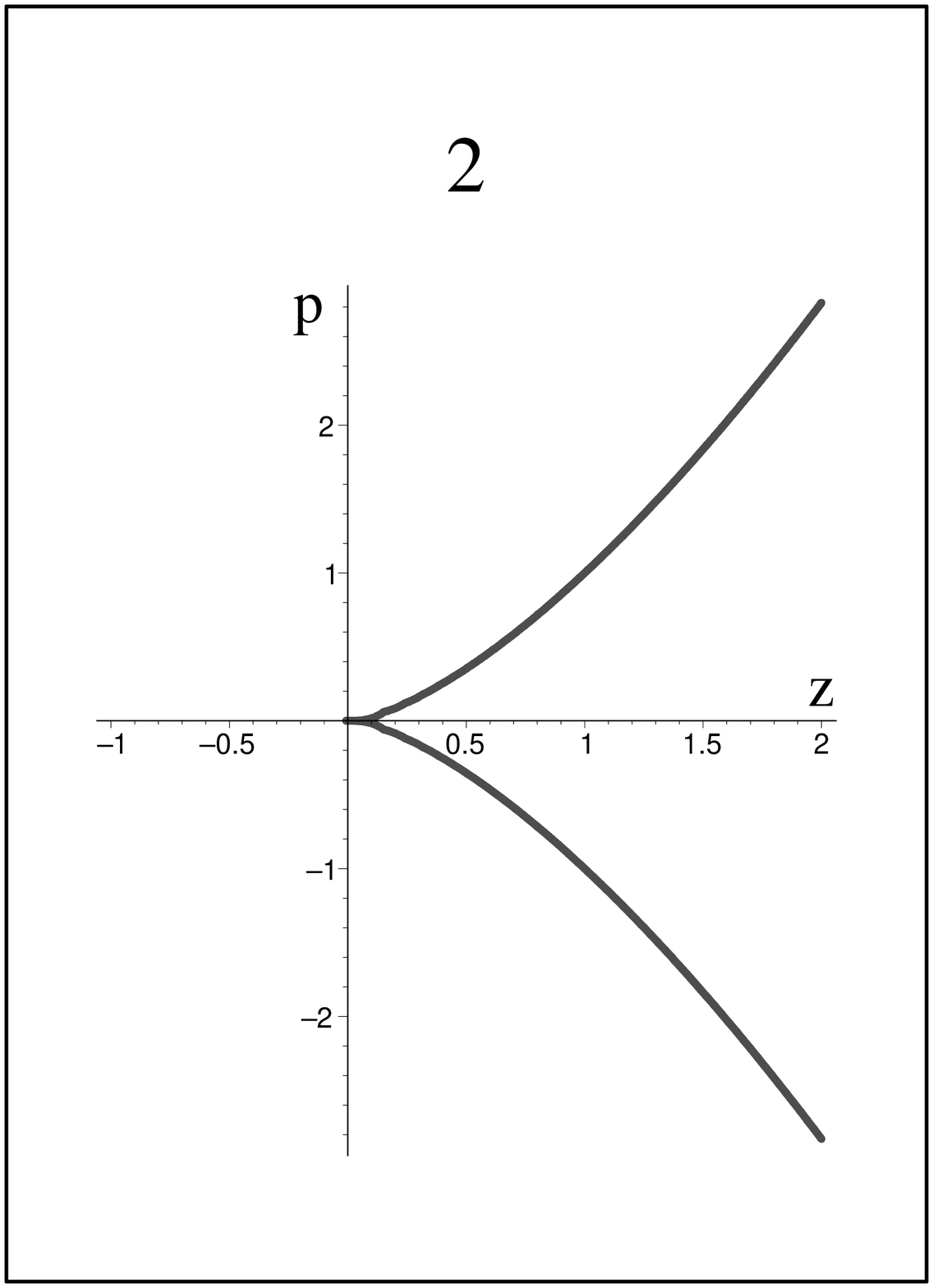} 
& &
\includegraphics[width=4cm, height=5cm]{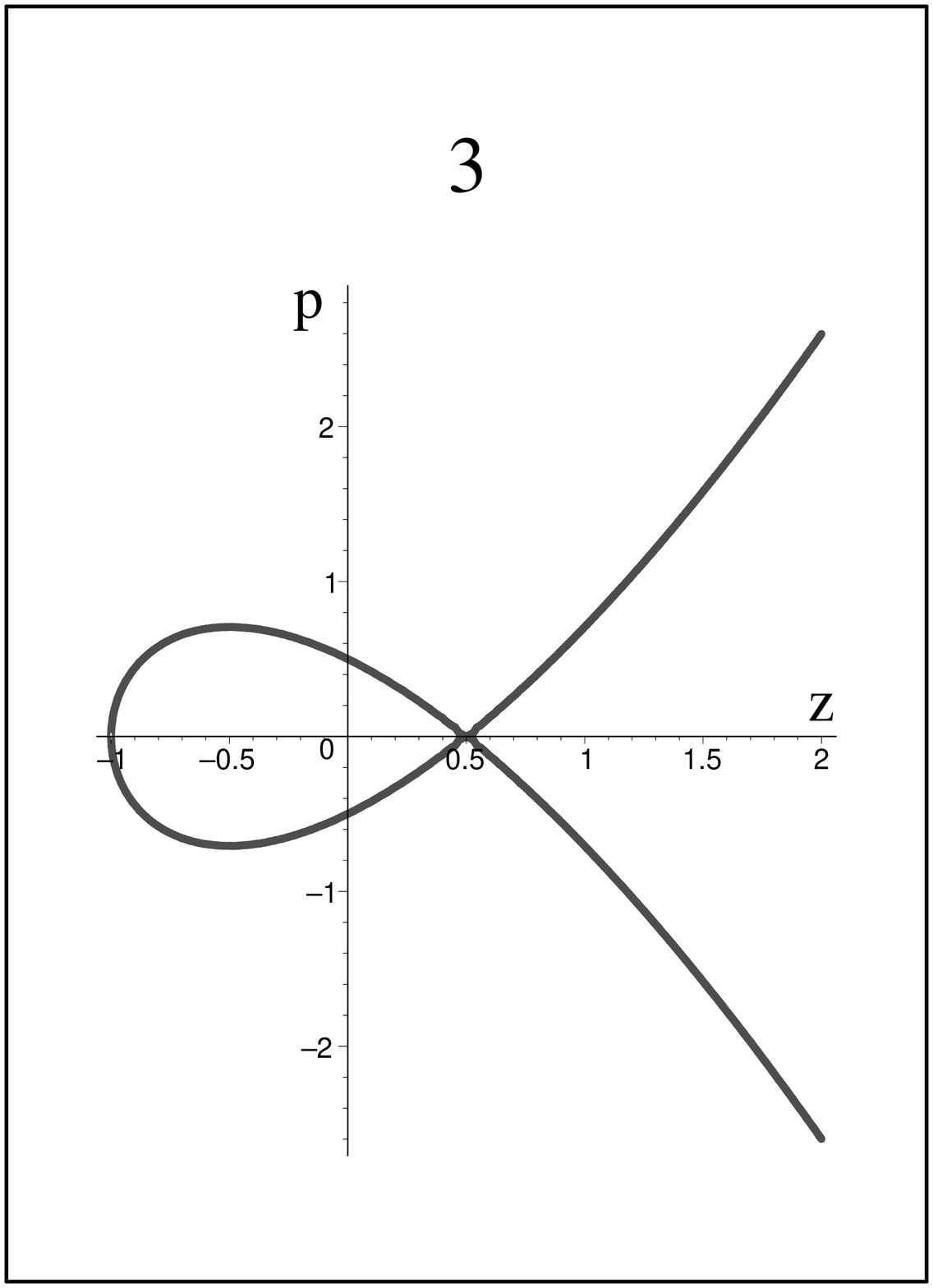}
\end{tabular}
\caption{First plot: $y=1.5\ y_0, x=1/3$ ($u$ negative), second plot: $y= y_0, x=1/3$ ($u$ null), third plot: $y=0.5\ y_0, x=1/3$ ($u$ positive). In the first plot there is also an isolated point in $(z,p)=(-0.5,0)$}
\label{fig-cusptc}
\end{figure}

The parameterization (\ref{ratparsingcurveBH}) clearly indicates the interrelation between the degenerated cubic curve (\ref{singcubicBH}) and the  Burgers-Hopf (dispersionless Korteweg-de Vries) equation
\begin{equation}
 \label{BHeqn}
u_x=3uu_y.
\end{equation}
Equation (\ref{BHeqn}) is known to be equivalent  to the compatibility condition for two Hamilton-Jacobi equations (see e.g. \cite{KG,K2,Z})
\begin{equation}
\label{compcondBH}
\eqalign{
 (\partial_y S)^2 -2u =z, \\
\partial_x S = (\partial_y S)^3-3u\partial_y S. 
}
\end{equation}
In our approach equation (\ref{BHeqn}) describes the Hamiltonian deformations of the parabola $f=z-q^2+2u(y,x)$ generated by the Hamiltonian $H=q^3-3u(y,x)q$ or coisotropic deformations of the common points of the curves (\cite{KO})
\begin{equation}
\label{coisoBH}
\eqalign{
 f=z-q^2+2u(y,x)=0,\\
g=p+q^3-3uq=0.
}
\end{equation}
with the standard Poisson bracket $\{f,g\}=\partial_yf\partial_qg-\partial_yg\partial_qf
+ \partial_xf\partial_pg-\partial_xg\partial_pf$. 
 Thus the auxiliary problem (\ref{coisoBH}) for equation (\ref{BHeqn}) represents nothing else than the parameterization (\ref{ratparsingcurveBH}) of the cubic curve (\ref{singcubicBH}). So, any solution $u(x,t)$ of the Burgers-Hopf equation
(\ref{BHeqn}) provides us with the family of cubic curves (\ref{singcubicBH}) parameterized by the variables $x$ and $y$.
The solution
\begin{equation}
\label{BHsoln}
 u=\frac{y_0-y}{3x}
\end{equation}
 of equation (\ref{BHeqn}) ($y_0$ is a constant) describes simple Hamiltonian deformation of the degenerated cubic (\ref{singcubicBH}) shown in figure (\ref{fig-cusptc}).
The deformation (\ref{BHsoln}) connects singular points of  different regimes shown in the figures 
\ref{PE+triv},\ref{PE-triv},\ref{PE0triv}, i.e. curve with a separate point, cusp and node. This observation shows the relevance of the Burgers-Hopf (dKdV) equation for the description of the singular sector in the process of deformation of elliptic curves. Within the study of the Laplace growth process this fact has been observed earlier in papers \cite{BAZW,KMWZ,LTW}. \par
The inverse process of desingularization on the degenerated cubic curve, i.e. the passage from the curve (\ref{singcubicBH}) to the curve (\ref{ellcurve}) can be viewed at least in two different ways. One is based on the Birkhoff stratification for the Burgers-Hopf hierarchy. In this approach the passage from (\ref{singcubicBH}) to (\ref{ellcurve}) is associated with the transition from the big cell of the Sato Grassmannian to the first stratum \cite{KK}. Another approach discussed in \cite{BAZW,KMWZ,LTW} suggests the dispersive regularization. Possible interconnection between these two approaches still remains an open problem.
\section{Deformations of plane quintic}
\par
\begin{figure}[h!]
\centering
\begin{tabular}{ccccc}
 \includegraphics[width=4cm, height=5cm]{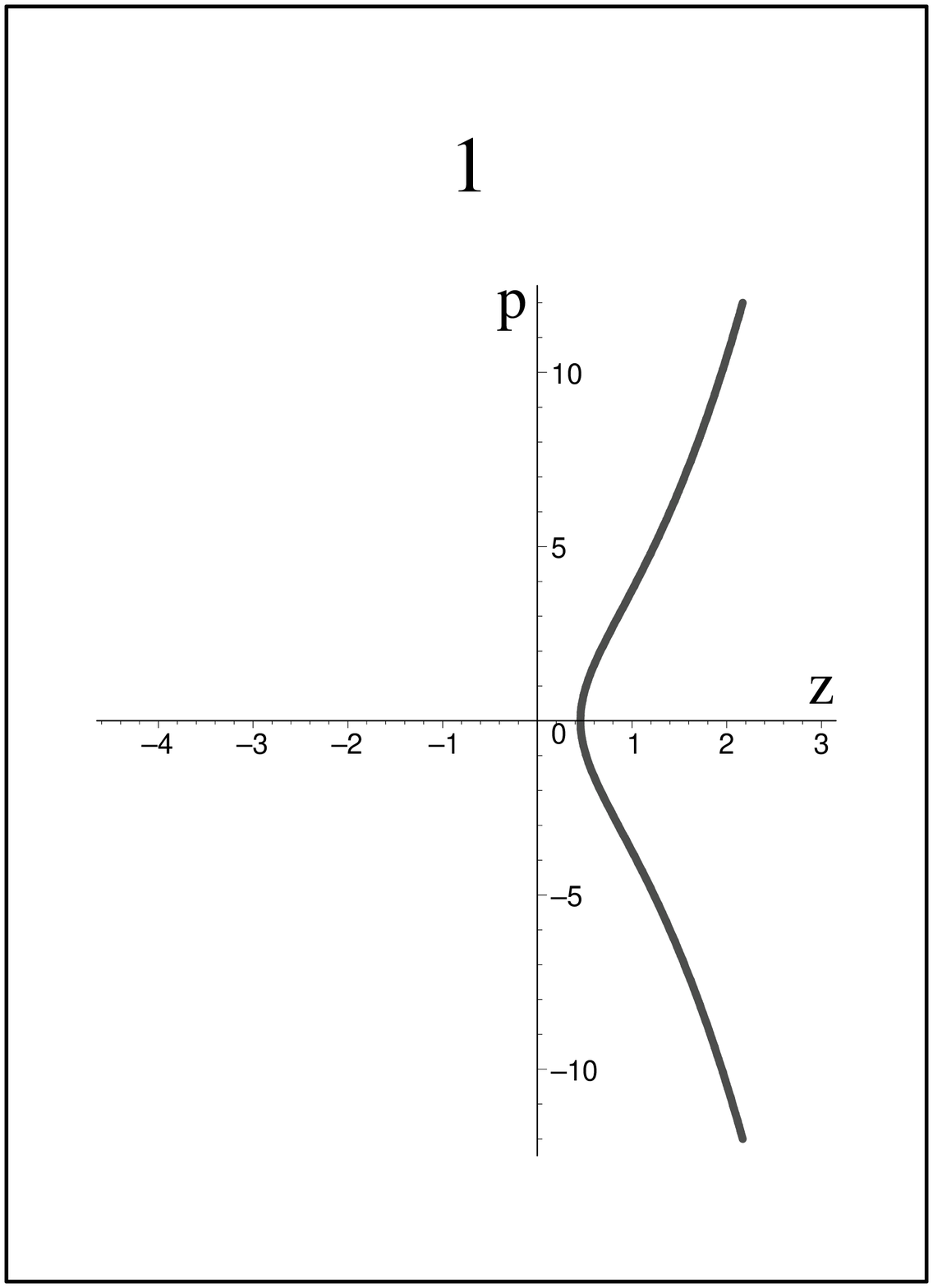}
& &
\includegraphics[width=4cm, height=5cm]{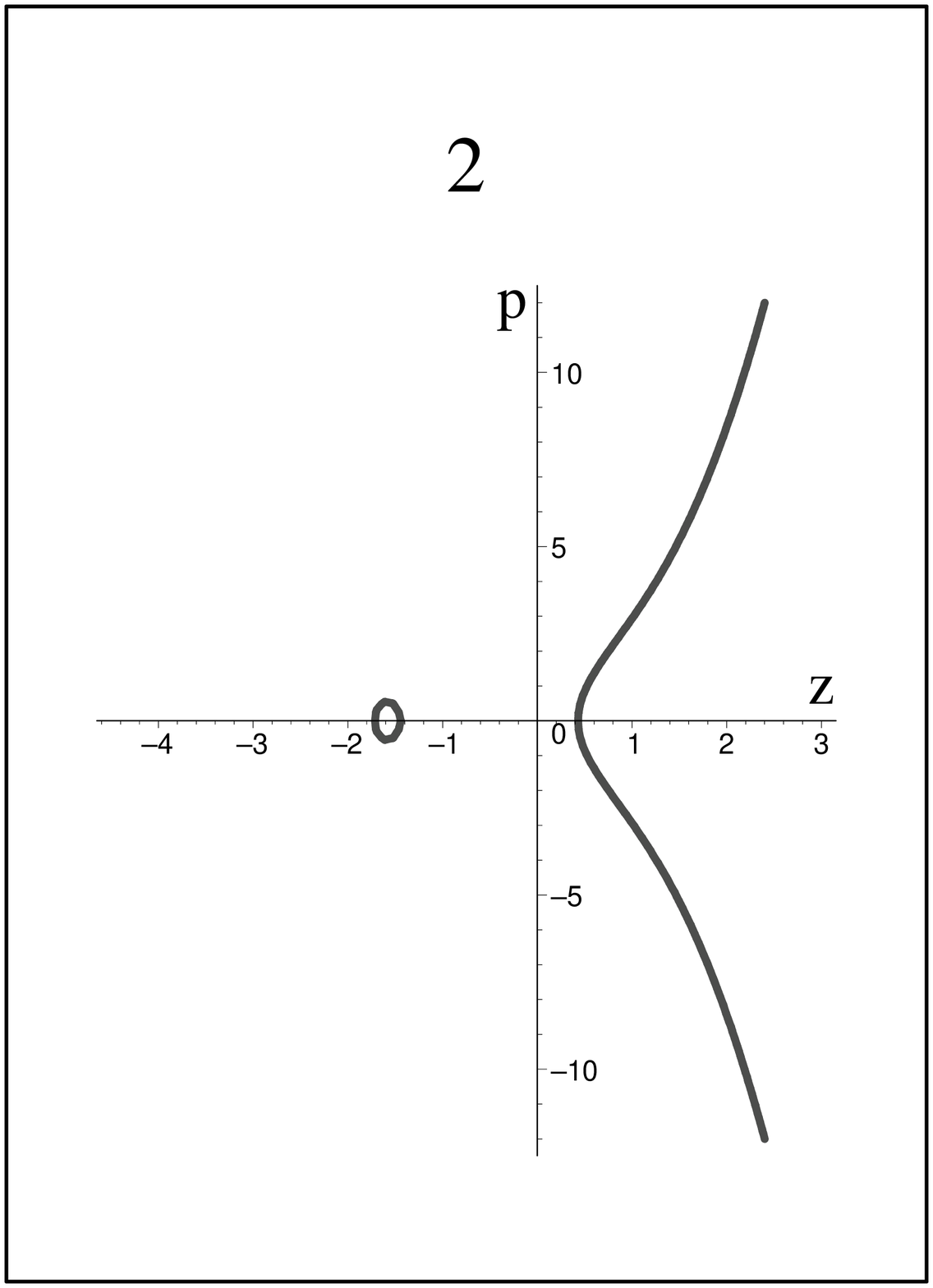}
& &
\includegraphics[width=4cm, height=5cm]{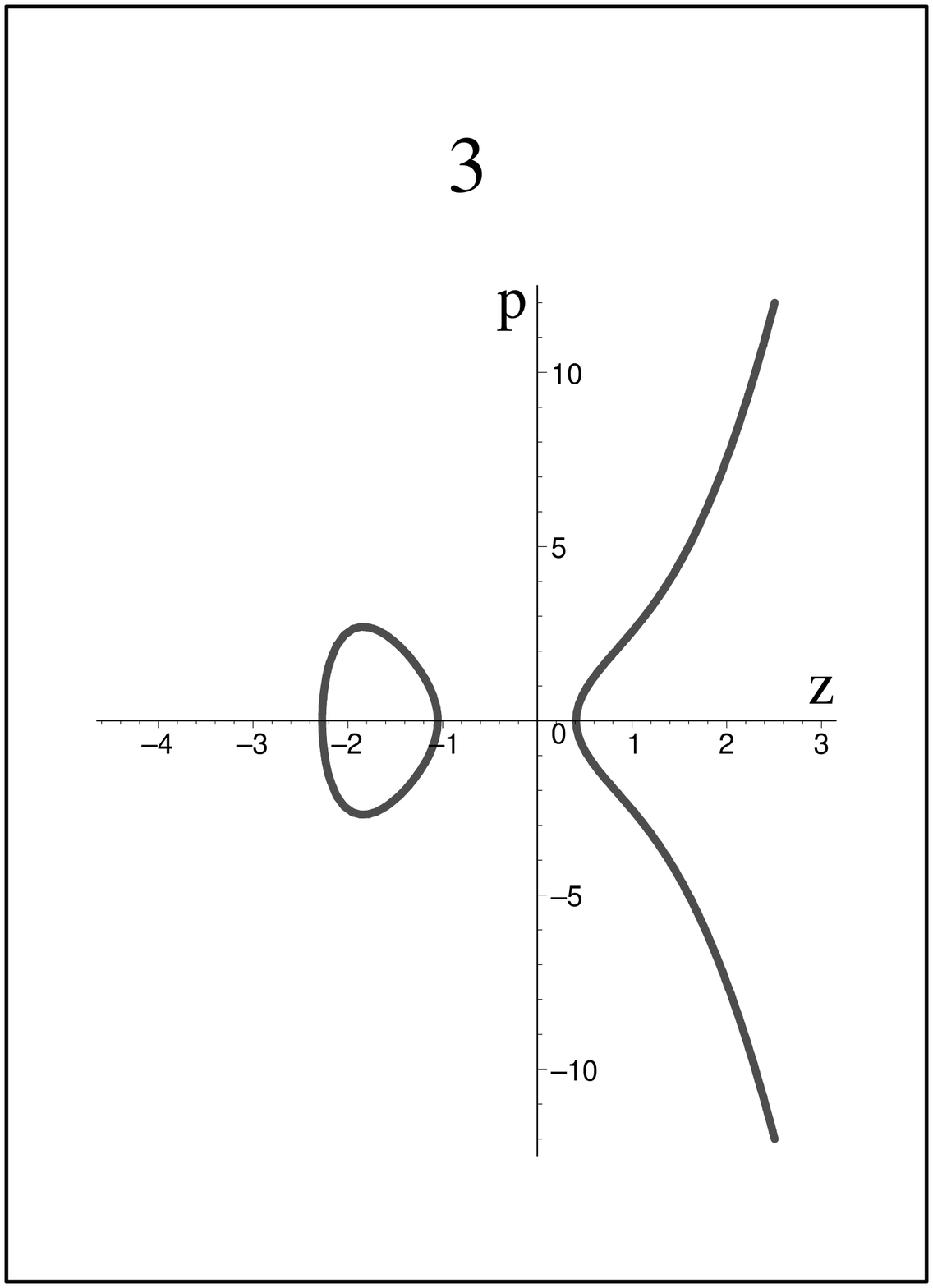}
\\
\includegraphics[width=4cm, height=5cm]{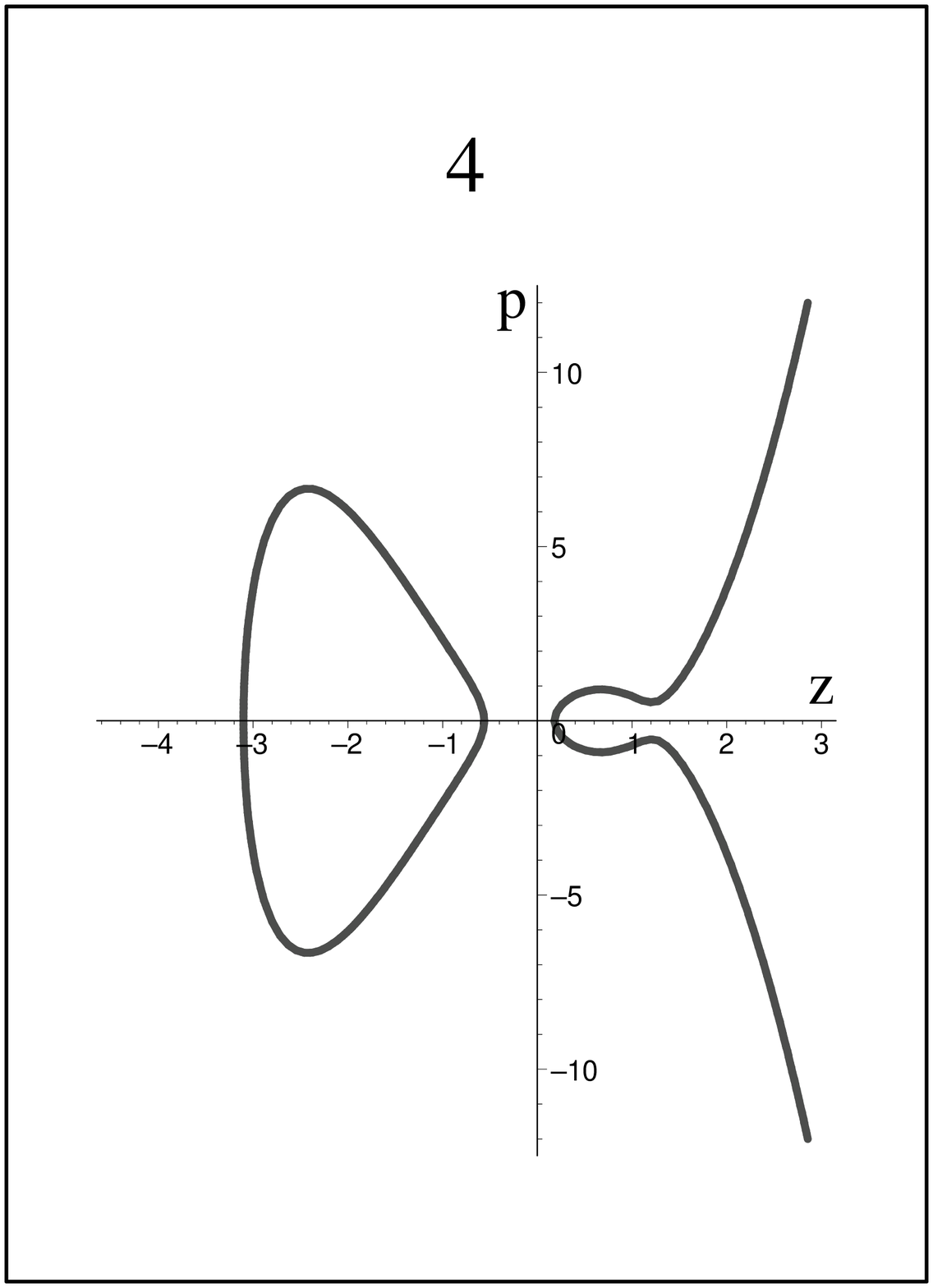} 
& &
\includegraphics[width=4cm, height=5cm]{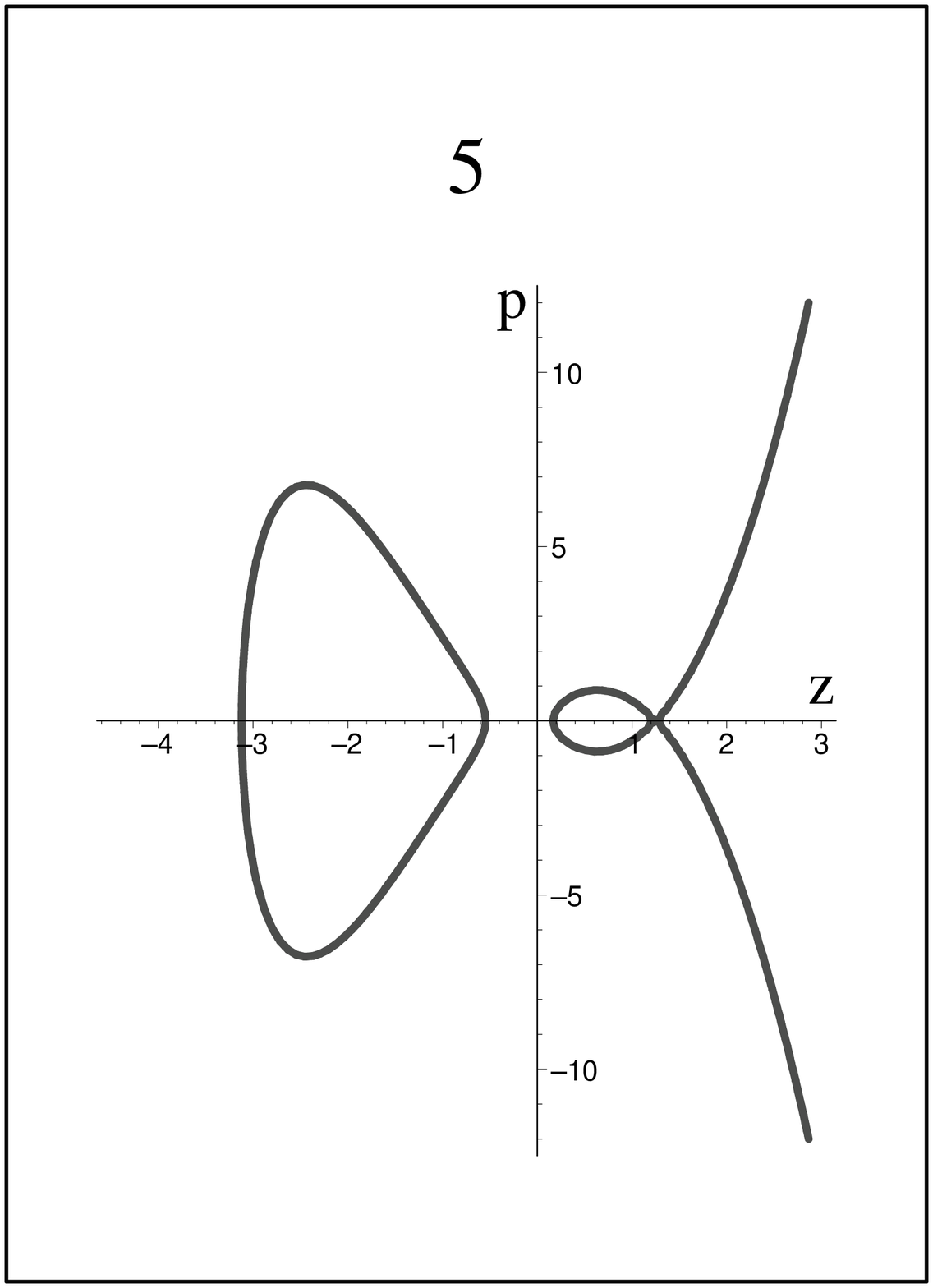}
& &
\includegraphics[width=4cm, height=5cm]{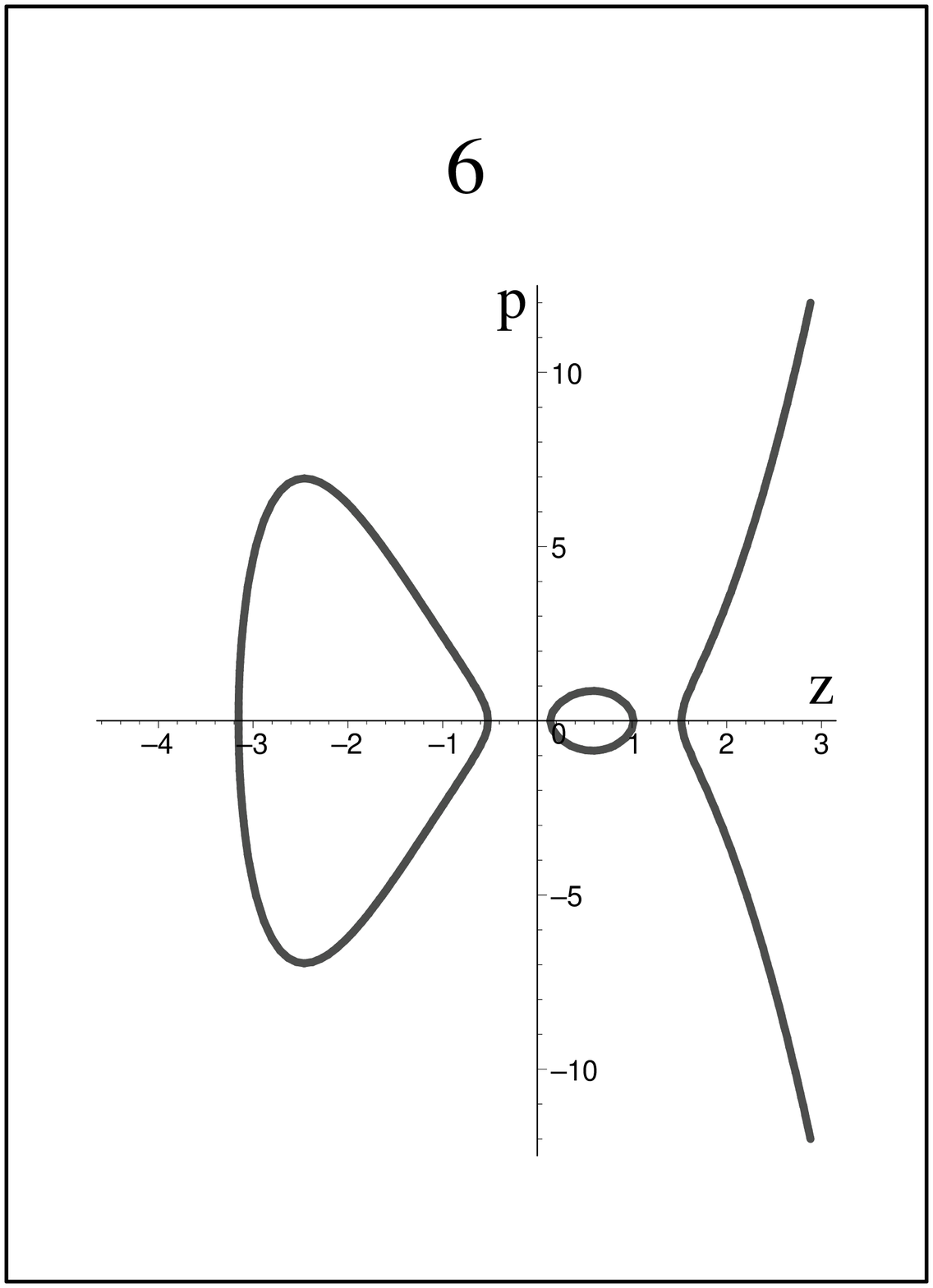}
\\
\includegraphics[width=4cm, height=5cm]{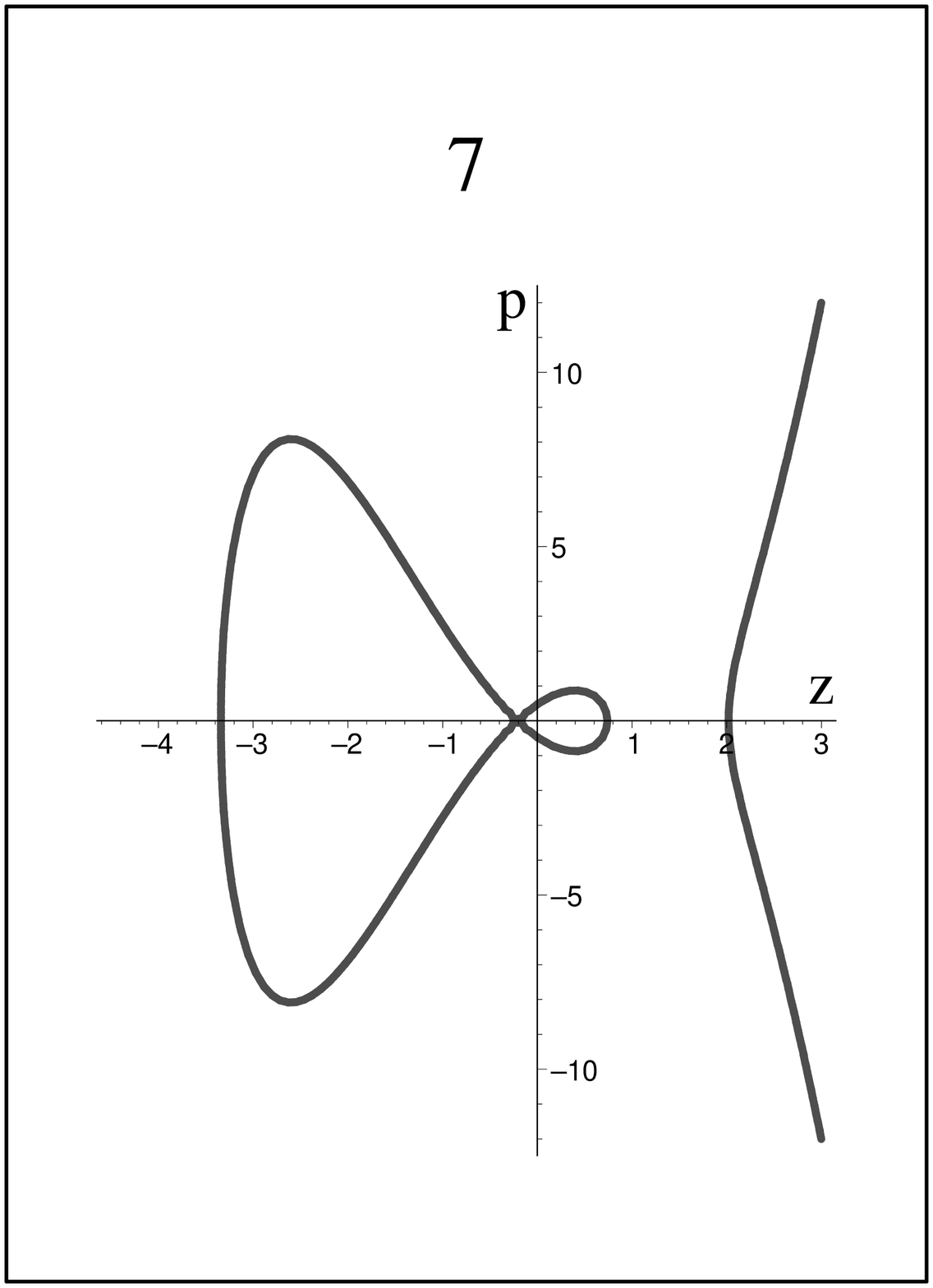}
& &
\includegraphics[width=4cm, height=5cm]{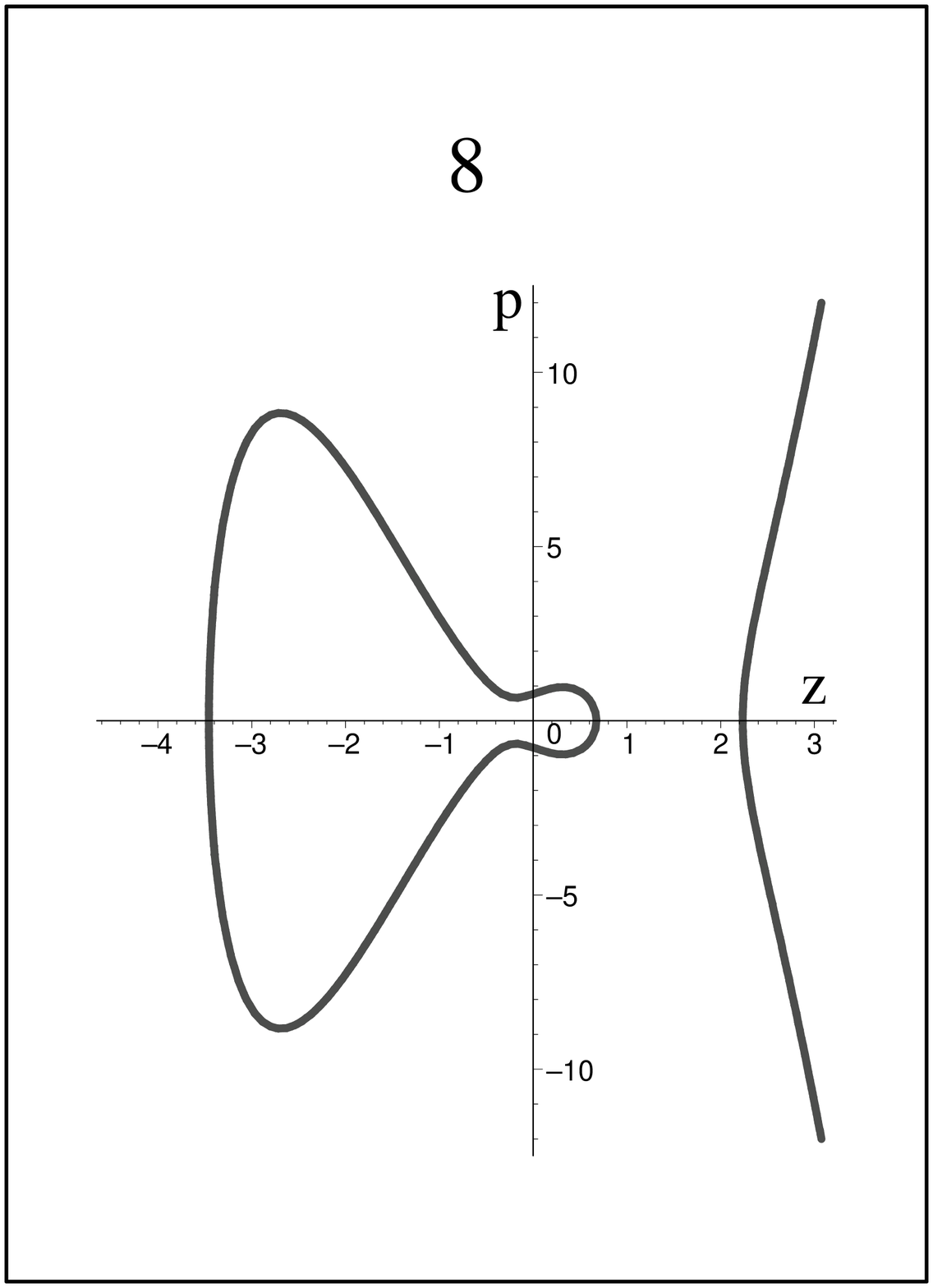}
& &
\includegraphics[width=4cm, height=5cm]{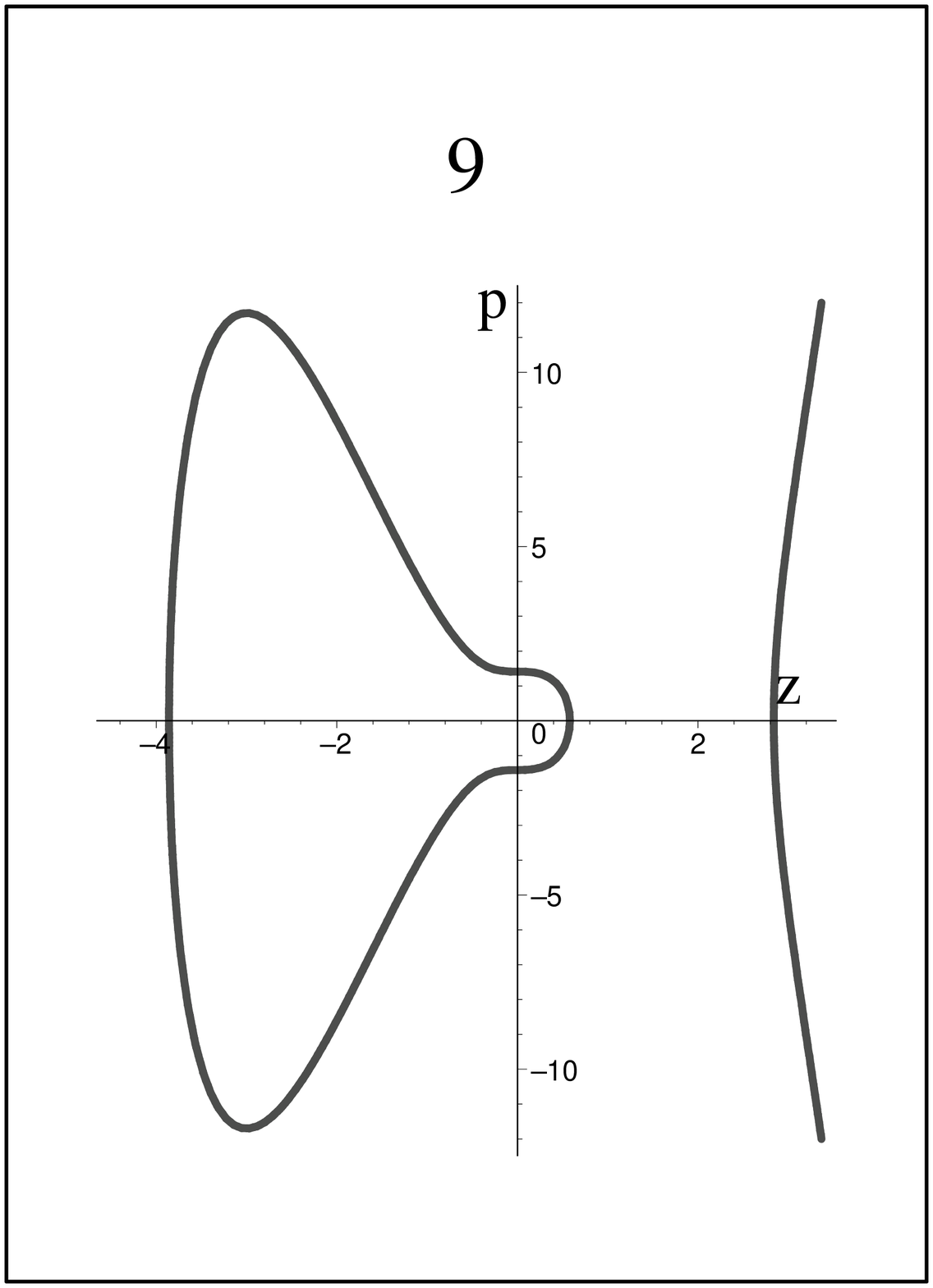}
\end{tabular}
\caption{First plot: $x=7, t=0$, second plot: $x=7, t=-3.6$, third plot: $x=7, t=-5$, fourth plot: $x=7, t=-9$, fifth plot: $x=7, t=-9.11$, sixth plot: $x=7, t=-9.3$, seventh plot: $x=7, t=-10.45$, eight plot: $x=7, t=-11.2$,ninth plot: $x=7, t=-14$.}
\label{fig:2bolle-lin-g2}
\end{figure}

 The analysis of the process of formation of singularities and bubbles presented in sections  \ref{sect-defdef}-\ref{sect-singevo} can be extended to high order (higher genus) plane algebraic curves. Such analysis becomes more involved but, on the other hand, it shows the presence of new phenomena. For instance, for the fifth order curve given by 
\begin{equation}
\label{g2}
 f=p^2-(z^5+\sum_{i=0}^4 u_iz^i)=0
\end{equation}
in addition to the transition regime described above one has a novel regime of creation of two bubbles either one after another  or one from another or simultaneously.
The choice of the Hamiltonian similar to that used in the genus $1$ case, i.e.
\begin{equation}
 H=\left( \frac{u_4}{2}-z\right)p
\end{equation}
gives rise to the following system $(x_1=x)$ (see e.g. \cite{KK})
\begin{equation}
 \label{KdV-g2}
\eqalign{ 
\partial_t u_4 = {u_3}_x-\frac{2}{2}{u_4}_x u_4, \\
\partial_t u_i = {u_{i-1}}_x-\frac{1}{2}{u_{i}}_x u_4-{u_4}_xu_{i}, \qquad i=1,2,3\ ,\\
\partial_t u_0 = -\frac{1}{2}{u_0}_x u_4-{u_4}_x u_0.
}
\end{equation}
In this case equation (\ref{defdefcurve}) becomes
\begin{equation}
 \partial_t f + \{f,H\}=-{u_4}_x f.
\end{equation}
The simplest polynomial solution of this system is
\begin{equation}
 \label{solnlin-KdVg2}
\eqalign{
u_4 = C_4, \\
u_3 = C_3+ A_2 t, \\ 
u_2 = C_2+ \left(A_1 -\frac{1}{2}A_2 C_4\right) t +A_2 x,\\ 
u_1 = C_1+ \left(A_0 -\frac{1}{2}A_1 C_4\right) t   +A_1 x,\\
u_0 = C_0 -\frac{1}{2} A_0C_4 t + A_0 x. 
}
\end{equation}
Even this very simple example depicted in figure \ref{fig:2bolle-lin-g2} in the particular case 
$A_0=A_1=A_2=C_4=1,\ C_0=-12,\ C_1=C_2=0,\ C_3=3$ shows a new phenomenon, i.e. the formation of two bubbles. 
There are also various regimes of formation of singularities and bubbles (included the cusp $(5,2)$) described by the system (\ref{KdV-g2}).\par
For higher order hyperelliptic curves of genus $g$ one observes processes of formation of $g$ bubbles.\par
The Burgers-Hopf (dKdV) hierarchy, again, is relevant for description of the singular sectors of these transition regimes. The process of regularization of corresponding singular curves via the transition to higher Birkhoff strata in Sato Grassmannian and its connection with results of the papers \cite{BAZW,KMWZ,LTW} will be discussed elsewhere.
\section{Conclusion}
\par Approach formulated in the present paper allows us also to introduce a novel class of deformations for the, so called, quadrature (algebraic) domains on the plane. Quadrature domains ($\mathcal{D}$) on the plane are those for which an integral of any function $\phi$ over this domain is the linear superposition of values of the $\phi$ and its derivatives in a finite number of points (see e.g. \cite{AS,Sak}). Such domains show up in many problems of mathematics and fluid mechanics (see e.g. \cite{GuPu,KMWZ,Ric,Sak,MPT}). A particular property of quadrature domains is that their boundaries $\partial \mathcal{D}$ are algebraic curves. Any Hamiltonian deformation of the boundary $\partial \mathcal{D}$ generates a special class of deformations of quadrature domains. One can refer to such deformations as the Hamiltonian (or coisotropic) deformations of quadrature domains. The study of the properties of such deformations, for instance, the analysis of deformations of the quadrature domain data is definitely of interest.\par
 Finally we note that the Hamiltonian deformations can be defined for surfaces and hypersurfaces too. Indeed, let a hypersurface in $\mathbb{C}^n$ be defined by the equation
\begin{equation}
 \label{hypersurf}
f(p_1,\dots,p_n)=0.
\end{equation}
Introducing deformation parameters $x_1,\dots,x_n,t$, one may define Hamiltonian deformations of a hypersurface (\ref{hypersurf}) by the formulae (\ref{defdefcurve}),(\ref{stdPoiss}),(\ref{gendefLiouv}) passing from two to $n$ variables $p_j$ and $x_j$. For algebraic surfaces ($n=3$) and hypersurfaces such deformations are described by the systems of equations of hydrodynamical type. Properties of such systems and corresponding Hamiltonian deformations are worth to study.

\end{document}